\documentclass[useAMS,usenatbib]{mn2e}

\usepackage{graphicx}
\usepackage{subfigure}
\usepackage{paralist}
\usepackage{amssymb}
\usepackage{bm}
\usepackage{bbding}
\usepackage[fleqn]{amsmath}
\usepackage{multirow}
\usepackage{epsfig}
\usepackage{epstopdf}
\usepackage[colorlinks,
            linkcolor=blue, 
            anchorcolor=blue, 
            citecolor=blue, 
            ]{hyperref}
\title[Hamiltonian models]{Multi-harmonic Hamiltonian models with applications to first-order resonances}
\author[Lei \& Li]{Hanlun Lei$^{1,2}$\thanks{E-mail: leihl@nju.edu.cn}, Jian Li$^{1,2}$\thanks{E-mail: ljian@nju.edu.cn}\\
$^{1}$ School of Astronomy and Space Science, Nanjing University, Nanjing 210023, China\\
$^{2}$ Key Laboratory of Modern Astronomy and Astrophysics in Ministry of Education, Nanjing University, Nanjing 210023, China}

\begin{document}

\date{Accepted. Received; in original form}

\pagerange{\pageref{firstpage}--\pageref{lastpage}} \pubyear{2020}

\maketitle
\label{firstpage}

\renewcommand{\arraystretch}{1.3}

\begin{abstract}

In this work, two multi-harmonic Hamiltonian models for mean motion resonances are formulated and their applications to first-order resonances are discussed. For the $k_p$:$k$ resonance, the usual critical argument $\varphi = k \lambda - k_p \lambda_p + (k_p - k) \varpi$ is taken as the resonant angle in the first model, while the second model is characterized by a new critical argument $\sigma = \varphi / k_p$. Based on canonical transformations, the resonant Hamiltonians associated with these two models are formulated. It is found that the second Hamiltonian model holds two advantages in comparison to the first model: (a) providing a direct correspondence between phase portraits and Poincar\'e sections, and (b) presenting new phase-phase structures where the zero-eccentricity point is a visible saddle point. Then, the second Hamiltonian model is applied to the first-order inner and outer resonances, including the 2:1, 3:2, 4:3, 2:3 and 3:4 resonances. The phase-space structures of these first-order resonances are discussed in detail and then the libration centers and associated resonant widths are identified analytically. Simulation results show that there are pericentric and apocentric libration zones where the libration centers diverge away from the nominal resonance location as the eccentricity approaches zero and, in particular, the resonance separatrices do not vanish at arbitrary eccentricities for both the inner and outer (first-order) resonances.

\end{abstract}

\begin{keywords}
celestial mechanics--planets and satellites: dynamical evolution
and stability--methods: analytical
\end{keywords}

\section{Introduction}
\label{Sect1}

Mean motion resonance (MMR) is a fundamental mechanism in celestial mechanics and it occurs when the mean motion frequencies of two planets moving around a central star are close to the ratio of integers. In the region where a certain MMR is active, the gravitational interactions between planets are enhanced, making the associated dynamics be quite complicated. It is known that MMR plays a crucial role in the long-term stability of planetary systems, because the MMR provides a protection mechanism so that the distance between planets can not reach the possible minimum \citep{murray1999solar, wan2003resonance}. In addition, MMRs are thought as the dynamical source of some phenomena arising in practical systems, such as the existence of the Hecuba gap and the Hilda group in the asteroid belt \citep{murray1986structure, murray1999solar}. Thus, for a certain MMR, it is of significance to understand (a) the place where the resonance takes, (b) the size that measures the resonance zone and (c) the global structure in phase space. Usually, there are analytical and numerical approaches to explore the dynamics of MMRs.

From the analytical point of view, there are several fundamental models of resonance. With $(r,R)$ as a pair of conjugate variables, the pendulum model with the Hamiltonian function
\begin{equation*}
{\cal H}=\frac{1}{2} R^2 - \cos{r}
\end{equation*}
is adopted as the basic model of the so-called ideal resonance problem \citep{garfinkel1966formal}. \citet{henrard1983second} called the pendulum model the first fundamental model of resonance. Considering the fact that the force amplitude of the cosine term of $r$ is proportional to the eccentricity or inclination, \citet{henrard1983second} introduced a second fundamental model of resonance (called `SFMR') with the Hamiltonian function as
\begin{equation*}
{\cal H} = -3(\delta + 1) R + R^2 - 2\sqrt{2R} \cos{r}
\end{equation*}
where $\delta$ is a free parameter. A large varieties of resonance problems covering from the orbit-orbit to orbit-spin resonance problems in celestial mechanics can be approximated to the standard model `SFMR', which is specified by the parameter $\delta$. Later, \citet{breiter2003extended} pointed out that the model `SFMR' cannot reproduce the separatrix bifurcations (or known as `saddle connections' and 'heteroclinic bridges'). In order to extend the model `SFMR', \citet{breiter2003extended} introduced an extended fundamental model of resonance (EFMR) with the Hamiltonian function as
\begin{equation*}
{\cal H} = R^3 + \frac{1}{2}u R^2 + v R + \sqrt{2R} \cos{r}
\end{equation*}
where $u$ and $v$ are two free parameters. Besides, there are some other analytical models of resonance, e.g. the fundamental model of high-order resonances \citep{lemaitre1984high}, the third fundamental model of resonance \citep{shinkin1995integrable}, the second fundamental model of resonance with asymmetric equilibria \citep{jancart2002second}, and so on.

Among these fundamental models, there is a common feature: they hold one degree of freedom with the resonant angle $r$ as the unique angular coordinate. In other words, all of them are totally integrable (the solution can be expressed by means of elliptic integral). The resonant motion occurs on the level curves of Hamiltonian, so that the phase portraits could reveal the global behaviors. In the phase portraits, there are stable equilibria (corresponding to libration centers) and unstable equilibria (corresponding to saddle points). The isoline of Hamiltonian stemming from the saddle points play the role of dynamical separatrix, dividing the phase space into domains of libration and circulation. Usually, the libration zone centered at a libration center is bounded by its nearby separatrix, thus the size of libration zone can be measured by the distance between a pair of separatrices.

\citet{winter1997resonanceI} reviewed various analytical models including the pendulum model (the first fundamental model of resonance) and the second fundamental model of resonance, and the authors applied these analytical models to the major first-order interior resonances. Regarding the 2:1, 3:2 and 4:3 resonances, \citet{winter1997resonanceI} found that both the analytical models predict the overlap of nearby libration zones at low eccentricities (leading to chaotic behaviors in low-eccentricity domains). Similar results can be found in \citet{murray1999solar} (see Fig. 8.7 in the textbook). Regarding the inner 2:1 resonance, \citet{morbidelli2002modern} adopted the `SFMR' to discuss the resonance problems and described that (a) the location of the libration center diverge on the left side of the nominal resonance location as $e \to 0$ and (b) one separatrix vanishes when the motion integral is smaller than a critical value (or the eccentricity is smaller than 0.2). Due to the absence of one separatrix, the resonant width is undefined in small-eccentricity regions \citep{morbidelli2002modern}. The absence of one separatrix is also observed by \citet{ramos2015resonance} for the inner 2:1 resonance. In addition, there are some other analytical or semi-analytical works on the issue of mean motion resonances performed in various environments \citep{gallardo2019strength, gallardo2020three, lei2019three}.

From the numerical viewpoint, in the planar circular restricted three-body problem, the location of libration center and the associated resonant width can be identified by analyzing the Poincar\'e surfaces of sections, as performed in \citet{malhotra1996phase,wang2017mean,malhotra2018neptune,lan2019neptune,malhotra2020divergence} and \citet{winter1997resonanceI,winter1997resonanceII}. In the works of \citet{malhotra1996phase} and \citet{winter1997resonanceI,winter1997resonanceII}, the Poincar\'e sections are defined by means of $y=0$ and $\dot y >0$ where $y$ and $\dot y$ are the state variables measured in the rotating reference frame, and the Poincar\'e sections are usually presented in the $(x,\dot x)$ plane or in the $(M,a)$ plane (here $M$ is the mean anomaly and $a$ is the semimajor axis). By using both the numerical method based on Poincar\'e sections and analytical approach (i.e. `SFMR'), \citet{winter1997resonanceI} compared the numerical and analytical results of resonant widths for the inner 2:1, 3:2 and 4:3 resonances (see Fig. 11 in their work) and found some discrepancies.

Concerning the production of Poincar\'e sections, Malhotra and her collaborators made an important improvement \citep{wang2017mean}: recording the test particles state vectors at every successive perihelion passage (actually, the section is defined by means of $\dot r = 0$ and $\ddot r > 0$, where $r$ is the radial distance of the test particle relative to the central star). For convenience, they introduced an angular separation, denoted by $\psi = \varpi - \lambda_p$ ($\varpi$ is the longitude of pericenter and $\lambda_p$ is the mean longitude of the planet), to measure the relative angle between the test particle at the perihelion and the planet. As stated by \citet{malhotra2020divergence}, such a change is very important because it yields a more direct visualization and physical interpretation for the resonance zones arising in the Poincar\'e surfaces of section. Regarding the first-order inner resonances, \citet{malhotra2020divergence} analyzed the structures arising in the sections and numerically identified the widths of pericentric and apocentric libration zones. For those first-order inner resonances, they reported two interesting and novel conclusions: (a) the resonance separatrix does not vanish at low eccentricities  and (b) the ``bridges'' of libration zones exist between adjacent resonances.

Motivated by the results obtained by analyzing Poincar\'e sections in \citet{malhotra2020divergence}, we may ask: (a) can we reproduce all the numerical results for first-order inner resonances through analytical methods? (b) what will happen for the first-order outer resonances? To answer these questions, we formulate two multi-harmonic Hamiltonian models based on the Laplacian expansion of planetary disturbing function for the $k_p$:$k$ resonances: (i) in the first model the usual critical argument defined by $\varphi = k \lambda - k_p \lambda_p + (k_p - k)\varpi$ is taken as the resonant angle (thus this model is in accordance to the previous fundamental models and the only difference lies in the presence of higher-harmonic terms), and (ii) in the second model the new critical argument defined by $\sigma = \varphi/k_p$ is taken as the resonant angle. Comparing the phase structures produced from these two Hamiltonian models, we find an important advantage of the second Hamiltonian model is that we can make a direct correspondence between the Poincar\'e sections produced in \citet{malhotra2020divergence} and phase portraits in the analytical model (this is a key motivation of the present work). Then, we apply our new (second) Hamiltonian models to both the first-order inner and outer resonances with a Jupiter-mass planet and, indeed, we achieve our goals: (a) we can reproduce numerical results given in \citet{malhotra2020divergence} for the first-order inner resonances by means of analytical method and (b) analytical results are produced for the first-order outer resonances. It should be noted that, besides the first-order resonances, our Hamiltonian models are also applicable for high-order mean motion resonances.

The remaining part of this work is organized as follows. In Section \ref{Sect2}, the Hamiltonian function of the planar circular restricted three-body problem is briefly introduced and, in Section \ref{Sect3}, two multi-harmonic Hamiltonian models are formulated for mean motion resonances. In Section \ref{Sect4}, the Hamiltonian models are directly compared and their features are discussed in detail. The (second) multi-harmonic Hamiltonian model is applied to the first-order inner and outer resonances in Section \ref{Sect5}. At last, the summary and discussion are provided in Section \ref{Sect6}.

\section{Hamiltonian function}
\label{Sect2}

In this study, we concentrate on the motion of a test particle (e.g. an asteroid in our Solar system) in the planar circular restricted three-body problem with the Sun and a giant planet as the massive and second primaries. In this approximation, all the objects considered are in a common plane and the planet moves around the Sun in a circular orbit. The motion of the test particle is governed by the gravitational attractions generated by the primaries. The test particle moves around the Sun in an osculating Keplerian orbit perturbed by the gravitational attraction coming from the planet. For convenience of description, we denote the Sun as the central body and the planet as the perturber.

Usually, the time and space variables used in the entire work are normalized by taking the total mass of the Sun and planet as the unit of mass, the distance between the Sun and planet as the unit of length and the orbital period of the planet divided by $2 \pi$ as the unit of time. Under the system of normalized units, both the universal gravitational constant $\cal {G}$ and the mean motion frequency of the planet $n_p$ become unitary (i.e. ${\cal G} = 1$ and $n_p = 1$ in normalized units).

In the following investigations, we describe the motion of the objects involved under a Sun-centered inertial coordinate system with the orbit of the planet as the fundamental plane. Under this reference frame, the planet moves on a unitary circle (i.e. $a_p = 1$ in normalized units) and its position is determined by the mean anomaly $\lambda_p = M_p + \varpi_p$ where $M_p$ is the mean anomaly and $\varpi_p$ is the longitude of pericentre. In addition, the orbit of the test particle is described by the classical elements: the semimajor axis $a$, eccentricity $e$, longitude of pericenter $\varpi$ and the mean anomaly $M$ (or the true anomaly $f$).

In this study, the Sun--Jupiter system is taken as the basic model to perform practical simulations (under other Sun--planet systems simulations can be performed in a similar manner). In this system, the length unit is 5.2 $\rm{au}$, the time unit is 688.995 $\rm{d}$. The normalized mass of the Sun is $m_0 = 0.9990461188$ and the normalized mass of Jupiter becomes $m_p = 1 - m_0$.

\subsection{Expansion of disturbing function}
\label{Sect2-1}

In the Sun-centered inertial coordinate system, the motion of the test particle is governed by the disturbing function, given by \citep{murray1999solar}
\begin{equation}\label{Eq1}
{\cal R} = {\cal G}{m_p}\left( {\frac{1}{\Delta } - \frac{r}{r_p^2}\cos \psi } \right),
\end{equation}
where $r_p$ is the distance between the planet and the central star ($r_p = a_p = 1$ holds in the CRTBP with normalized units) and $\Delta$ is the relative distance between the planet and the test particle, expressed by
\begin{equation*}
\Delta  = \sqrt {r_p^2 + {r^2} - 2r r_p \cos \psi } = \sqrt {1 + {r^2} - 2r\cos \psi }
\end{equation*}
with $r$ as the distance of the test particle relative to the Sun and $\psi$ as the separation angle between the radius vectors of the test particle and planet, given by
\begin{equation*}
\psi  = f + \varpi  - {\lambda _p}.
\end{equation*}

Following the procedure discussed in \citet{murray1999solar} and \citet{ellis2000disturbing}, in the planar configuration the disturbing function represented by equation (\ref{Eq1}) can be expanded in a formal series of the orbital elements as follows:
\begin{equation}\label{Eq2}
\begin{aligned}
{\cal R} & = {\cal G}{m_p}\sum\limits_{n = 0}^\infty  {\sum\limits_{j =  - \infty }^\infty  {\sum\limits_{m = 0}^n {\sum\limits_{s =  - \infty }^\infty  {{{\left( { - 1} \right)}^{n - m}}{A_{n,j}\left( \alpha \right)} {n \choose m} X_s^{m,j}\left( e \right)} } } } \\
& \times \cos \left[ {s\lambda  + \left( {j - s} \right)\varpi  - j{\lambda _p}} \right]\\
& - \frac{{\cal G}{m_p}}{a_p} \alpha \sum\limits_{s =  - \infty }^\infty  {X_s^{1,1}\left( e \right)\cos \left[ {s\lambda  + \left( {1 - s} \right)\varpi  - {\lambda _p}} \right]}
\end{aligned}
\end{equation}
where the semimajor axis ratio ($\alpha = a/a_p$) related function ${A_{n,j}\left( \alpha \right)}$ is defined by means of the Laplace coefficients $b_{1/2}^j \left( \alpha \right)$ in the following manner:
\begin{equation*}
{A_{n,j}}\left( \alpha  \right) = \frac{1}{2}\frac{{{\alpha ^n}}}{{n!}}\left[ {\frac{{{{\rm d}^n}}}{{{\rm d}{\alpha ^n}}}b_{{1 \mathord{\left/
 {\vphantom {1 2}} \right.
 \kern-\nulldelimiterspace} 2}}^{\left( j \right)}(\alpha )} \right]
\end{equation*}
and the Hansen coefficients $X_s^{m,j}\left( e \right)$ are functions of the eccentricity, calculated by \citep{hughes1981computation}
\begin{equation*}
X_c^{a,b}\left( e \right) = {e^{\left| {c - b} \right|}}\sum\limits_{s = 0}^\infty  {Y_{s + t,s + u}^{a,b}{e^{2s}}},
\end{equation*}
with $t=\max(0,c-b)$ and $u=\max(0,b-c)$. In particular, $Y_{s+t,s+u}^{a,b}$ is the Newcomb operator, which can be computed in a recurrence manner \citep{hughes1981computation, murray1999solar}.

\subsection{Hamiltonian function}
\label{Sect2-2}

For convenience, we introduce the modified Delaunay variables to study the resonance dynamics \citep{morbidelli2002modern},
\begin{equation}\label{Eq3}
\begin{aligned}
&\Lambda   = \sqrt {\mu a} ,\quad \lambda  = M + \varpi,\\
&P  = \sqrt {\mu a} \left( {1 - \sqrt {1 - {e^2}} } \right),\quad p =  - \varpi,\\
&{\Lambda _p} ,\quad {\lambda _p} = {M_p} + {\varpi _p},
\end{aligned}
\end{equation}
where $\mu = {\cal G} m_0$. In the disturbing function given by equation (\ref{Eq2}), all the classical elements $a$, $e$, $\varpi$, $\lambda$ and $\lambda_p$ are replaced by the modified Delaunay's variables. As a result, the Hamiltonian function, governing the motion of the test particle, can be written as follows:
\begin{equation}\label{Eq4}
\begin{aligned}
{\cal H} &= - \frac{\mu }{{2a}} + {n_p}{\Lambda _p} - {\cal R}\left( {a,e,\lambda ,\varpi ,{\lambda _p}} \right)\\
&=  - \frac{{{\mu ^2}}}{{2{\Lambda ^2}}} + {n_p}{\Lambda _p} - {\cal R}\left( {\Lambda ,P,\lambda ,p,{\lambda _p}} \right),
\end{aligned}
\end{equation}
where $n_p$ is equal to unity in normalized units, $\Lambda _p$ is the conjugate momentum of $\lambda_p$, and the expression of disturbing function ${\cal R}$ is provided by equation (\ref{Eq2}). Evidently, the dynamical model specified by the Hamiltonian given by equation (\ref{Eq4}) is of three degree of freedom with $\lambda$, $p (=  - \varpi)$ and $\lambda_p$ as angular coordinates.

\section{Multi-harmonic Hamiltonian models}
\label{Sect3}

Based on the Hamiltonian function given in the previous section, we intend to establish two Hamiltonian models with multiple harmonics of the critical argument for the $k_p$:$k$ resonances. The first Hamiltonian model takes the `usual critical argument' $\varphi$ as the resonant angle (the corresponding model corresponds to the classical fundamental models and the difference lies in its presence of high-harmonic terms in the Hamiltonian function), while the second model takes a new critical argument, denoted by $\sigma = \frac{\varphi}{k_p}$, as the resonant angle.

\subsection{The first multi-harmonic Hamiltonian model}
\label{Sect3-1}

Concerning a test particle located inside the $k_p$:$k$ resonance with planet, the usual resonant angle is introduced by
\begin{equation}\label{Eq5}
\varphi  = k\lambda  - {k_p}{\lambda _p} + \left( {{k_p} - k} \right)\varpi,
\end{equation}
so that the mean anomaly of the test particle, $\lambda$, can be expressed by means of $\varphi$, $\lambda_p$ and $\varpi$ as follows:
\begin{equation}\label{Eq6}
\lambda  = \frac{1}{k}\varphi  + \frac{{{k_p}}}{k}{\lambda _p} - \frac{{{k_p} - k}}{k}\varpi.
\end{equation}
For first-order resonances, it holds $\left| {k_p - k} \right| = 1$. By putting equation (\ref{Eq6}) into equation (\ref{Eq2}) and combining the modified Delaunay's variables, the disturbing function can be expressed as ${\cal R} (\Lambda, P, \varphi, \varpi, \lambda_p)$.

When the test particle is located inside the $k_p$:$k$ resonance, the associated resonant angle $\varphi$ becomes a slow angular variable compared to $\lambda_p$, so that the disturbing function can be separated into short-period terms containing the fast variable $\lambda_p$ and long-period terms without $\lambda_p$. When we are studying the resonant dynamics, it is usual to remove those short-period terms from the disturbing function by means of averaging theory \citep{gallardo2006Atlas},
\begin{equation}\label{Eq7}
{{\cal R}^{{*}}} = \frac{1}{{2k\pi }}\int\limits_0^{2k\pi } {{\cal R}{\rm d}{\lambda _p}},
\end{equation}
which leads to the averaged resonant disturbing function.

Replacing the expression of disturbing function given by equation (\ref{Eq2}) in equation (\ref{Eq7}), we can easily obtain the analytical expression of the resonant disturbing function. In particular, when $k_p = 1$ (corresponding to the 1:$k$-type resonances), the resonant disturbing function ${\cal R}^*$ can be written as
\begin{small}
\begin{equation*}
\begin{aligned}
{{\cal R}^{{*}}} &= {\cal G}{m_p}\sum\limits_{n = 0}^\infty  {\sum\limits_{j =  - \infty }^\infty  {\sum\limits_{m = 0}^n {{{\left( { - 1} \right)}^{n - m}}{A_{n,j}}\left( \alpha  \right) {n \choose m} X_{jk}^{m,j}\left( e \right)\cos \left( {j\varphi } \right)} } } \\
& - \frac{{\cal G}{m_p}}{a_p} {\alpha} X_k^{1,1}\left( e \right)\cos \varphi
\end{aligned}
\end{equation*}
\end{small}
and, when $k_p \ne 1$ (in this case the indirect part of disturbing function has no contribution to the resonant disturbing function), the resonant disturbing function ${\cal R}^*$ can be obtained as
\begin{equation*}
\begin{aligned}
{{\cal R}^{{*}}} &= {\cal G}{m_p}\sum\limits_{n = 0}^\infty  {\sum\limits_{\scriptstyle {j =  - \infty  \to \infty} \hfill\atop
\scriptstyle\bmod (j,{k_p}) = 0\hfill} {\sum\limits_{m = 0}^n {{{\left( { - 1} \right)}^{n - m}}{A_{n,j}}\left( \alpha  \right) {n \choose m} } } } \\
&\times X_{jk/{k_p}}^{m,j}\left( e \right)\cos \left( {\frac{j}{{{k_p}}}\varphi } \right)
\end{aligned}
\end{equation*}
where $\bmod (j,{k_p}) = 0$ means that $j$ is divisible by $k_p$. For the purpose of simplification, we denote the resonant disturbing function truncated at order $N$ in eccentricity by a compact form as follows:
\begin{equation}\label{Eq8}
{{\cal R}^{\rm{*}}} = \sum\limits_{n = 0}^N {{{\cal C}_n}\cos (n\varphi )}
\end{equation}
where $N$ is the number of harmonics of the angle $\varphi$ and the coefficients ${\cal C}_n$ are related to the action variables $\Lambda$ and $P$ (or, equivalently, the elements $a$ and $e$). As a result, the averaged resonant Hamiltonian can be expressed by
\begin{equation}\label{Eq9}
\begin{aligned}
{{\cal H}^*} &=  - \frac{{{\mu ^2}}}{{2{\Lambda ^2}}} + {n_p}{\Lambda _p} - {{\cal R}^*}\\
&=  - \frac{{{\mu ^2}}}{{2{\Lambda ^2}}} + {n_p}{\Lambda _p} - \sum\limits_{n = 0}^N {{{\cal C}_n}\cos (n\varphi )}.
\end{aligned}
\end{equation}
Evidently, the number $N$ determines the accuracy of the resonant Hamiltonian. The influence of $N$ upon the phase portraits of mean motion resonances is to be discussed in Section \ref{Sect4} in detail.

In the following study, the dynamical model with $N=1$ is called the one-harmonic model, the one with $N=2$ is called the two-harmonic model and the ones with $N \ge 3$ are called the $N$-harmonic models. Alternatively, the averaged resonant disturbing function given by equation (\ref{Eq7}) can be obtained by means of direct numerical integration \citep{gallardo2006Atlas}, and the corresponding dynamical model is called the numerical resonant model.

To formulate the resonant model, we need to introduce a new set of canonical variables,
\begin{equation}\label{Eq10}
\begin{aligned}
&{\Phi _1} = \frac{1}{k}\Lambda ,\quad {\varphi _1} = k\lambda  - {k_p}{\lambda _p} - ({k_p} - k)p = \varphi,\\
&{\Phi _2} = P + \frac{{{k_p} - k}}{k}\Lambda ,\quad {\varphi _2} = p,\\
&{\Phi _3} = {\Lambda _p} + \frac{{{k_p}}}{k}\Lambda ,\quad {\varphi _3} = {\lambda _p},
\end{aligned}
\end{equation}
which can be transformed from the set of modified Delaunay's variables through the following generating function:
\begin{equation*}
{\cal S} = k\lambda {\Phi _1} + {\lambda _p}\left( {{\Phi _3} - {k_p}{\Phi _1}} \right) + p\left[ {{\Phi _2} - ({k_p} - k){\Phi _1}} \right].
\end{equation*}
Under the new set of variables defined by equation (\ref{Eq10}), the resonant Hamiltonian given by equation (\ref{Eq9}) becomes
\begin{equation}\label{Eq11}
{{\cal H}^*} =  - \frac{{{\mu ^2}}}{{2{{\left( {k{\Phi _1}} \right)}^2}}} - {k_p}{n_p}{\Phi _1} - \sum\limits_{n = 0}^N {{{\cal C}_n}\left( {{\Phi _1},{\Phi _2}} \right)\cos (n{\varphi _1})},
\end{equation}
where the constant terms have been eliminated from the resonant Hamiltonian. Evidently, the resonant model determined by equation (\ref{Eq11}) has a single degree of freedom with $\varphi_1 (= \varphi)$ as the angular coordinate, so that the dynamical model becomes totally integrable. In addition, the angular variables $\varphi_2$ and $\varphi_3$ are absent from the resonant Hamiltonian (i.e., $\varphi_2$ and $\varphi_3$ are cyclic coordinates of the current resonant model), thus their conjugate momenta $\Phi_2$ and $\Phi_3$ become the motion integral, given by
\begin{equation}\label{Eq12}
{\Phi _2} = P + \frac{{{k_p} - k}}{k}\Lambda  = \sqrt {\mu a} \left( {\frac{{{k_p}}}{k} - \sqrt {1 - {e^2}} } \right) = {\rm const}
\end{equation}
and
\begin{equation}\label{Eq13}
{\Phi _3} = {\Lambda _p} + \frac{{{k_p}}}{k}\sqrt {\mu a} = {\rm const}.
\end{equation}
The motion integral given by equation (\ref{Eq12}) means that, in the long-term evolution, the Keplerian energy of the test particle exchanges with its angular momentum, implying that there is a coupled oscillation between the semimajor axis and eccentricity in the long-term evolution.

\begin{figure}
\centering
\includegraphics[width=0.235\textwidth]{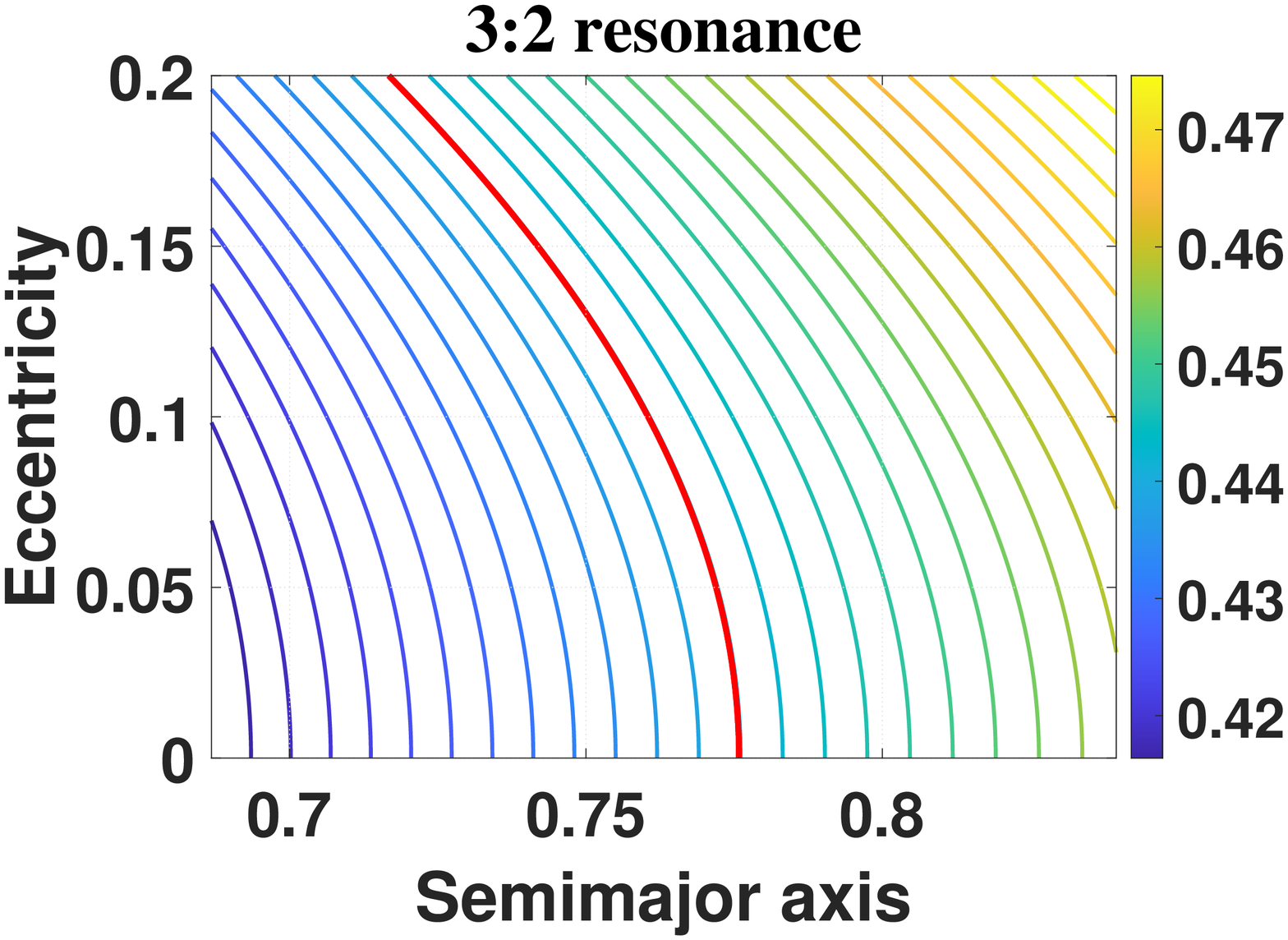}
\includegraphics[width=0.235\textwidth]{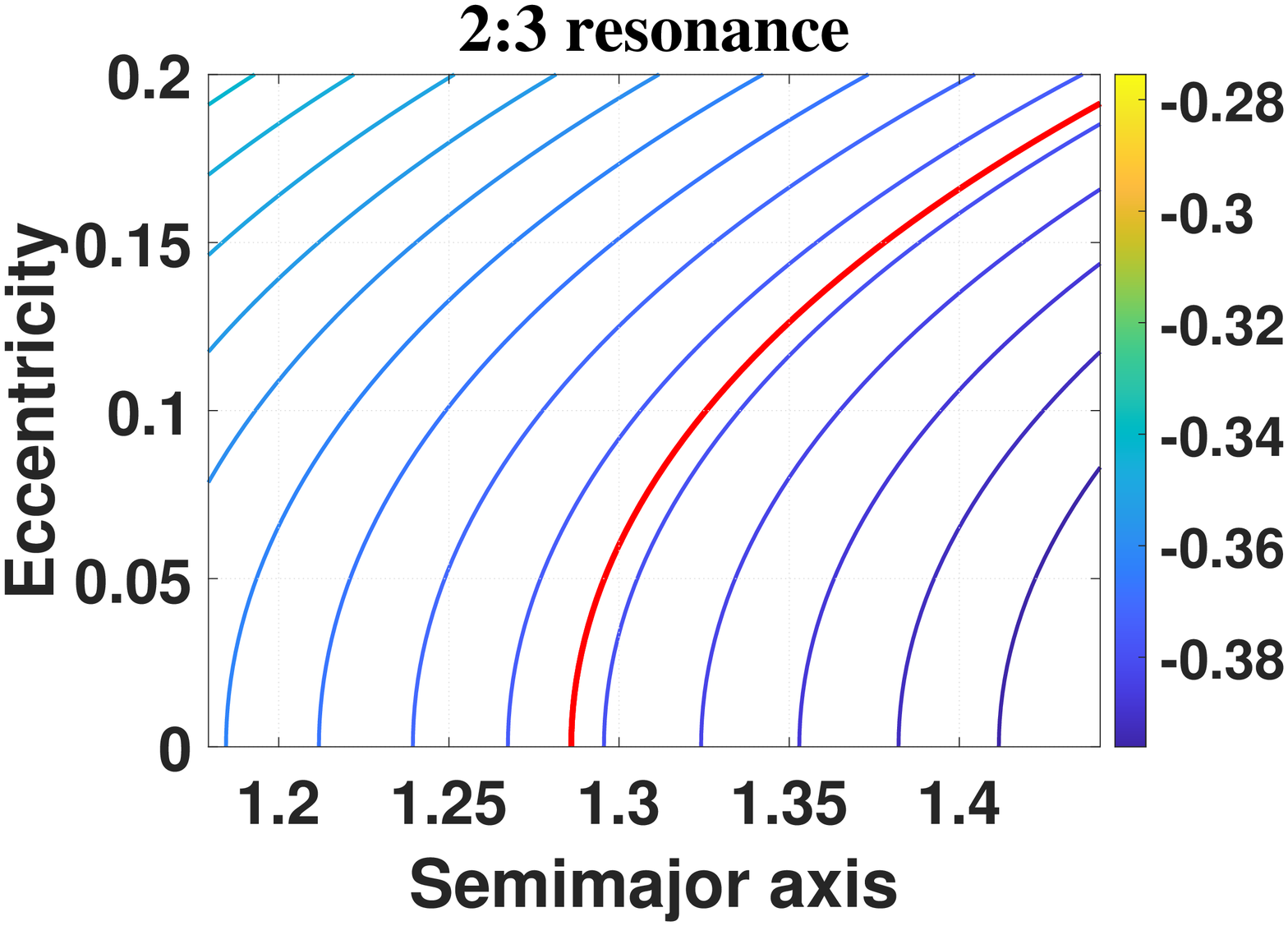}
\caption{Level curves of the motion integral ${\Phi_2} (= \Gamma _2) = P + \frac{{{k_p} - k}}{k}\Lambda  = \sqrt {\mu a} \left( {\frac{{{k_p}}}{k} - \sqrt {1 - {e^2}} } \right)$ in the space spanned by the semimajor axis and eccentricity for the inner 3:2 (\emph{left panel}) and outer 2:3 (\emph{right panel}) resonances. The motion integral $\Phi_2$ is defined by equation (\ref{Eq12}) in the first model and the motion integral $\Gamma_2$ is defined by equation (\ref{Eq19}) in the second model. The red lines shown in both panels corresponds to ${\Phi_2} (=\Gamma _2) = 0.4404$ (i.e., $a_{\max} = 0.7758$) and ${\Phi_2} (=\Gamma _2) = -0.378$ (i.e., $a_{\min} = 1.2860$), which are to be used in Figs. \ref{Fig1A} and \ref{Fig1B}. Normalized units are used for the semimajor axis.}
\label{Fig1}
\end{figure}

In Fig. \ref{Fig1}, the level curves of $\Phi_2$ are presented in the $(a,e)$ plane for the inner 3:2 and outer 2:3 resonances with a Jupiter-mass planet. For the inner 3:2 resonance, the magnitude of $\Phi_2$ is greater than zero, while the magnitude of $\Phi_2$ for the outer 2:3 resonance is smaller than zero. The motion integral shows that the test particle can only move along the isoline of $\Phi_2$ determined by its initial condition. In other words, only one of the elements ($a$ and $e$) is independent when $\Phi_2$ is given.

In the expression of the motion integral, if we assume $e=0$, we could obtain the maximum value of semimajor axis denoted by $a_{\max}$ for inner resonances and obtain the minimum value of semimajor axis denoted by $a_{\min}$ for outer resonances. Thus, for the inner resonances where $k_p > k$, there is a one-to-one correspondence between $a_{\max}$ and $\Phi_2$ and, for the outer resonances where $k_p < k$, there is also a one-to-one correspondence between $a_{\min}$ and $\Phi_2$. For the purpose of intuition, we will also use $a_{\max}$ (or $a_{\min}$) to stand for the motion integral in the following discussions.

It should be noted that the Hamiltonian model with $N=1$ or $N=2$ discussed in the current work has been widely used in previous analytical studies \citep{henrard1983mechanism, henrard1983second, lemaitre1984high, beauge1994asymmetric, winter1997resonanceI, winter1997resonanceII, morbidelli2002modern, jancart2002second, breiter2003extended, ramos2015resonance}. Compared to the SFMR (the second fundamental model for resonance) discussed in \citet{henrard1983mechanism} and the EFMR (the extended fundamental model of resonance) discussed in \citet{breiter2003extended}, our Hamiltonian model discussed here includes multiple harmonics of the resonant angle $\varphi$ in the Hamiltonian (in practice, the number of harmonics of $\varphi$ is controlled by $N$). It is to be noted that, in order to reduce to the fundamental models of resonance including `SFMR' and `EFMR', the resonant Hamiltonian needs to be expanded around the libration center by Taylor series, and then the expansion is truncated at order 2 in the model `SFMR' and order 3 in the model `EFMR'.

\begin{figure}
\centering
\includegraphics[width=0.235\textwidth]{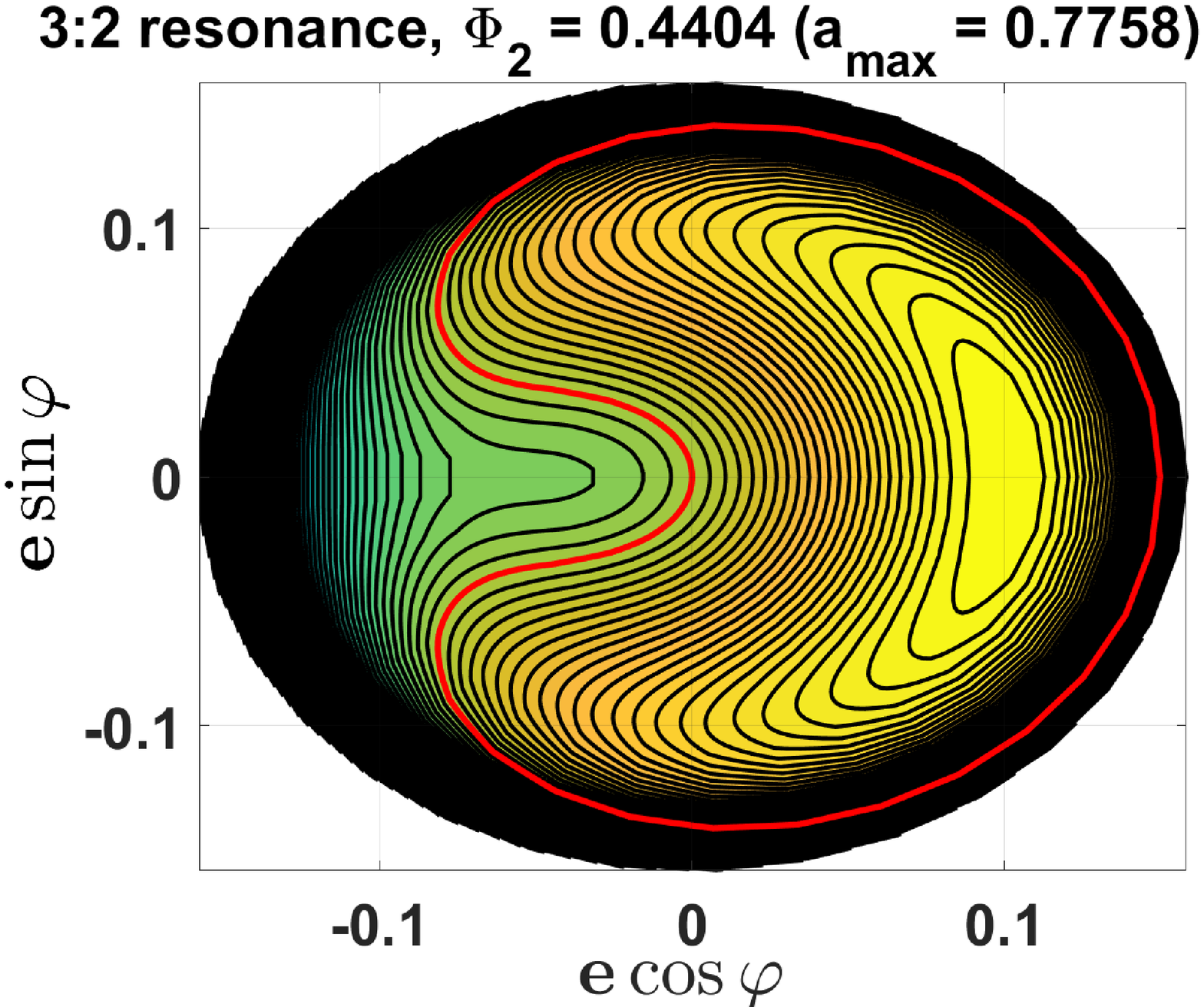}
\includegraphics[width=0.235\textwidth]{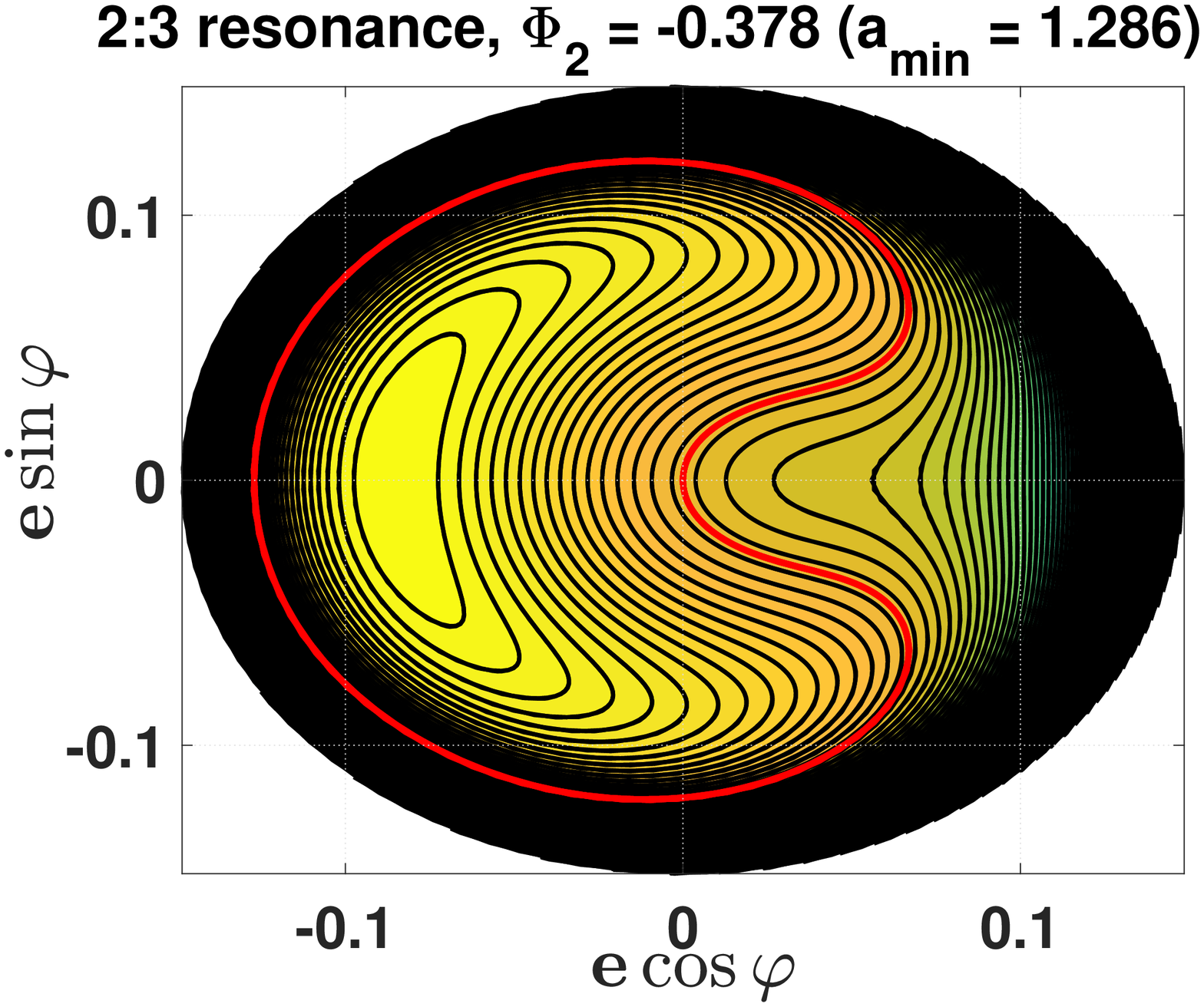}
\caption{Level curves of the resonant Hamiltonian (phase portraits) of the 3:2 resonance characterized by $\Phi_2 = 0.4404$ (i.e., $a_{\max} = 0.7758$) (\emph{left panel}) and the 2:3 resonance characterized by $\Phi_2 = - 0.378$ (i.e., $a_{\min} = 1.2860$) (\emph{right panel}). The level curves stemming from the coordinate center ($e=0$) are marked in red lines, which divide the entire phase space into regions of libration and circulation. For the 3:2 resonance, the libration center is located at $\varphi = 0$ and, for the 2:3 resonance, the libration center is located at $\varphi = \pi$.}
\label{Fig1A}
\end{figure}

To understand the global dynamics, we need to analyze the phase-space structures. By plotting the level curves of the resonant Hamiltonian in the $(\sigma_1 = \sigma, \Gamma_1)$ plane, it is possible to obtain the phase portrait characterized by the motion integral $\Phi_2$ (i.e., $a_{\max}$ or $a_{\min}$). Considering the relationship between $\Gamma_{1,2}$ and the elements $a$ and $e$, we could present the phase portraits in the space $(e \cos{\varphi}, e \sin{\varphi})$, as commonly adopted by previous works. In Fig. \ref{Fig1A}, the phase portraits are reported for the 3:2 resonance characterized by $\Phi_2 = 0.4404$ (i.e., $a_{\max} = 0.7758$) in the left panel and for the 2:3 resonance characterized by $\Phi_2 = -0.378$ (i.e., $a_{\min} = 1.2860$) in the right panel. For the 3:2 resonance, the resonant angle is defined by $\varphi = 2 \lambda - 3 \lambda_p + \varpi$ and, for the 2:3 resonance, the resonant angle is $\varphi = 3 \lambda - 2 \lambda_p - \varpi$. The red lines shown in both panels of Fig. \ref{Fig1A} represent the level curves passing through the coordinate center, which divide the whole phase spaces into the libration domain (the region inside the critical level curve) and circulation domain (the region outside the critical level curve). Thus, the level curve passing through the coordinate center plays the role of dynamical separatrix. However, in the phase portraits shown in Fig. \ref{Fig1A}, the coordinate center (i.e. the zero-eccentricity point) is not a visible equilibrium point, as pointed out by \citet{morbidelli2002modern}.

In addition, it is observed from Fig. \ref{Fig1A} that (a) for the 3:2 or 2:3 resonance with the currently considered motion integral, there is only one stationary point arising in the phase portrait, (b) for the 3:2 resonance shown in the left panel, the libration center is located at $\varphi = 0$, and (c) for the 2:3 resonance shown in the right panel, the libration center is located at $\varphi = \pi$.

The phase portraits for the first-order resonances shown in Fig. \ref{Fig1A} are in agreement with previous results, e.g. Fig. 9.2 in \citet{morbidelli2002modern}, Fig. 4 in \citet{winter1997resonanceI}, Fig. 7 in \citet{henrard1983second} and Fig. 1 in \citet{jancart2002second}.

\subsection{The second multi-harmonic Hamiltonian model}
\label{Sect3-2}

In recent several years, for the purpose of numerically producing Poincar\'e sections in the planar circular restricted three-body problem, \citet{wang2017mean} introduced the angle $\psi = \varpi - \lambda_p$, which specifies the angular separation of the planet and the test particle when the latter is at the pericentre. They demonstrated that, when the test particle is at the pericentre, the usual resonant angle $\varphi = k \lambda - k_p \lambda_p + (k_p - k) \varpi$ is equal to $k_p$ times the separation angle $\psi$ for the $k_p$:$k$ resonances (i.e. $\varphi = k_p \psi$ when $\varphi$ is evaluated at the pericentre). In other words, during each libration period of the usual resonant angle $\varphi$, there are $k_p$ points appearing in the Poincar\'e section.

The Poincar\'e sections measured by the angular separation $\psi$ have been successfully applied in characterizing the dynamics of mean motion resonances, as performed by Malhotra and her collaborators in a series of works, e.g. \citet{wang2017mean, malhotra2018neptune, lan2019neptune} and \citet{malhotra2020divergence}.

Inspired by the fact that there are $k_p$ points arising in the Poincar\'e section during each libration period of $\varphi$, we introduce a new critical argument, denoted by $\sigma = {\varphi}/{k_p}$, to formulate the Hamiltonian model for mean motion resonances. In this new model, we can see that, during each libration period of $\sigma$, there is only one point appearing in the Poincar\'e section. As a consequence, it is possible to establish a one-to-one correspondence between the points appearing in the Poincar\'e section and the points arising in the phase portrait for a certain mean motion resonance. In other words, the phase portraits obtained in this resonant model can be directly compared with the Poincar\'e sections produced in \citet{malhotra2020divergence}. In particular, it becomes possible for us to study the resonant widths numerically identified by \citet{malhotra2020divergence} in an analytical manner. This is a key feature of our resonant model.

According to the aforementioned discussions, we define a new critical argument as
\begin{equation}\label{Eq14}
\sigma  = \frac{1}{{{k_p}}}\varphi  = \frac{1}{{{k_p}}}\left[ {k\lambda  - {k_p}{\lambda _p} + ({k_p} - k)\varpi } \right].
\end{equation}
Under the consideration of the relationship between $\varphi$ and $\sigma$ ($\varphi = k_p \sigma$), the averaged resonant disturbing function given by equation (\ref{Eq8}) can be written as
\begin{equation}\label{Eq15}
{{\cal R}^{{*}}} = \sum\limits_{n = 0}^N {{{\cal C}_n}\cos (n{k_p}\sigma )},
\end{equation}
and the resonant Hamiltonian becomes
\begin{equation}\label{Eq16}
{{\cal H}^*} =  - \frac{{{\mu ^2}}}{{2{\Lambda ^2}}} + {n_p}{\Lambda _p} - \sum\limits_{n = 0}^N {{{\cal C}_n}\cos (n{k_p}\sigma )},
\end{equation}
where the coefficients ${\cal C}_n$ are the same as the ones arising in equation (\ref{Eq8}).

To formulate the resonant model with $\sigma$ as the critical argument, we need to introduce the following set of variables,
\begin{equation}\label{Eq17}
\begin{aligned}
{\Gamma _1} &= \frac{{{k_p}}}{k}\Lambda ,\quad {\sigma _1} = \frac{1}{{{k_p}}}\left[ {k\lambda  - {k_p}{\lambda _p} - ({k_p} - k)p} \right] = \sigma,\\
{\Gamma _2} &= P + \frac{{{k_p} - k}}{k}\Lambda ,\quad {\sigma _2} = p,\\
{\Gamma _3} &= {\Lambda _p} + \frac{{{k_p}}}{k}\Lambda ,\quad {\sigma _3} = {\lambda _p}.
\end{aligned}
\end{equation}
It is not difficult to check that this change of variables is a canonical transformation with the following generating function,
\begin{equation*}
{\cal S} = \frac{k}{{{k_p}}}\lambda {\Gamma _1} + {\lambda _p}\left( {{\Gamma _3} - {\Gamma _1}} \right) + p\left[ {{\Gamma _2} - (1 - \frac{k}{{{k_p}}}){\Gamma _1}} \right].
\end{equation*}
Under the new set of variables, the resonant Hamiltonian given by equation (\ref{Eq16}) can be organized as follows:
\begin{equation}\label{Eq18}
{{\cal H}^{{*}}} =  - \frac{{{\mu ^2}}}{{2{{\left( {\frac{k}{{{k_p}}}{\Gamma _1}} \right)}^2}}} - {n_p}{\Gamma _1} - \sum\limits_{n = 0}^N {{{\cal C}_n}\left( {{\Gamma _1},{\Gamma _2}} \right)\cos (n{k_p}{\sigma _1})},
\end{equation}
where the constant terms have been eliminated from the resonant Hamiltonian. Obviously, the dynamical model determined by the Hamiltonian given by equation (\ref{Eq18}) is of one degree of freedom with $\sigma_1 = \sigma$ as the unique angular coordinate, so that the system is totally integrable. In addition, the angular coordinates $\sigma_2$ and $\sigma_3$ are cyclic, indicating that their conjugate momenta become the motion integral of system, given by
\begin{equation}\label{Eq19}
{\Gamma _2} = P + \frac{{{k_p} - k}}{k}\Lambda  = \sqrt {\mu a} \left( {\frac{{{k_p}}}{k} - \sqrt {1 - {e^2}} } \right) = {\rm const}
\end{equation}
and
\begin{equation}\label{Eq20}
{\Gamma _3} = {\Lambda _p} + \frac{{{k_p}}}{k}\sqrt {\mu a}  = {\rm const}.
\end{equation}
Obviously, the motion integral given by equations (\ref{Eq19}) and (\ref{Eq20}) are the same as the ones given by equations (\ref{Eq12}) and (\ref{Eq13}), respectively. Similarly, we can also use $a_{\max}$ (or $a_{\min}$) to stand for the motion integral $\Gamma_2$ for the inner resonances (or the outer resonances). The level curves of $\Gamma_2$ are reported in Fig. \ref{Fig1} in the $(a,e)$ plane for the inner 3:2 and outer 2:3 resonances.

It is noted that similar expressions of the motion integral $\Gamma_2$ (or $\Phi_2$) appeared in the multi-harmonic Hamiltonian model can be found in many previous works, e.g. \citet{beauge1994asymmetric}, \citet{gomes1997orbital}, \citet{morbidelli2002modern} and \citet{ramos2015resonance}. The relationship between the Jacobi constant which is used in \citet{malhotra2020divergence} to characterize Poincar\'e sections and the motion integral $\Gamma_2$ (or $\Phi_2$) is discussed in Appendix \ref{A_0}.

\begin{figure}
\centering
\includegraphics[width=0.235\textwidth]{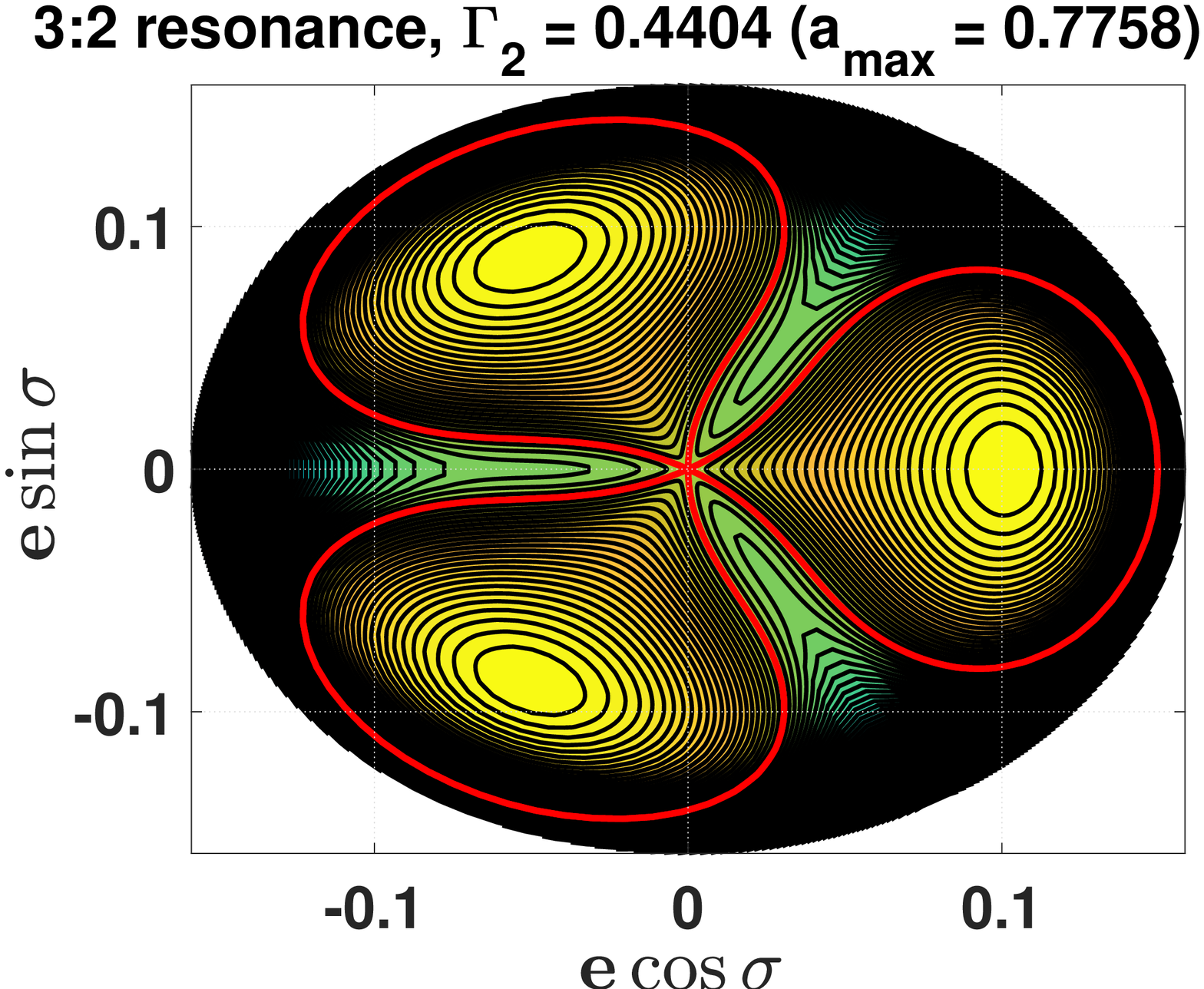}
\includegraphics[width=0.235\textwidth]{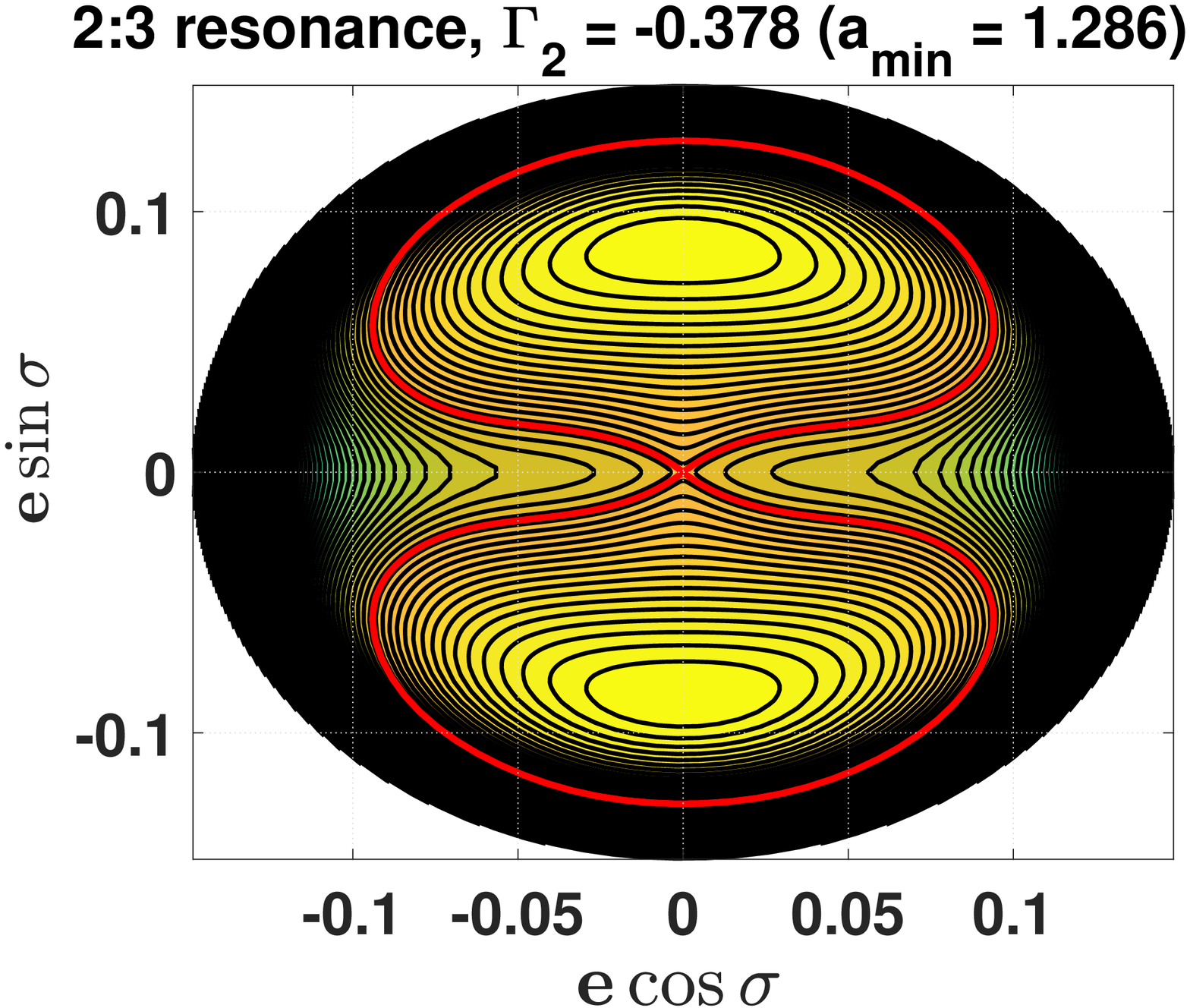}
\caption{Level curves of the resonant Hamiltonian (phase portraits) of the 3:2 resonance characterized by $\Gamma_2 = 0.4404$ (i.e., $a_{\max} = 0.7758$) (\emph{left panel}) and the 2:3 resonance characterized by $\Gamma_2 = - 0.378$ (i.e., $a_{\min} = 1.2860$) (\emph{right panel}). The level curves passing through the coordinate center are marked in red lines. For the 3:2 resonance shown in the left panel, there are three libration islands centered at $\sigma = 0, 2\pi/3, 4\pi/3$ (corresponding to period-3 fixed points in the Poincar\'e section). For the 2:3 resonance shown in the right panel, there are two islands of resonance, centered at $\sigma = \pm \pi/2$ (corresponding to period-2 fixed points in the Poincar\'e section). In both panels, the coordinate center ($e=0$) is a saddle point of the resonant model, so that the level curves stemming from the zero-eccentricity point play the role of dynamical separatrix, dividing the entire phase space into regions of libration and circulation.}
\label{Fig1B}
\end{figure}

To understand the global dynamics of mean motion resonances, it is necessary to analyze the phase portraits. Regarding the inner 3:2 resonance and outer 2:3 resonance, in Fig. \ref{Fig1B} we plot the level curves of the resonant Hamiltonian in the $(e \cos{\sigma}, e \sin{\sigma})$ plane with the same motion integral used in Fig. \ref{Fig1A}.

It is observed from Fig. \ref{Fig1B} that (a) there are three typical libration centers in the phase portrait of the 3:2 resonance located at $\sigma = 0, 2\pi/3, 4\pi/3$ (corresponding to period-3 fixed points in the Poincar\'e section), (b) there are two typical libration centers in the phase portrait of the 2:3 resonance located at $\sigma = \pm \pi/2$ (corresponding to period-2 fixed points in the Poincar\'e section), and (c) in both panels, the coordinate center at $e = 0$ corresponds to a saddle point of the resonant model and the level curve stemming from it plays the role of dynamical separatrix, dividing the whole phase space into domains of libration and circulation. It is noted that, when the motion integral $\Gamma_2$ is varied, the number of fixed points (i.e., the number of resonance islands) will change.

\section{Model validation}
\label{Sect4}

In the previous section, two multi-harmonic Hamiltonian models of mean motion resonances have been formulated. In the first model, the resonant angle is given by $\varphi = k \lambda - k_p \lambda_p + (k_p - k)\varpi$ (this is the classical critical argument) and, in the second model, the resonant angle is given by $\sigma = \varphi / k_p$.

In Section \ref{Sect4-1}, we make a comparison between the phase-space structures produced in the first and second resonant models for some specific resonances. In Section \ref{Sect4-2}, focusing on the second Hamiltonian model, we compare the Hamiltonian models truncated at different orders in eccentricity to the associated numerical resonant model with an aim at exploring the influence of $N$ (the number of harmonics) upon the structures of phase portraits.

\subsection{Comparisons between the first and second Hamiltonian models}
\label{Sect4-1}

\begin{figure}
\centering
\includegraphics[width=0.235\textwidth]{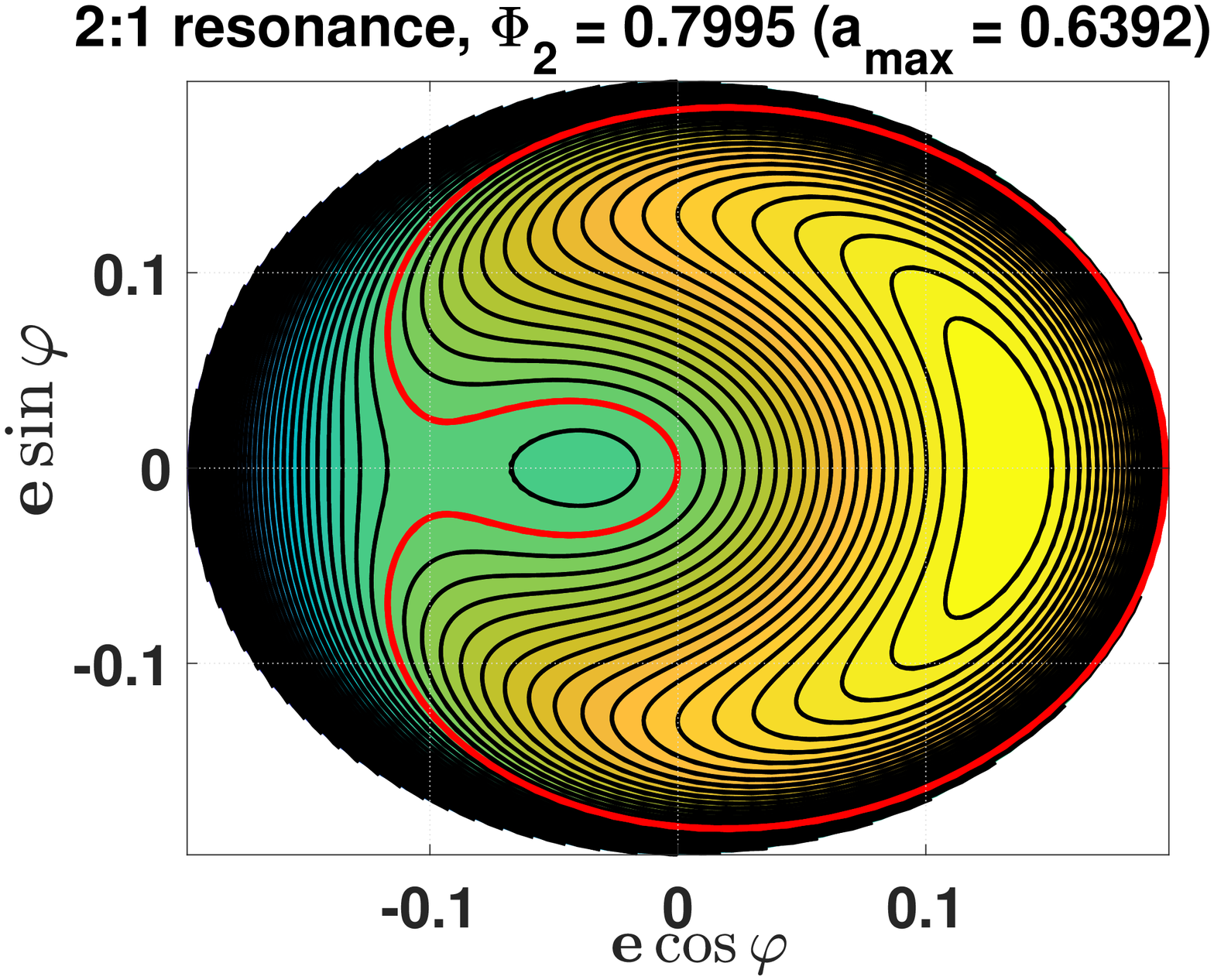}
\includegraphics[width=0.235\textwidth]{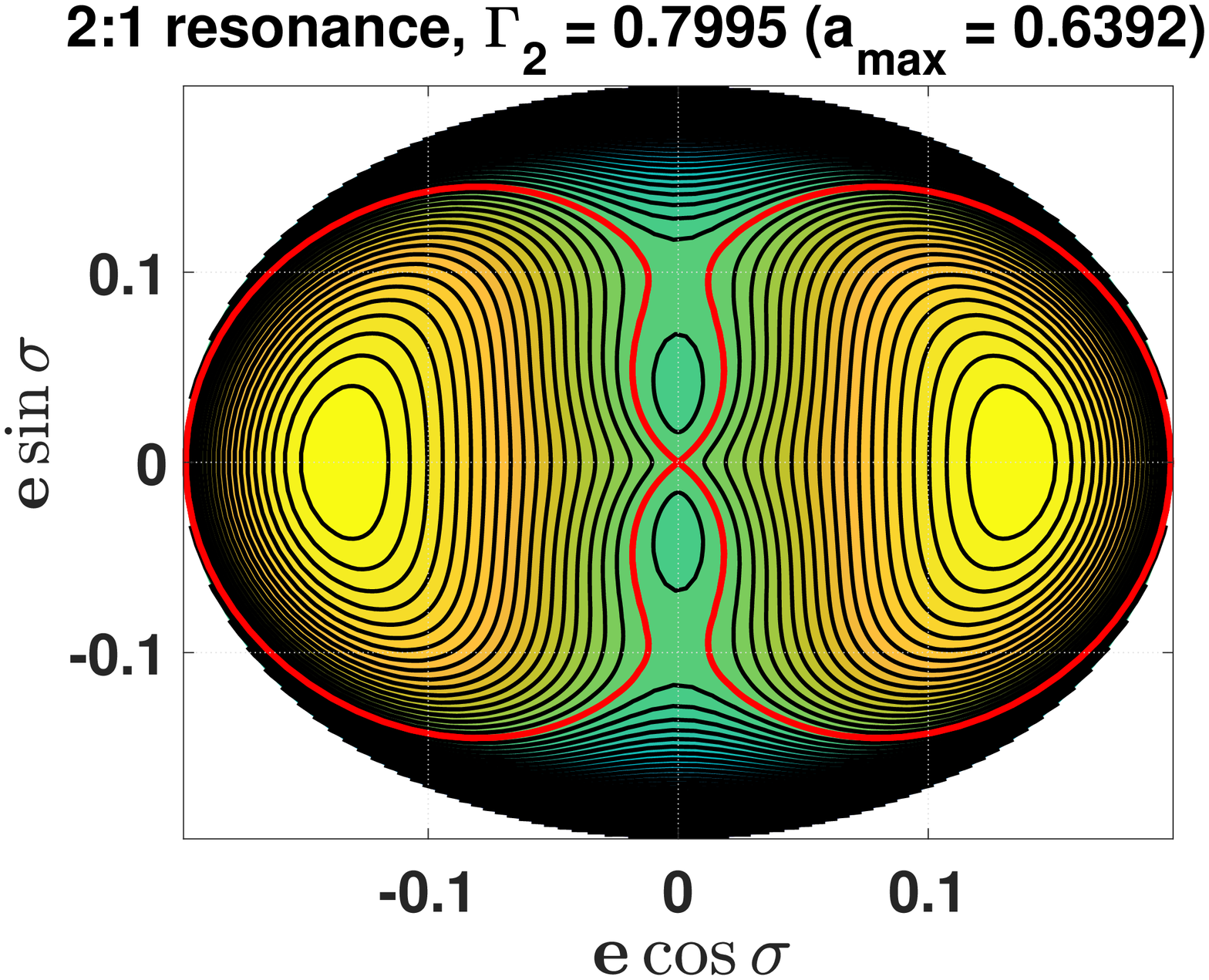}\\
\includegraphics[width=0.235\textwidth]{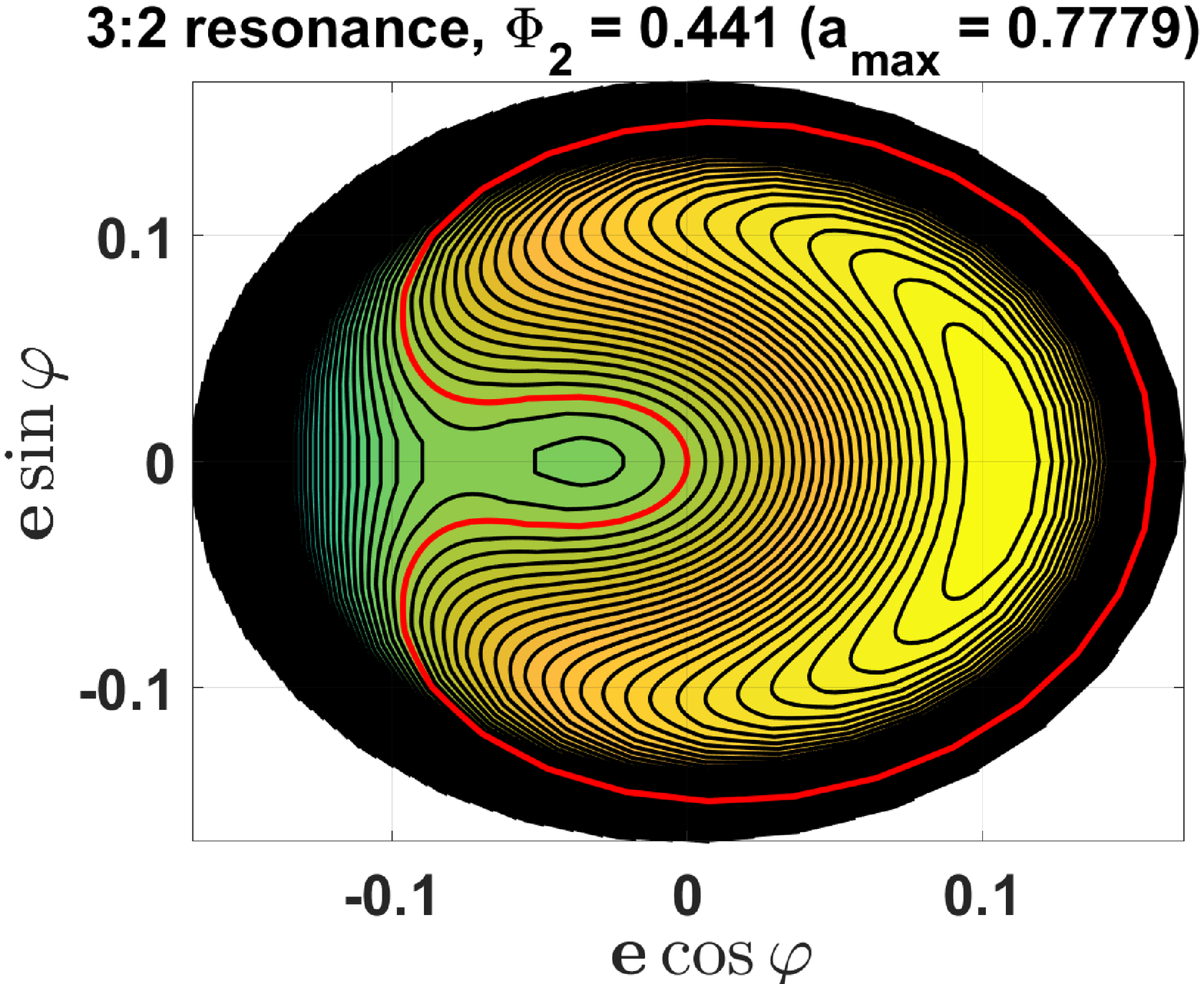}
\includegraphics[width=0.235\textwidth]{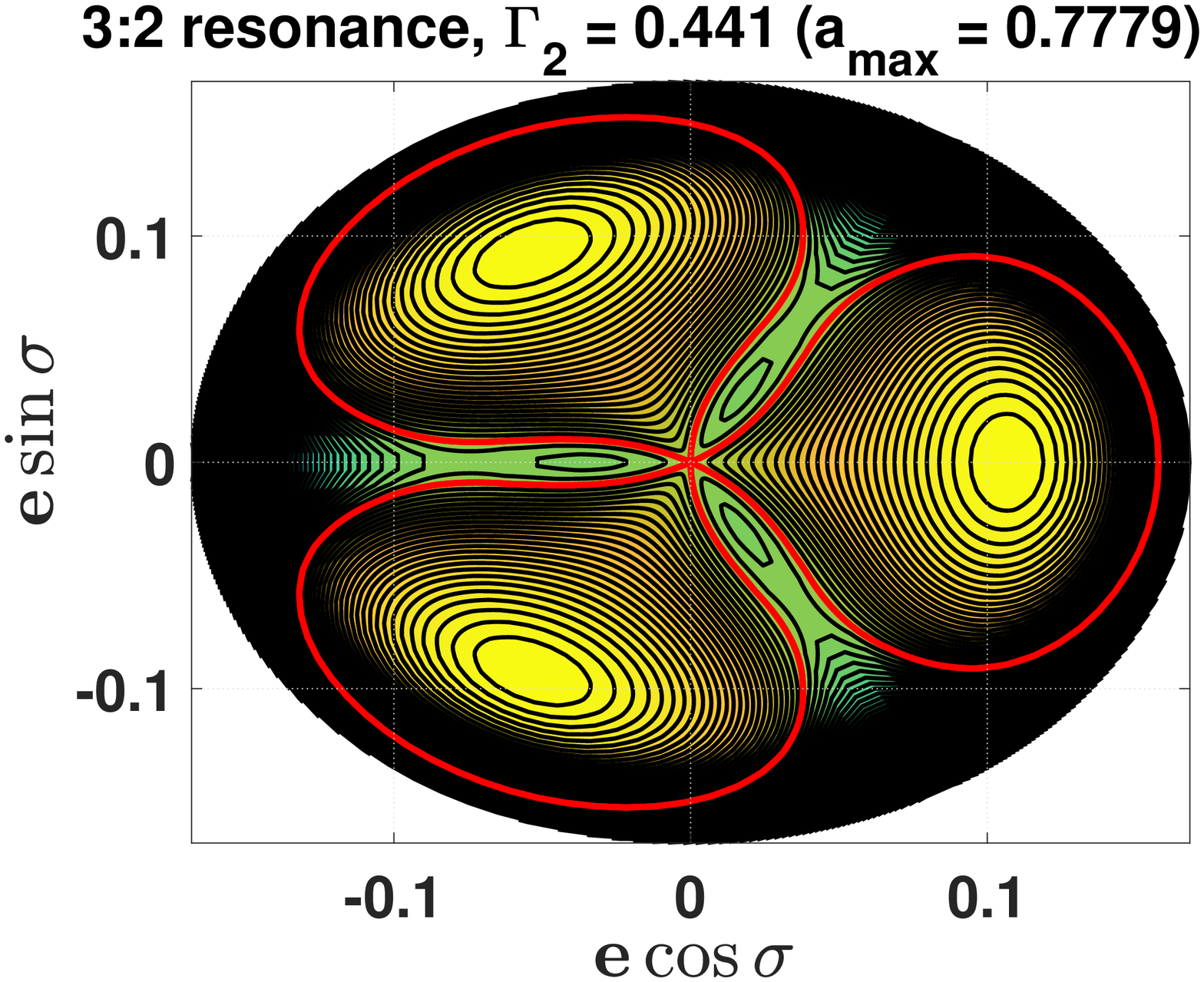}
\caption{Level curves of the resonant Hamiltonian associated with the inner 2:1 resonance characterized by the motion integral $\Gamma_2 = \Phi_2 = 0.7995$ (i.e., $a_{\max} = 0.6392$) and the 3:2 resonance specified by $\Gamma_2 = \Phi_2 = 0.441$ (i.e., $a_{\max} = 0.7779$). The panels in the left-hand column correspond to the phase-space structure in the resonant model with the usual critical argument $\varphi$ (the first resonant model), and the panels in the right-hand column correspond to the phase-space structure in the resonant model with the critical argument $\sigma$ (the second resonant model). For the 2:1 resonance, the usual critical argument is defined by $\varphi = \lambda - 2 \lambda_p + \varpi$ and the new critical argument is $\sigma = \varphi/2$. For the 3:2 resonance, the usual critical argument is defined by $\varphi = 2 \lambda - 3 \lambda_p + \varpi$ and the new critical argument is $\sigma = \varphi/3$. In all plots, the level curves passing through the zero-eccentricity point (i.e. the coordinate center) are marked in red lines.}
\label{Fig2}
\end{figure}

In this section, a direct comparison is made between the phase-space structures produced from the first Hamiltonian model and the ones generated from the second Hamiltonian model for the inner 2:1 and 3:2 resonances, as shown in Fig. \ref{Fig2}. In both models, the truncated order in eccentricity is taken as $N=10$ (our simulations show that the model with $N = 10$ is accurate enough to approximate the dynamics of mean motion resonances). The phase-space structures are presented in the $(k = e \cos \varphi, h = e \sin \varphi)$ plane for the first resonant model (see the panels in the left-hand column) and the phase-space structures are presented in the $(k = e \cos \sigma, h = e \sin \sigma)$ plane for the second resonant model (see the panels in the right-hand column). For the inner 2:1 resonance, the motion integral is taken as $\Phi_2 (=\Gamma_2) = 0.7995$ (i.e., $a_{\max} = 0.6392$) and, for the inner 3:2 resonance, the motion integral is taken as $\Phi_2 (=\Gamma_2) = 0.441$ (i.e., $a_{\max} = 0.7779$). In all plots of Fig. \ref{Fig2}, the red lines stand for the level curves of resonant Hamiltonian passing through the zero-eccentricity point (i.e. the coordinate center).

Observing from Fig. \ref{Fig2}, we can see that (a) as for the 2:1 and 3:2 resonances, their phase portraits present similar structures in the first model, while they are totally different in the second model, (b) there are three stationary solutions in the first resonant model, which is in agreement with the results given by \citet{morbidelli2002modern} and \citet{ramos2015resonance} at the same level of motion integral, (c) there are $3 k_p + 1$ stationary solutions in the second resonant model including $2k_p$ stable equilibria and $k_p + 1$ saddle points, (d) in the first model, the libration centers are located at $\varphi = 0, \pi$ and the saddle point is at $\varphi = \pi$ for both inner resonances, (e) in the second model, the libration centers are located at $\sigma = j \pi/k_p$ with $j=0,1,...,2k_p-1$ and the saddle points are located at $\sigma = j \pi/k_p$ with $j=1,3,...,2k_p-1$ as well as the zero-eccentricity point (i.e. the coordinate center), and (f) it is clear that one libration centre (or one saddle point) in the first model are split into $k_p$ libration centers (or $k_p$ saddle points) in the second model.

Next, let us consider the consistency between the Poincar\'e sections numerically produced in \citet{malhotra2020divergence} and the phase-space structures analytically produced in the second resonant model (note that the consistency is a key motivation of the new resonant model). For the inner 2:1 resonance, the phase-space structure associated with the second resonant model shown in the upper-right panel of Fig. \ref{Fig2} is in perfect agreement with the Poincar\'e section corresponding to the central panel of Fig. 2 in \citet{malhotra2020divergence} and, for the inner 3:2 resonance, the phase-space structure corresponding to the bottom-right panel of Fig. \ref{Fig2} is coincident with the Poincar\'e section given in the bottom-right panel of Fig. 5 in \citet{malhotra2020divergence} (please see their work for more details). As desired, a perfect consistency is found between the analytical results (phase portraits) and numerical results (Pincar\'e sections) based on our second resonant model. This correspondence is very helpful in understanding the numerical behaviors arising in the Poincar\'e sections.

In addition, it is observed from Fig. \ref{Fig2} that the zero-eccentricity point (i.e. the coordinate center) is a saddle point in the second resonant model, but it is not a visible equilibrium point in the first resonant model (or the classical resonant model).

In summary, the second Hamiltonian model formulated in this work holds two advantages in comparison to the first Hamiltonian model (or the classical resonant model): (a) providing a direct correspondence between Poincar\'e sections and phase portraits and (b) presenting the unfolded phase-space structures where the zero-eccentricity point (or coordinate center) becomes a visible saddle point at arbitrary motion integral. It is observed from the Poincar\'e sections produced by \citet{malhotra2020divergence} that the zero-eccentricity point is an unstable fixed point, which is in agreement with our analytical results. Considering these two advantages, from now on we will take the second Hamiltonian model as the fundamental model to study the dynamics of mean motion resonances in the following discussions, unless otherwise specified.

\subsection{Validation of the second Hamiltonian model}
\label{Sect4-2}

As stated in Section \ref{Sect3}, the accuracy of resonant Hamiltonian is specified by the number $N$ of harmonics because of the fact that the number $N$ stands for the truncated order in eccentricity in the disturbing function. Please refer to equation (\ref{Eq18}) for the expression of resonant Hamiltonian. In this section, we intend to compare the analytical dynamical models (the second Hamiltonian models) with different $N$ to the numerical model in order to explore the influence of $N$ upon the phase-space structures. In the numerical model, the resonant disturbing function is obtained by direct numerical integration for equation (\ref{Eq7}), as performed in \citet{gallardo2006Atlas}.

\begin{figure*}
\centering
\includegraphics[width=0.33\textwidth]{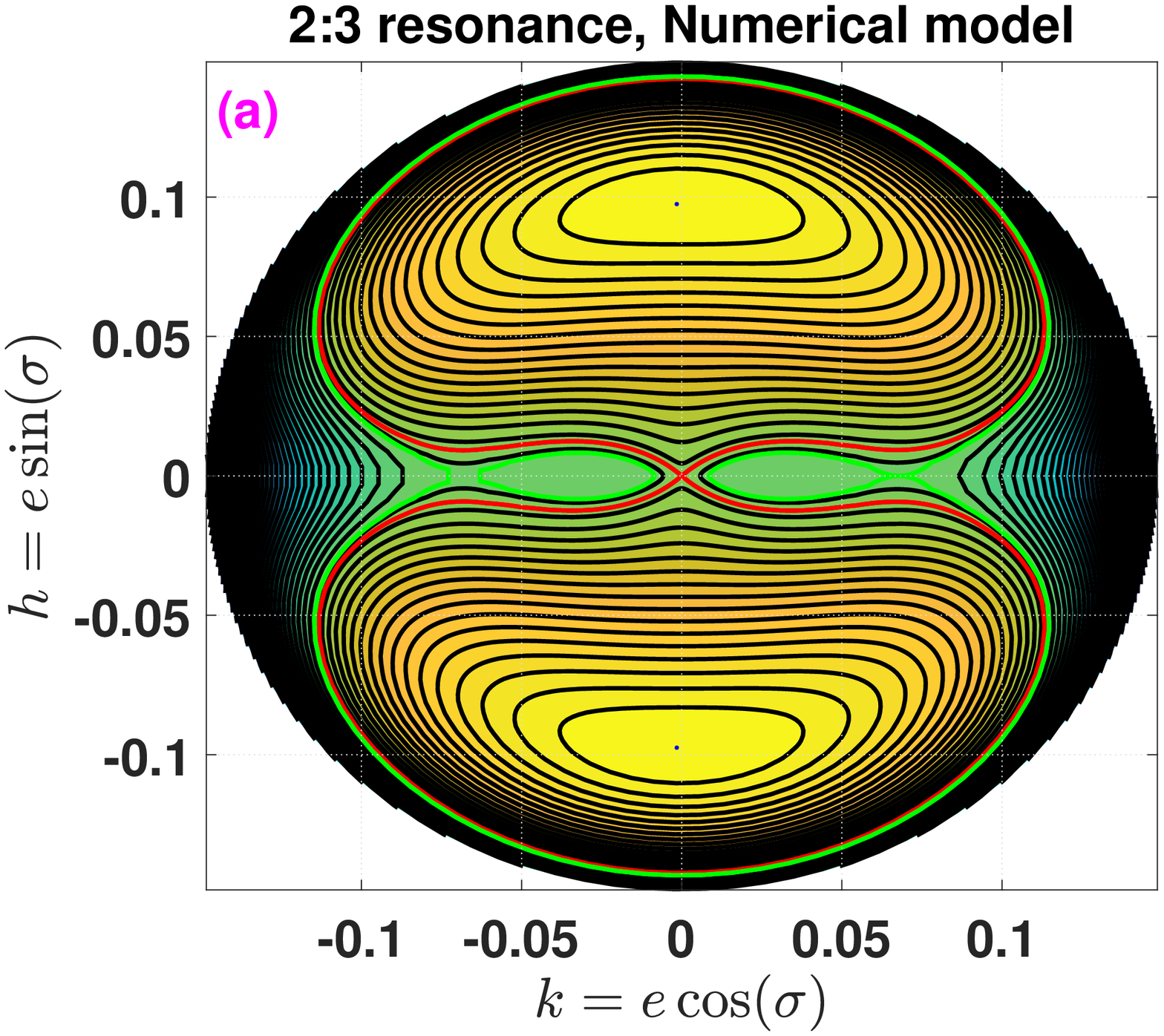}
\includegraphics[width=0.33\textwidth]{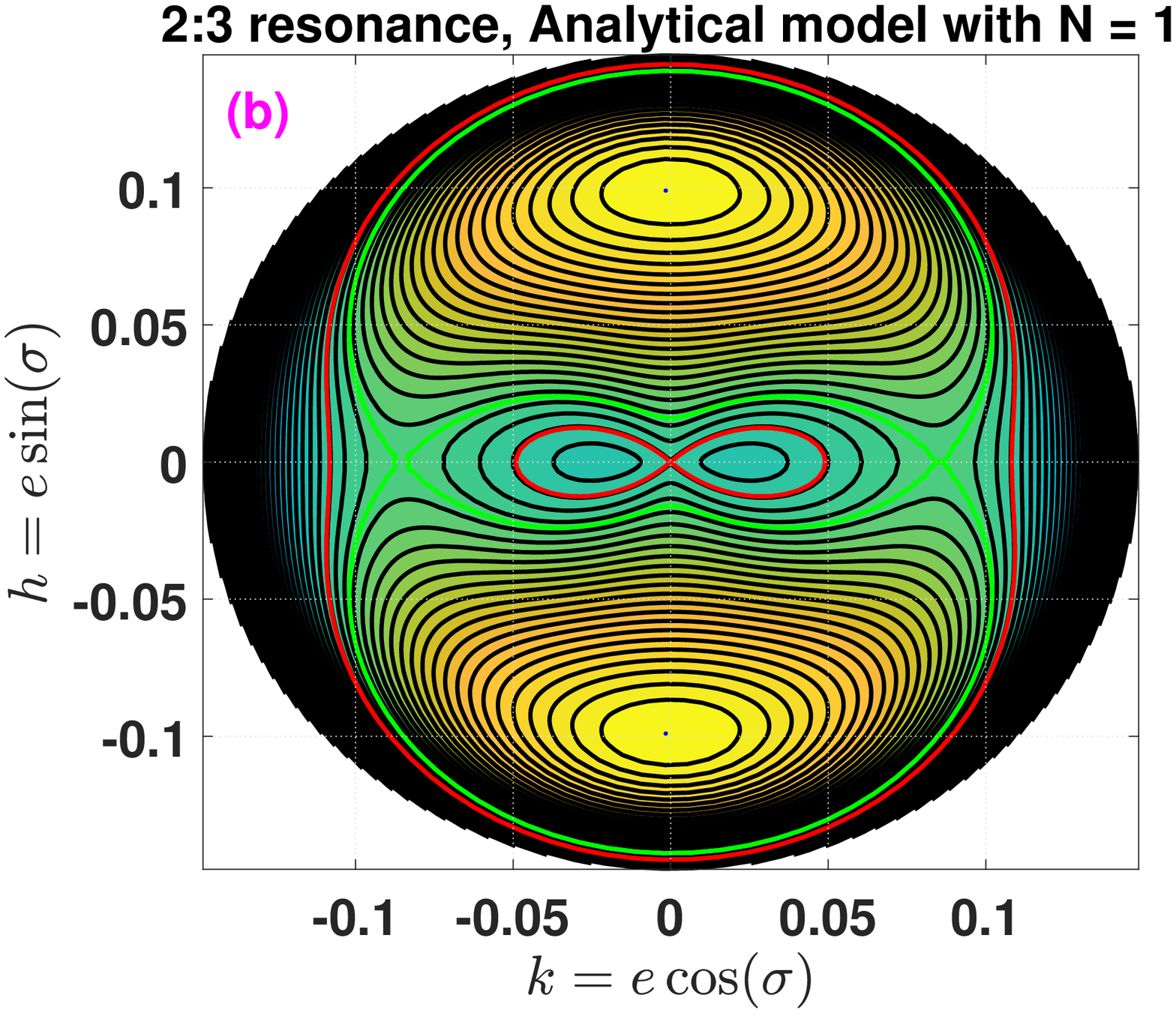}
\includegraphics[width=0.33\textwidth]{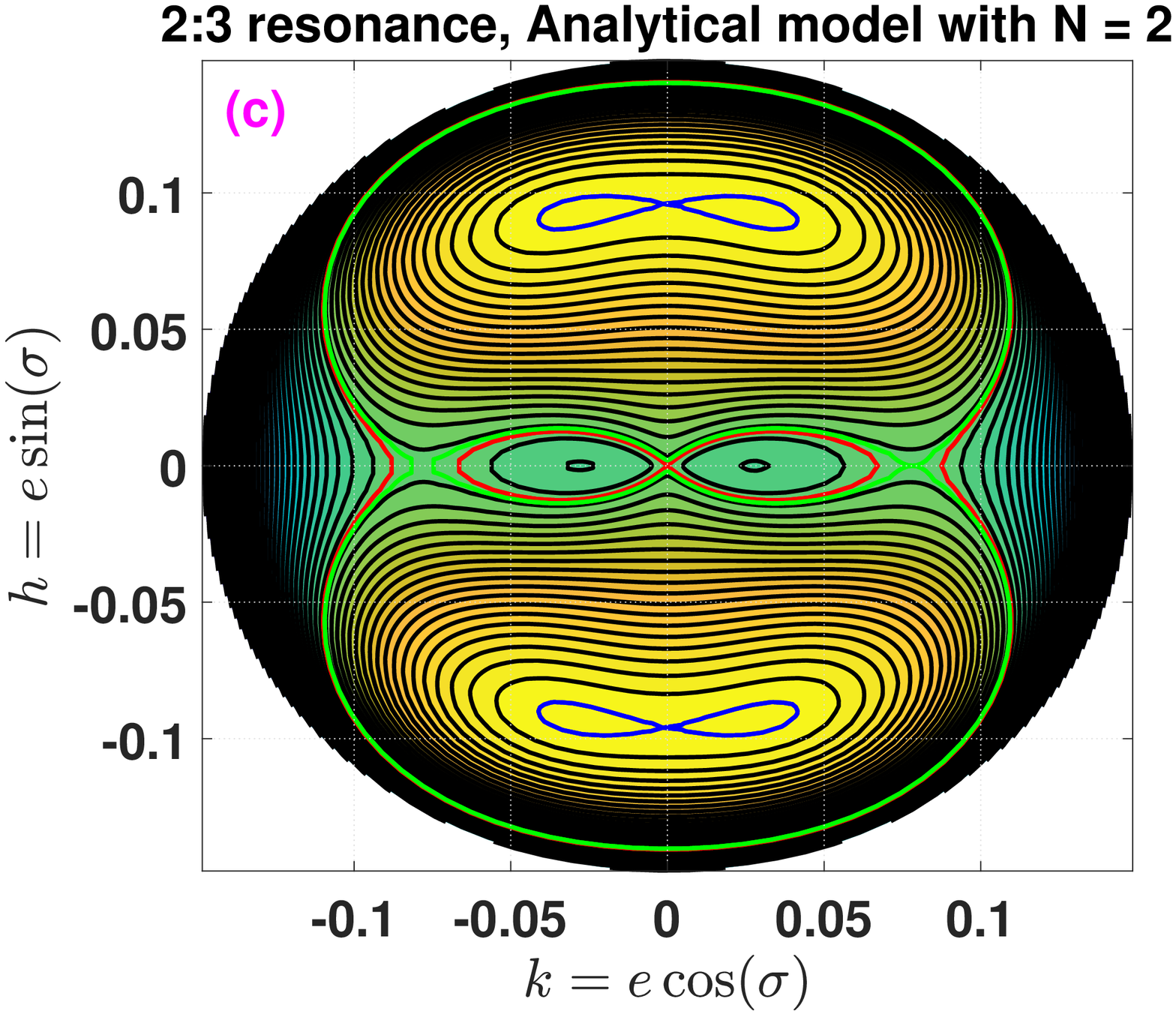}\\
\includegraphics[width=0.33\textwidth]{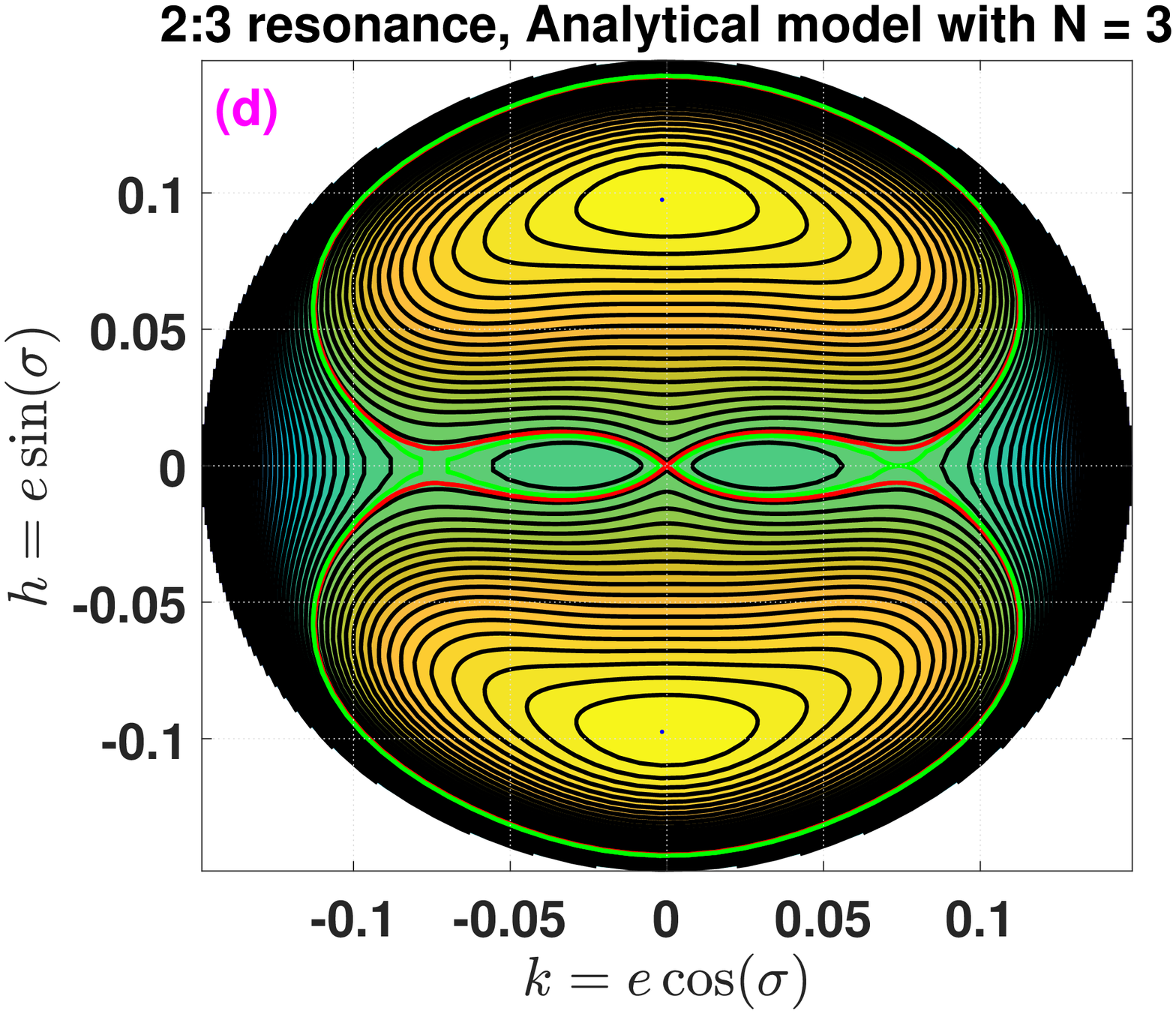}
\includegraphics[width=0.33\textwidth]{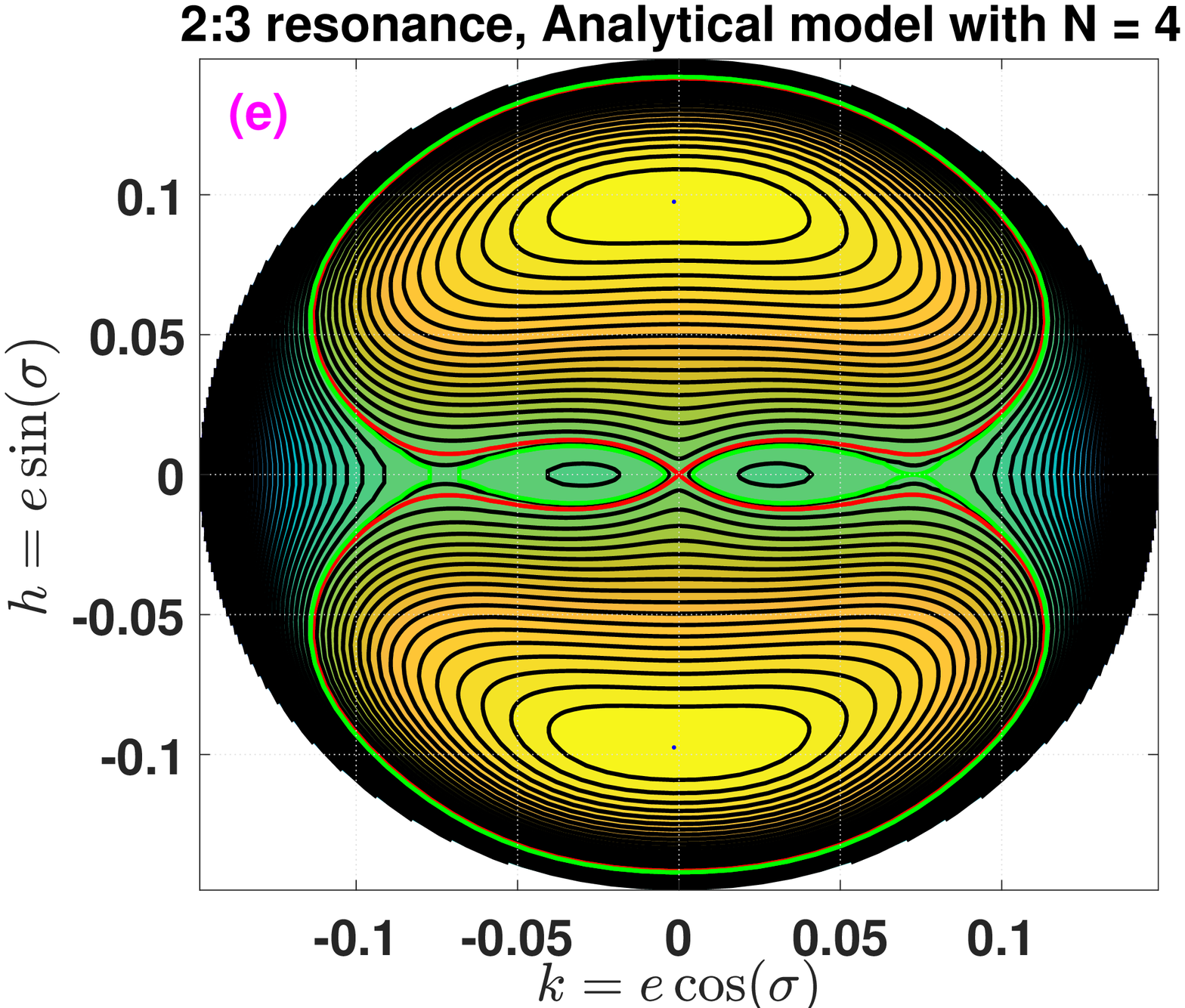}
\includegraphics[width=0.33\textwidth]{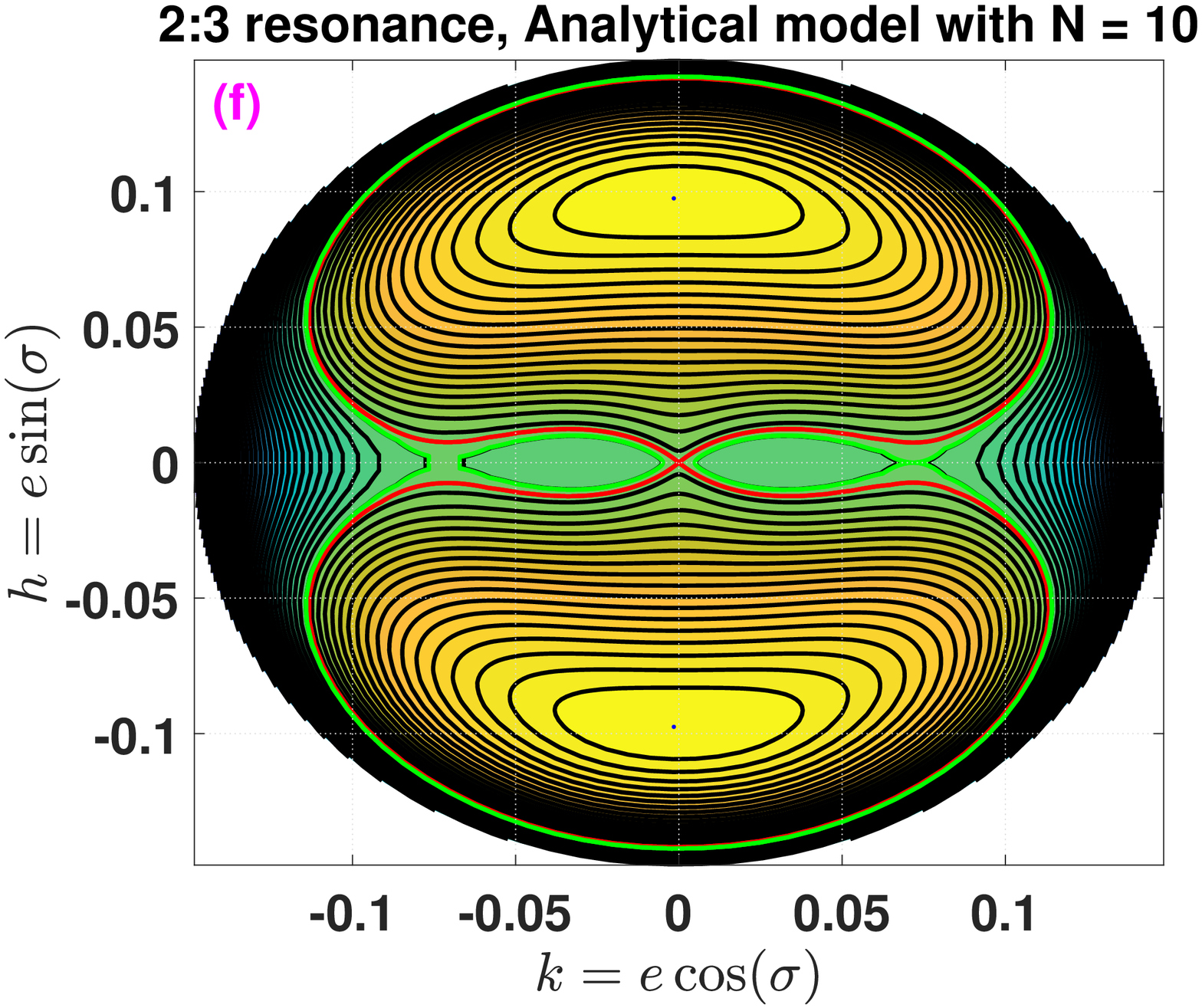}
\caption{Level curves of the resonant Hamiltonian associated with the outer 2:3 resonance specified by $\Gamma_2 = -0.3767$ (i.e., $a_{\min} = 1.2771$). The panel (a) corresponds to the numerical model where the resonant disturbing function is produced by means of direct numerical integration, the remaining panels are for the analytical models with different $N$ (see the text for details of analytical models). In all the plots, the dynamical separatrices passing through the zero-eccentricity point (i.e. the coordinate center) are marked in red lines and the ones passing through the other saddle point are shown in green lines. In particular, new asymmetric libration centres appear in the resonant model with $N=2$, as shown in panel (c) (it means that fake phase-space structures of the 2:3 resonance arise in this model). When $N \ge 3$, the asymmetric libration centres disappear from the phase portraits. In particular, the structure in the analytical model with $N=10$ is identical to the one in the numerical model (please compare panel (a) and panel (f)).}
\label{Fig3}
\end{figure*}

In Fig. \ref{Fig3}, we report the level curves of the (numerical and analytical) resonant Hamiltonian associated with the outer 2:3 resonance in the $(e \cos \sigma, e \sin \sigma)$ plane. In practical simulations, the motion integral is fixed at $\Gamma_2 = -0.3767$ (i.e., $a_{\min} = 1.2771$).

The panel (a) of Fig. \ref{Fig3} is for the phase portrait generated in the numerical model, and the remaining panels are for the phase portraits produced in the analytical models with $N=1,2,3,4$ and $N=10$. In all plots, we show the level curves passing through the zero-eccentricity and nonzero-eccentricity saddle points in green and red lines, which play the role of dynamical separatrices in the phase space.

In the phase portrait associated with the numerical model (see panel (a) of Fig. \ref{Fig3}), it is observed that (a) the libration center is located at $\sigma = 0, \pi/2, \pi, 3\pi/2$ and (b) there are three saddle points, two of them with non-zero eccentricity are located at $\sigma = 0, \pi$ and the last one is located at the coordinate center (i.e. the zero-eccentricity point). Similar to the notations used in \citet{malhotra2020divergence}, we call the libration islands centered at $\sigma = \pi/2, 3\pi/2$ (corresponding to $\varphi = \pi$) the apocentric zones and call the ones centered at $\sigma = 0, \pi$ (corresponding to $\varphi = 0$) the pericentric zones. It is clear to observe that (i) the apocentric libration zones are bounded by the separatrices stemming from the zero-eccentricity saddle point (i.e. red lines), and (ii) the pericentric libration zones are bounded by the separatrices stemming from the saddle points with non-zero eccentricity (i.e. green lines).

As for the analytical model with $N=1$ (see panel (b) of Fig. \ref{Fig3}), it is observed that the apocentric libration zones are bounded by the separatrices stemming from the saddle points with non-zero eccentricity (instead of the zero-eccentricity saddle point in the numerical model) and the pericentric libration zones are bounded by the separatrices stemming from the zero-eccentricity saddle point (instead of the nonzero-eccentricity saddle point in the numerical model). Evidently, this geometry of phase portrait is different from that shown in the numerical model, meaning that the analytical model with $N=1$ is not accurate enough to approximate the dynamics of first-order resonances, especially in the low-eccentricity regions.

Regarding the analytical model with $N=2$ (see panel (c) of Fig. \ref{Fig3}), the equilibrium points located at $\sigma = \pm \pi/2$ (these two points are stable equilibria in the numerical model) become saddle points and, surprisingly, asymmetric libration centers appear around them. It is noted that, in the numerical model, there are no asymmetric libration centers (see panel (a) of Fig. \ref{Fig3}). In other words, in the analytical model with $N=2$, incorrect dynamical structures arise. Due to the fake structures arising in the phase portraits, it means that the analytical model with $N=2$ is also not accurate enough to approximate the dynamics of first-order resonances.

It should be noted that, for the outer 2:3 resonance with the same motion integral as that used in Fig. \ref{Fig3}, the asymmetric libration centres in the analytical model with $N=2$ have been observed by \citet{beauge1994asymmetric}, who took the critical argument $\varphi$ as the resonant angle to formulate the resonant model (the same as the first resonant model with $N=2$ discussed in the current work). In the work of \citet{beauge1994asymmetric}, the author denoted the numerical model as the exact model, the analytical model with $N=1$ as `SFMR', the analytical model with $N=2$ as `F2' (please refer to Fig. 12 in his work for more details). As stated by \citet{beauge1994asymmetric}, the model `F2' has a better approximation to the real system for low eccentricities (compared to the model `SFMR'), but for high eccentricities it fails to reproduce the correct topology. The same problem has been detected by \citet{message1958search}. \citet{beauge1994asymmetric} explained that the possible reason for the disagreement between `SFMR' and `F2' is due to the problem of convergence of the disturbing function expansion, as predicted by Sundman's criterion \citep{ferraz1994convergence}. However, our practical simulations may lead us to an alternative explanation for such a strange phenomenon, as discussed below.

When the number $N$ is increased up to $N=3$, $N=4$ or $N=10$ (see panels (d--f) of Fig. \ref{Fig3}), asymmetric libration centers disappear and the geometry of the phase portrait becomes closer to the one in the numerical model as the number $N$ increases. In particular, the phase portrait in the analytical model with $N=10$ is identical to that in the numerical model (please compare the first and last panels of Fig. \ref{Fig3}).

Similar simulations are made for the first-order inner resonances including the 2:1, 3:2 and 4:3 resonances, and the phase-space structures associated with the numerical model and analytical models with $N=2$ and $N=10$ are reported in Appendix \ref{A_1} (see Fig. \ref{FigA1} for details). Interestingly, asymmetric libration centers (i.e. the incorrect topology) appear in the phase-space structures in the analytical model with $N=2$ for all the considered inner resonances, while they disappear from the phase-space structures in the analytical models with $N=10$. In particular, it is observed that the structures in the analytical model with $N=10$ are in perfect agreement with the ones in the numerical model.

According to the aforementioned discussions, we can see that the fake structures arising in the phase portraits of first-order resonances are due to the poor approximation of the disturbing function truncated at $N=1$ or $N=2$ (in other words, this problem can be avoided if the resonant Hamiltonian is truncated at a higher order in eccentricity), and thus we can conclude that the analytical models with $N \ge 3$ are required in order to approximate the dynamics of test particles inside first-order mean motion resonances.

In the following applications, we will take the analytical model (the second Hamiltonian model) with $N=10$ as the basic model to perform practical simulations, unless otherwise specified.

\section{Resonant width and applications}
\label{Sect5}

In this section, the analytical method about identifying the libration center and resonant width in the second Hamiltonian model is introduced, and then it is applied to the inner and outer first-order mean motion resonances with a Jupiter-mass planet.

\subsection{Libration center and resonant width}
\label{Sect5-1}

In the second Hamiltonian model, $\Gamma_2$ is the motion integral, and the equation of motion can be written as
\begin{equation}\label{Eq21}
{\dot \sigma _1} = \frac{{\partial {{\cal H}^{\rm{*}}}}}{{\partial {\Gamma _1}}},{\dot \Gamma _1} = \frac{{\partial {{\cal H}^{\rm{*}}}}}{{\partial {\sigma _1}}}.
\end{equation}
The equilibrium points of the resonant model can be obtained by solving the following stationary conditions:
\begin{equation}\label{Eq22}
\begin{aligned}
&{\dot \sigma_1} = \frac{{\partial {{\cal H}^{\rm{*}}}}}{{\partial {\Gamma _1}}} = 0,\\
&{\dot \Gamma_1} = \frac{{\partial {{\cal H}^{\rm{*}}}}}{{\partial {\sigma _1}}} = \sum\limits_{n = 1}^N {n{k_p}{{\cal C}_n}\left( {{\Gamma _1},{\Gamma _2}} \right)\sin (n{k_p}{\sigma _1})}  = 0.
\end{aligned}
\end{equation}
For a given motion integral $\Gamma_2$, we denote the equilibrium point as $(\sigma_1, \Gamma_1)=(\sigma_{10}, \Gamma_{10})$. The equations of motion given by equation (\ref{Eq21}) can be linearized around $(\sigma_{10}, \Gamma_{10})$, and the Jacobian matrix of the resulting linear system determines the stability of the equilibrium point.

As usual, the stable equilibrium points in the resonant model correspond to libration centers, and the unstable ones correspond to saddle points. For convenience of description, we denote the stable equilibrium points as $(\sigma_1, \Gamma_1)=(\sigma_{10}^s, \Gamma_{10}^s)$ and the unstable equilibrium points as $(\sigma_1, \Gamma_1)=(\sigma_{10}^u,\Gamma_{10}^u)$. According to the phase portraits shown in the previous section, the libration zones centered at $(\sigma_{10}^s, \Gamma_{10}^s)$ are bounded by the dynamical separatrices stemming from its closest saddle point $(\sigma_{10}^u,\Gamma_{10}^u)$. As a result, the resonant Hamiltonian of the separatrix passing through $(\sigma_{10}^u,\Gamma_{10}^u)$ is evaluated at $\sigma_1 = \sigma_{10}^s$ by
\begin{equation}\label{Eq23}
\begin{aligned}
{{\cal H}^{{*}}}\left( {{\sigma_1 = \sigma _{10}^u},{\Gamma_1 = \Gamma _{10}^u};{\Gamma _2}} \right) &= {{\cal H}^{{*}}}\left( {{\sigma_1 = \sigma _{10}^s},{\Gamma_1 = \Gamma _{\rm in}};{\Gamma _2}} \right)\\
&= {{\cal H}^{{*}}}\left( {{\sigma_1 = \sigma _{10}^s},{\Gamma_1 = \Gamma _{\rm out}};{\Gamma _2}} \right),
\end{aligned}
\end{equation}
where $\Gamma_{\rm out}$ stands for the outer boundary of $\Gamma_1$ and $\Gamma_{\rm in}$ stands for the inner boundary of $\Gamma_1$.

According to the relationship between $\Gamma_{1,2}$ and the elements $a$ and $e$, the boundary points specified by $(\Gamma_{\rm out}, \Gamma_2)$ and $(\Gamma_{\rm in}, \Gamma_2)$ can be equivalently represented in the element space by $(a_{\rm out},e_{\rm out})$ and $(a_{\rm in}, e_{\rm in})$, respectively. The distance between the inner and outer boundaries can measure the resonant width in the form of $\Delta {\Gamma_1}  = {\Gamma _{\rm out}} - {\Gamma _{\rm in}}$ or $(\Delta a, \Delta e)  = ({a _{\rm out}} - {a _{\rm in}}, {e _{\rm out}} - {e _{\rm in}})$.

\subsection{Applications to inner resonances}
\label{Sect5-2}

\begin{figure*}
\centering
\includegraphics[width=0.42\textwidth]{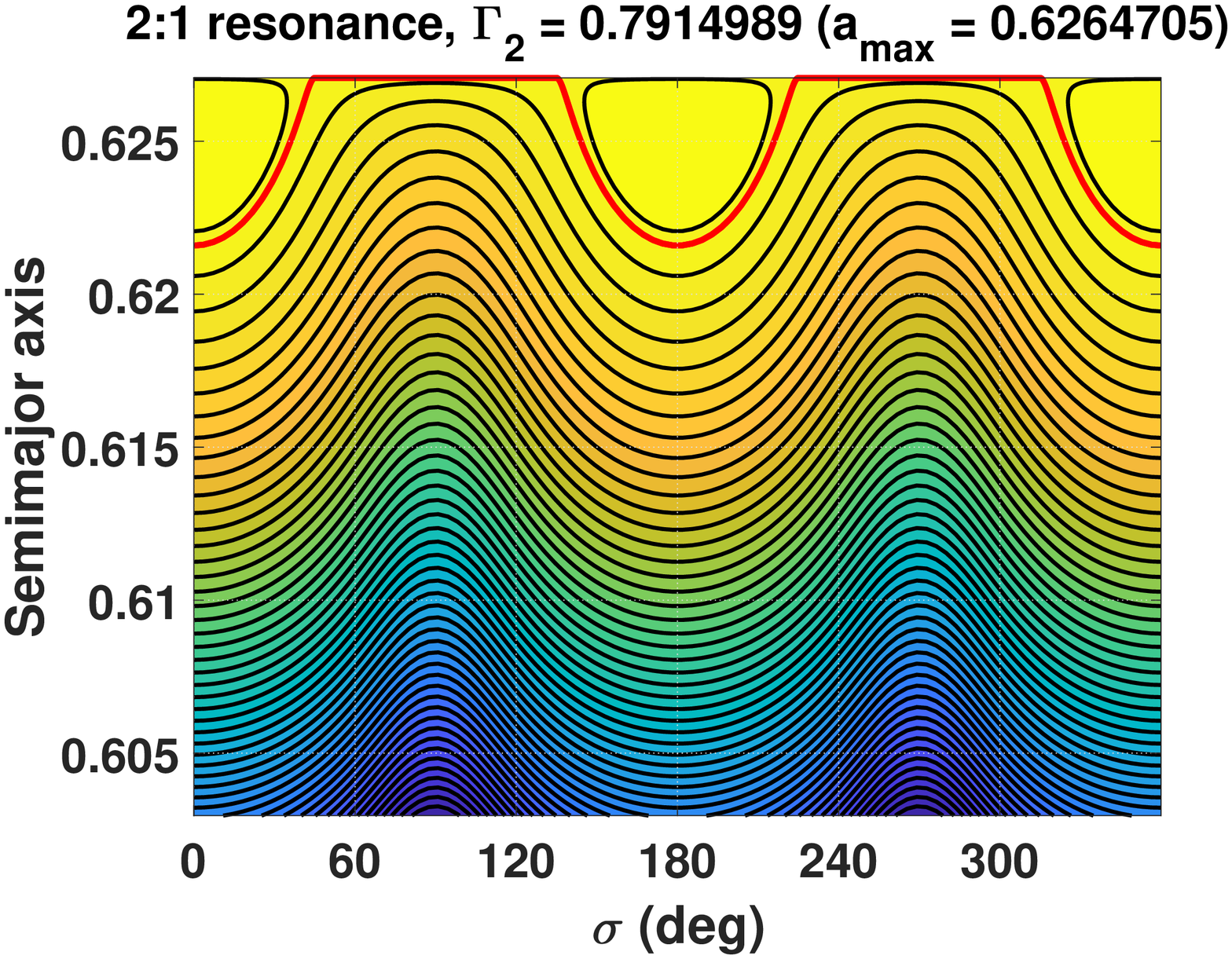}
\includegraphics[width=0.38\textwidth]{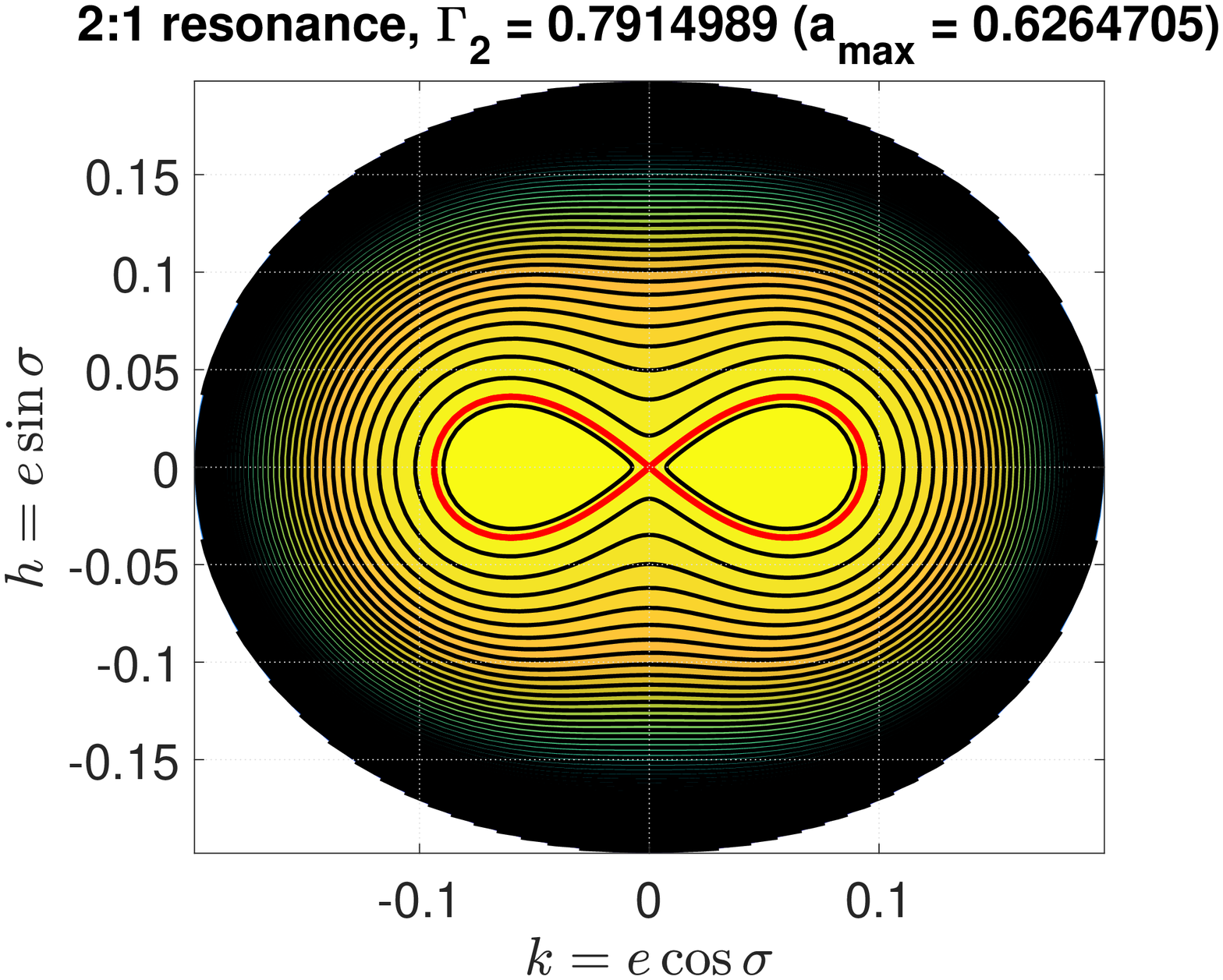}\\
\includegraphics[width=0.42\textwidth]{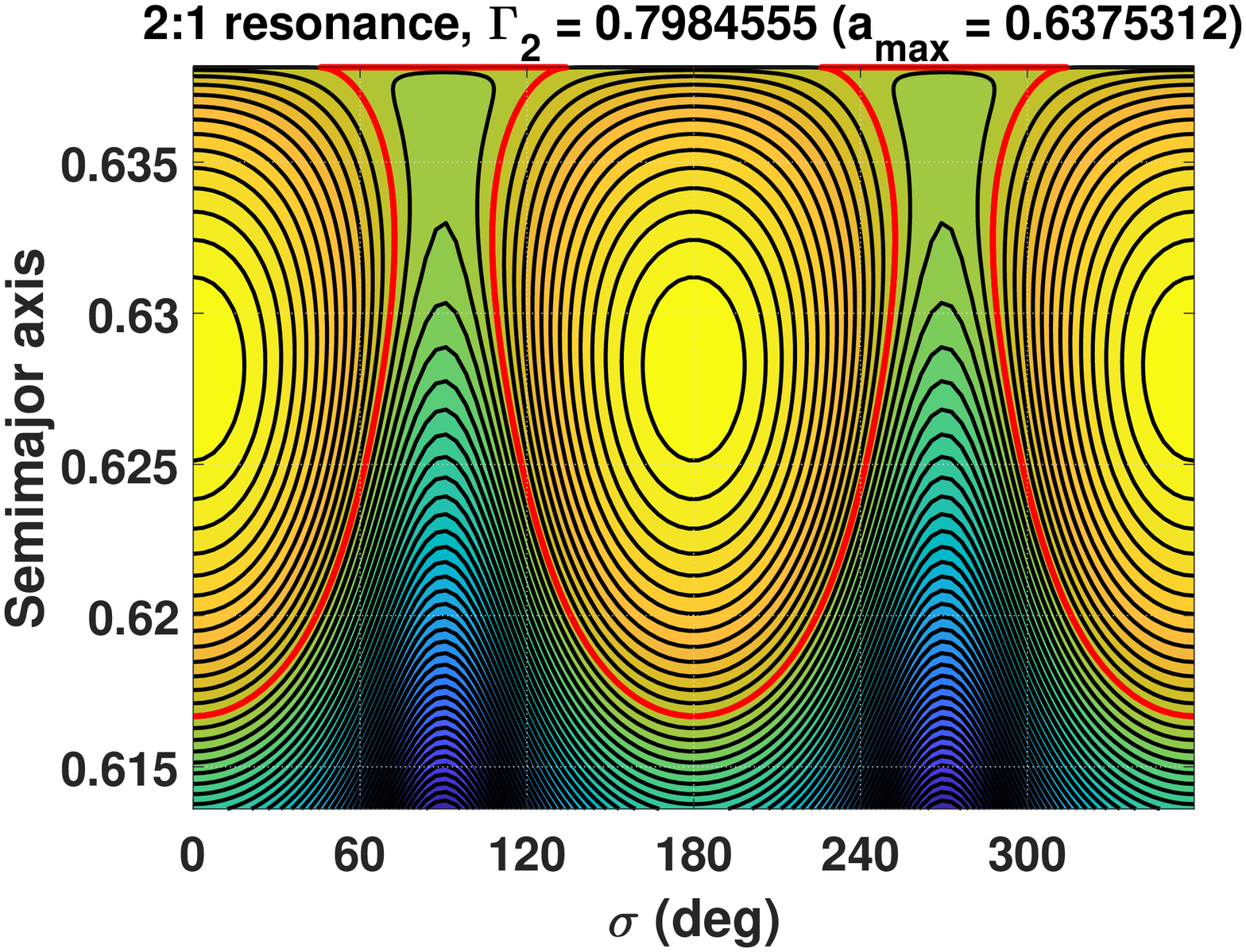}
\includegraphics[width=0.38\textwidth]{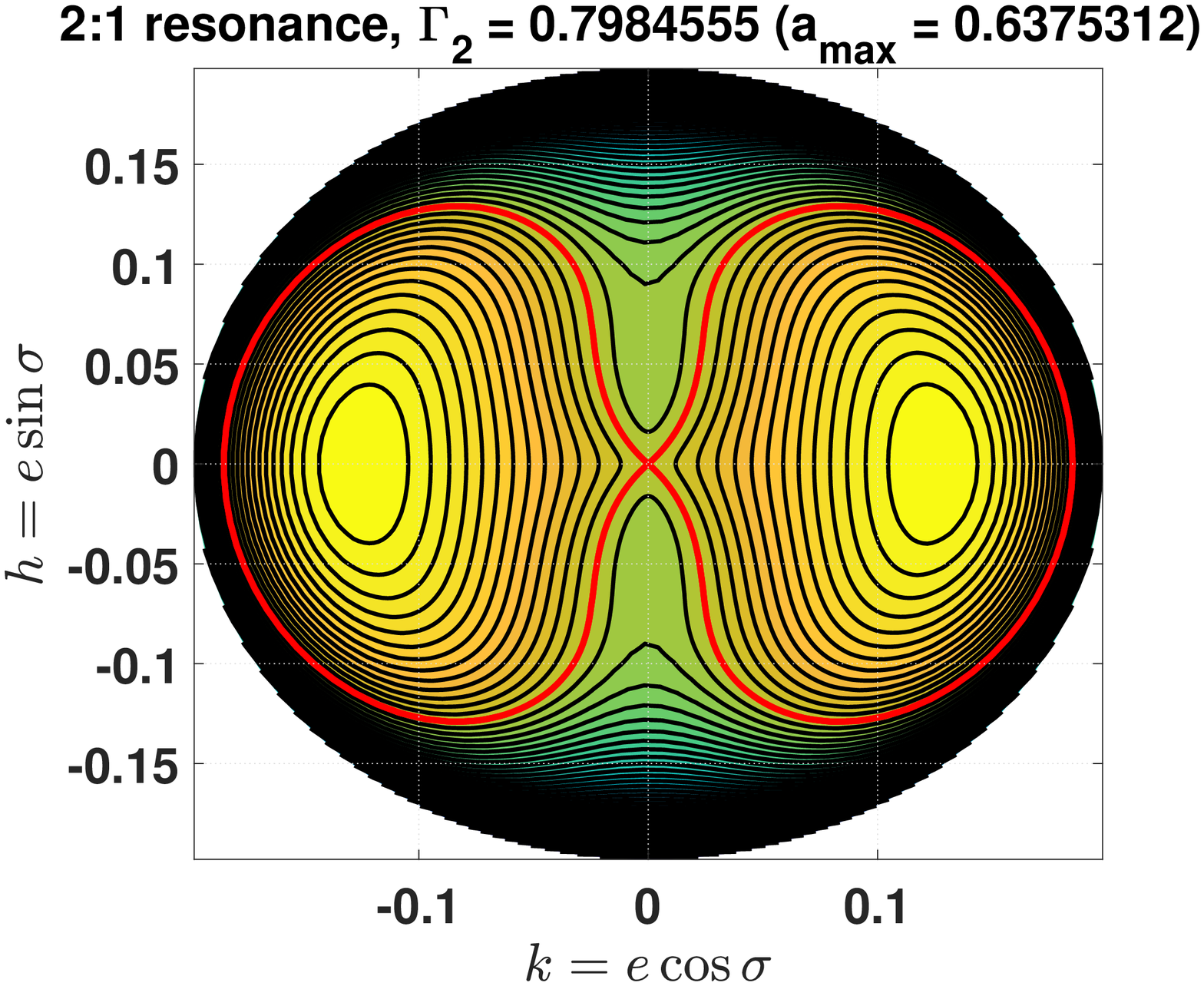}\\
\includegraphics[width=0.42\textwidth]{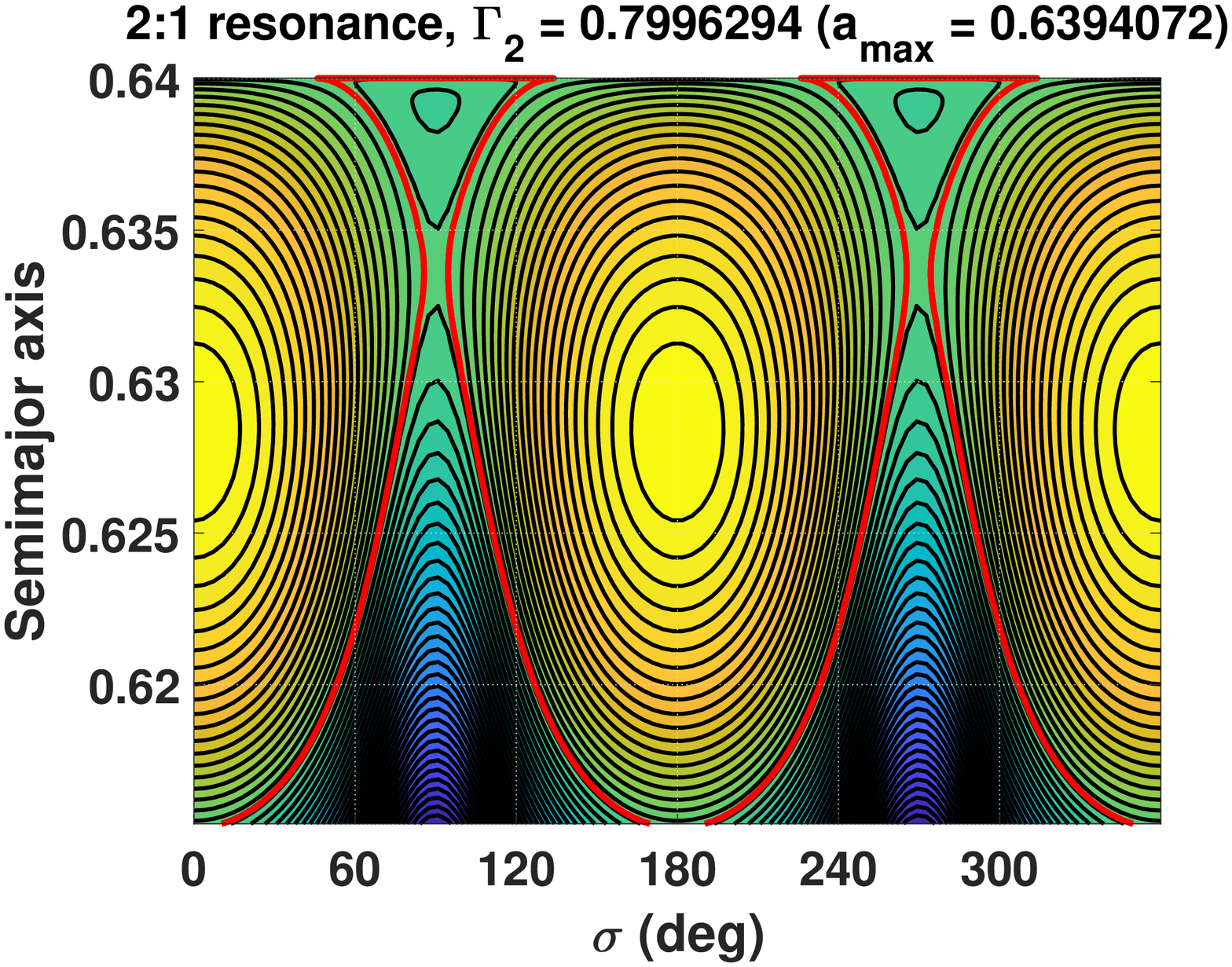}
\includegraphics[width=0.38\textwidth]{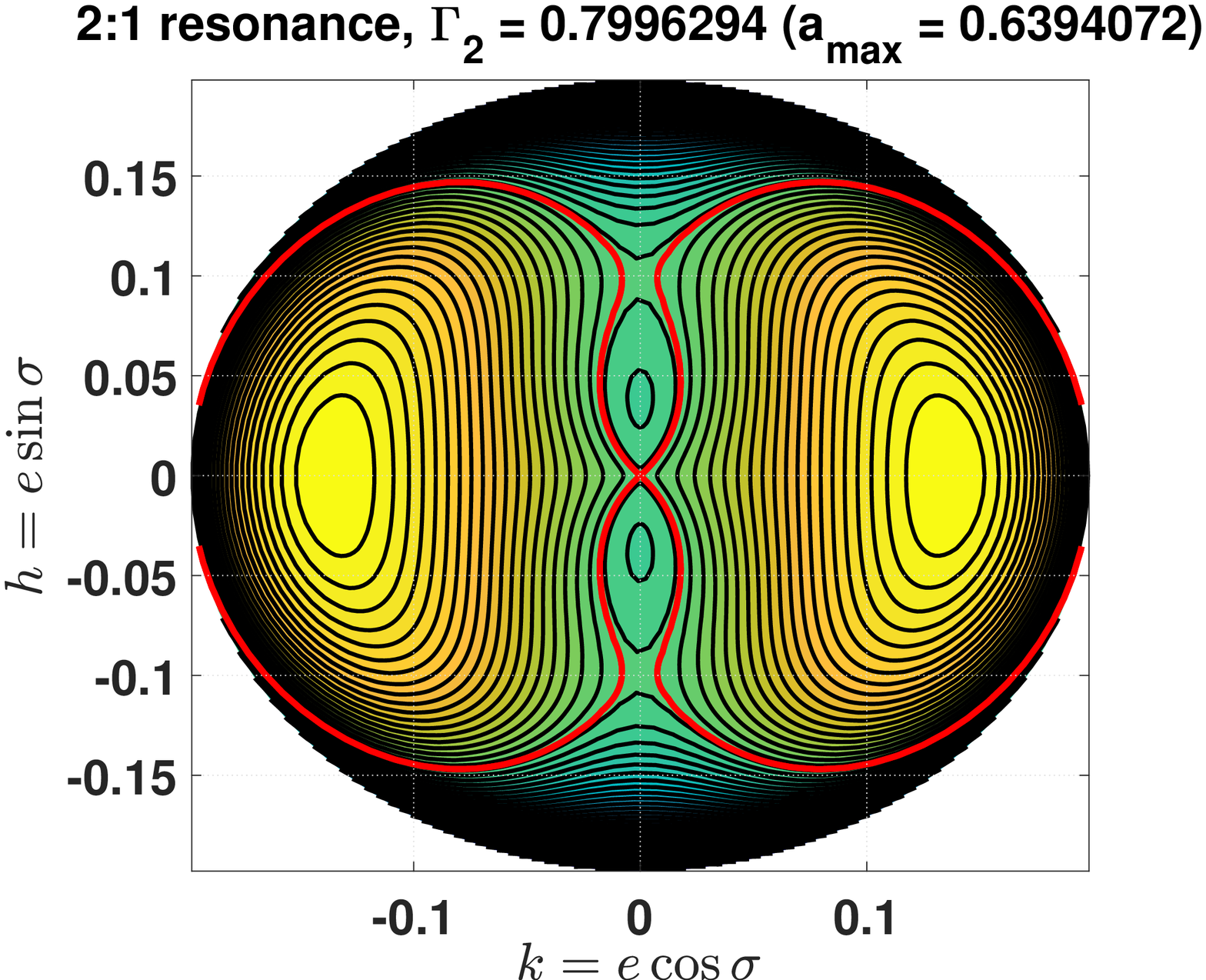}
\caption{Level curves of the resonant Hamiltonian associated with the inner 2:1 resonance specified by three values of the motion integral $\Gamma_2$ (i.e., three values of $a_{\max}$). The panels in the left column display the phase structures in the $(\sigma, a)$ plane and the panels in the right column correspond to the phase structures shown in the $(k,h)=(e \cos{\sigma}, e \sin{\sigma})$ plane. In each plot, the dynamical separatrix passing through the zero-eccentricity saddle point is marked in red line. For the inner 2:1 resonance, the critical motion integral is $\Gamma_2 = N_c = 0.7984555$ (corresponding to the middle panels). It is interesting to observe that the phase portraits shown here have the same topological structures as the Poincar\'e sections numerically produced by \citet{malhotra2020divergence} (see Fig. 2 in their work). Normalized units are used for the semimajor axis.}
\label{Fig4}
\end{figure*}

As for the first-order inner resonances, we mainly study the cases of 2:1, 3:2 and 4:3 resonances. With the inner 2:1 resonance as an example, Fig. \ref{Fig4} reports the level curves of the resonant Hamiltonian (phase portraits) in the $(\sigma, a)$ plane (see the panels in the left column) and in the $(e \cos{\sigma}, e \sin{\sigma})$ plane (see the panels in the right column) for three values of the motion integral $\Gamma_2$. In each plot, the level curve passing through the zero-eccentricity saddle point is denoted in red line.

For the 2:1 resonance, there is a critical value of $\Gamma_2$, denoted by $N_c = 0.7984555$ (the associated phase portraits are shown in the middle-row panels of Fig. \ref{Fig4}). When the motion integral satisfies $\Gamma_2 \le N_c$, there are two islands of resonance centered at $\sigma = 0$ and $\pi$ (corresponding to $\varphi = 0$ which belongs to pericentric branch) and one saddle point located at the zero-eccentricity point. When the motion integral increases higher than $N_c$, besides the libration centers at $\sigma = 0, \pi$, an additional pair of islands of resonance centered at $\sigma = \pm \pi/2$ (corresponding to $\varphi = \pi$ which belongs to apocentric branch) appear. In addition, in the case of $\Gamma_2 > N_c$, besides the zero-eccentricity saddle point, an additional pair of saddle points with non-zero eccentricity appear at $\sigma = \pm \pi/2$.

Evidently, when $\Gamma_2 \le N_c$, only the pericentric branch of libration centers exists and, when $\Gamma_2 > N_c$, both the pericentric and apocentric branches of libration centers exist. In other words, $N_c$ is the critical value of the motion integral, at which the bifurcation of equilibrium points occurs.

From Fig. \ref{Fig4}, we can observe that (a) when $\Gamma_2 \le N_c$ (see the panels in the first two rows) the pericentric libration zones centered at $\sigma = 0, \pi$ are bounded by the separatrix stemming from the zero-eccentricity saddle point, and (b) when $\Gamma_2 > N_c$ (see the panels in the last row) the pericentric libration zones centered at $\sigma = 0, \pi$ are bounded by the separatrix stemming from the zero-eccentricity saddle point and the apocentric libration zones centered at $\sigma = \pm \pi/2$ are bounded by the separatrices emanating from the other pair of saddle points with non-zero eccentricity.

As stated in Section \ref{Sect3}, the second resonant model with $\sigma = \varphi/k_p$ as the resonant angle holds such an advantage that it is possible to make a direct correspondence between the phase portraits of the analytical model and Poincar\'e sections. To this end, we could compare the phase-space structures shown in Fig. \ref{Fig4} with the Poincar\'e sections provided by \citet{malhotra2020divergence} (see the panels in the first two columns of Fig. 2 in their study). The phase portraits given in the current work are specified by the motion integral $\Gamma_2$, while the Poincar\'e sections made in \citet{malhotra2020divergence} are characterized by the Jacobi constant. Through comparisons, it is interesting to observe that the phase portraits in the current work are in agreement with the structures arising in the Poincar\'e sections.

For the other two inner resonances including the 3:2 and 4:3 resonances, their phase-space structures are presented in the $(e \cos{\sigma}, e \sin{\sigma})$ plane, as shown in Appendix \ref{A_2} (see Figs. \ref{FigA2} and \ref{FigA3} for details). In all these phase portraits, it is observed that the zero-eccentricity point is always a visible saddle point of the resonant model.

\begin{figure*}
\centering
\includegraphics[width=0.33\textwidth]{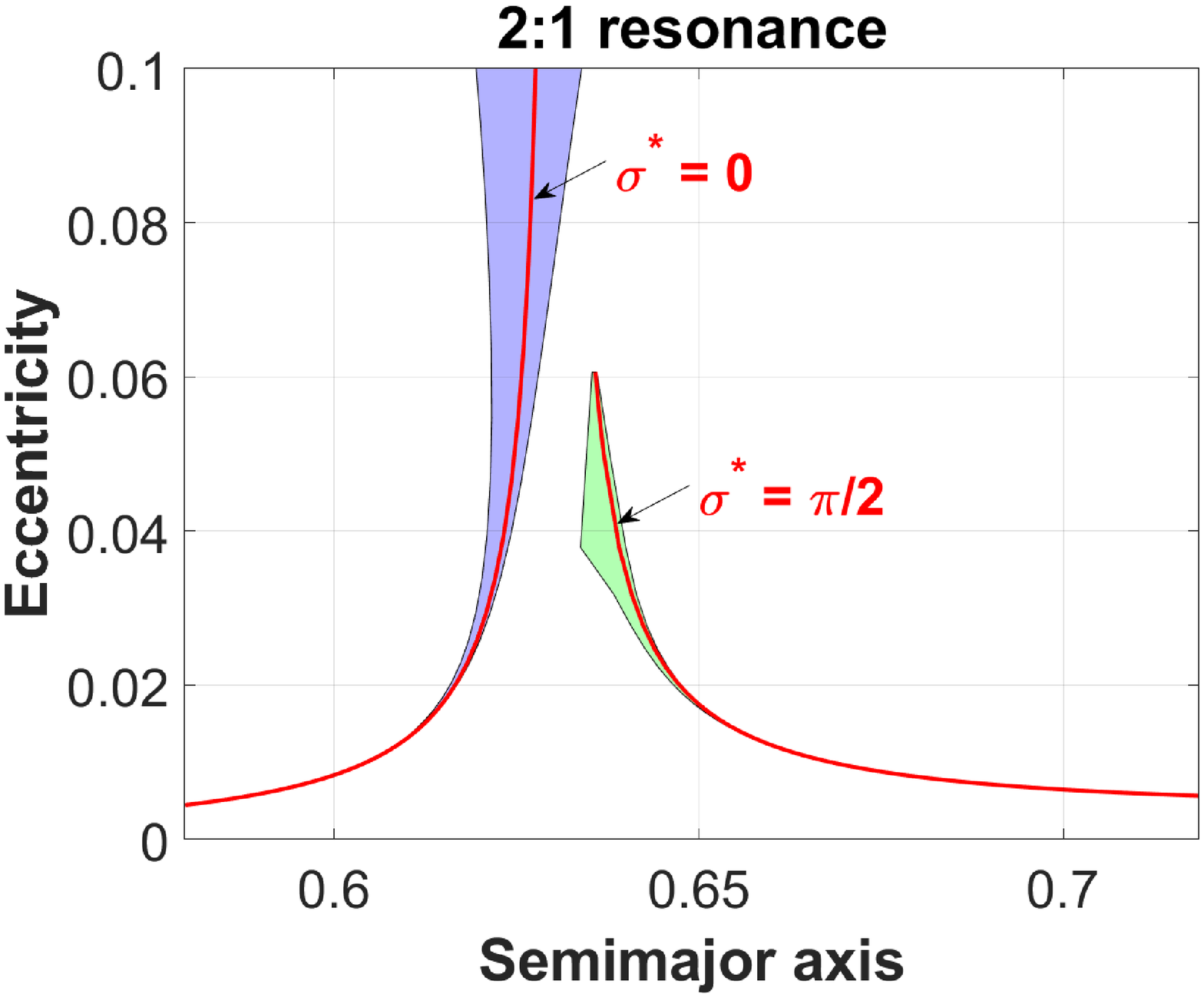}
\includegraphics[width=0.33\textwidth]{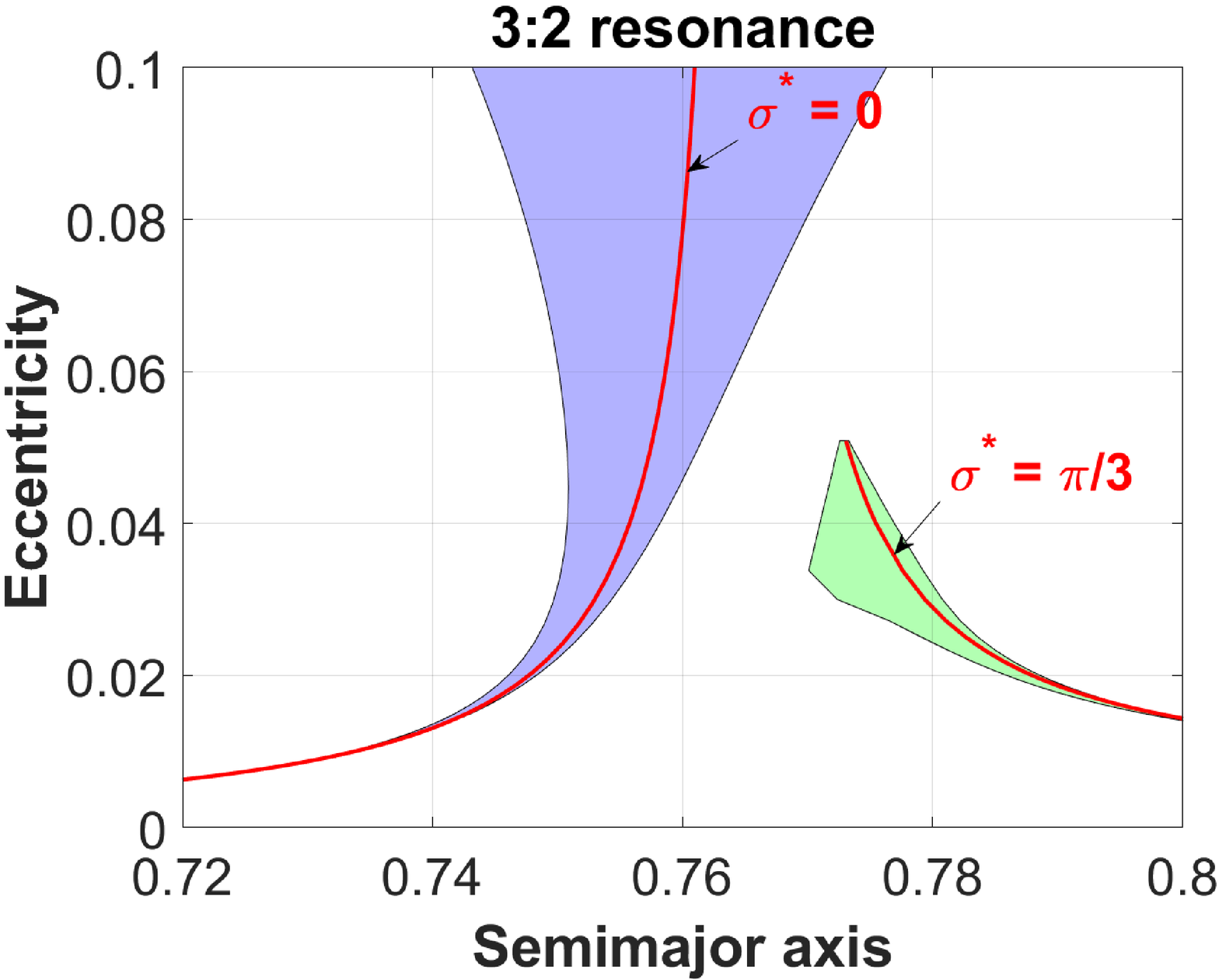}
\includegraphics[width=0.33\textwidth]{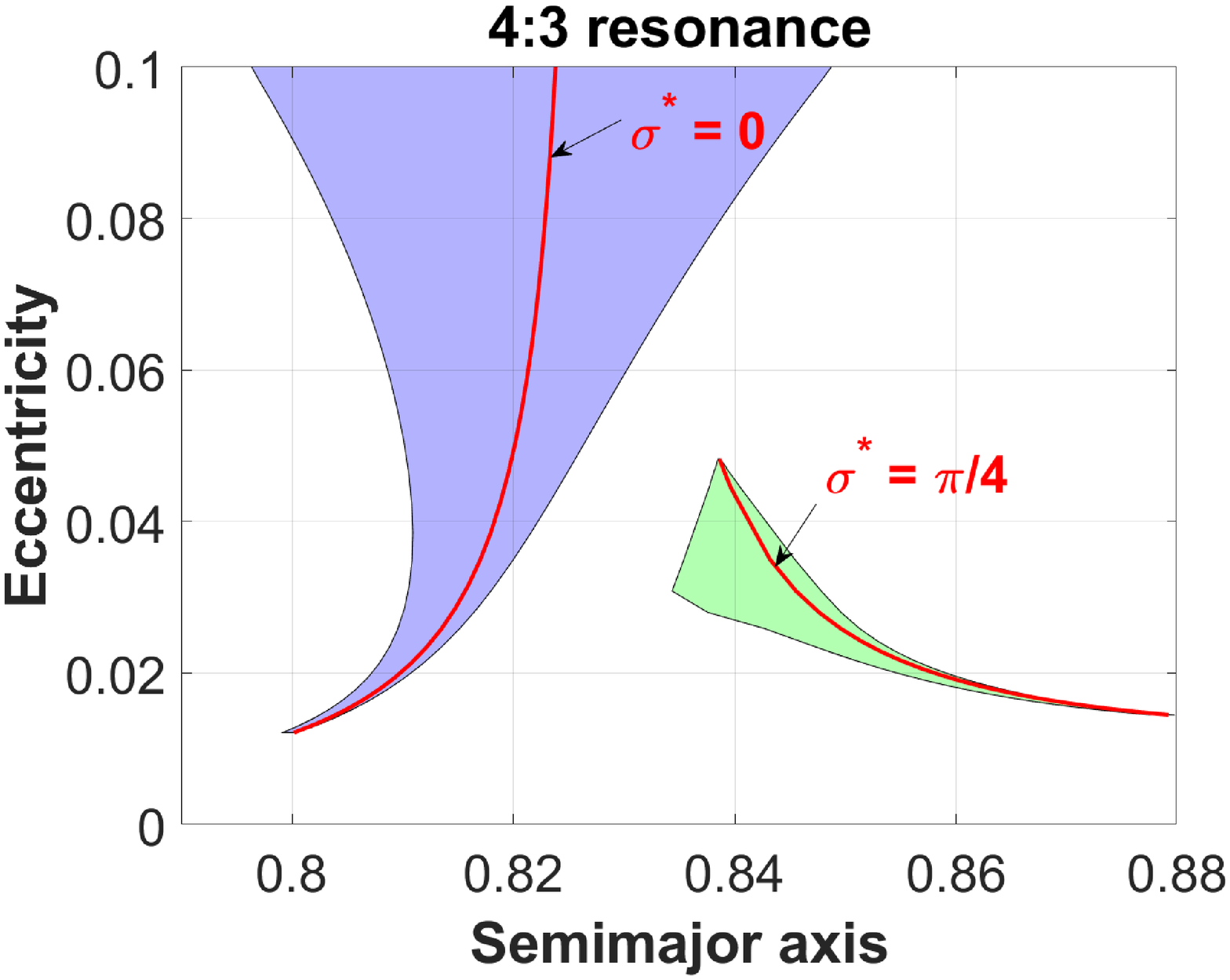}
\caption{The libration center and resonant width in terms of the variation of semimajor axis ($\Delta a$) of the inner 2:1, 3:2 and 4:3 resonances (from the left to right panels) for eccentricities covering from zero to 0.1. For each resonance, there are two families of libration centres, one is at $\sigma^* = 0$ (corresponding to $\varphi^* = 0$) which belongs to the pericentric branch and the other one is at $\sigma^* = \pi/{k_p}$ (corresponding to $\varphi^* = \pi$) which belongs to the apocentric branch. Normalized units are used for the semimajor axis.}
\label{Fig5}
\end{figure*}

Next, let's discuss the libration centers and resonant widths. Following the discussions carried out in Section \ref{Sect5-1}, the location of libration center and the associated resonant width in terms of the variation of semimajor axis can be analytically identified. The phase-space structures shown in Fig. \ref{Fig4} and Figs. \ref{FigA2} and \ref{FigA3} indicate that, for the $k_p$:$k$ resonance, there are $k_p$ libration centers in the pericentric branch (corresponding to $k_p$ islands of resonance) and, along the apocentric branch, there are $k_p$ libration centers in the case of $\Gamma_2 > N_c$ or zero libration center in the case of $\Gamma_2 \le N_c$. Theoretically speaking, all these $k_p$ libration centers in one branch have the same dynamical behaviors, so that we only consider one of them in each branch. In particular, in the pericentric branch, the libration center at $\sigma^* = 0$ is taken into account and, in the apocentric branch, the libration center at $\sigma^* = \pi/k_p$ is considered.

In Fig. \ref{Fig5}, the characteristic curves of two families of libration centers (including the pericentric and apocentric branches) are distributed in the $(a,e)$ plane and marked in red lines. The left panel is for the 2:1 resonance, the middle one for the 3:2 resonance and the right one for the 4:3 resonance. It is observed from Fig. \ref{Fig5} that, for each resonance considered, the characteristic curve of the pericentric branch extends from the nominal resonance location towards the left of the plot (i.e. being far away from the planet), while the characteristic curve of the apocentric branch goes from the nominal resonance location towards the right of the plot (i.e. being close to the planet).

For the libration center $(a_0,e_0)$, the dynamical separatrix passing through the saddle point with the closest Hamiltonian provides boundaries for the libration zones (see the phase-space structures in Fig. \ref{Fig4} for the detailed geometry), as stated in Section \ref{Sect5-1}. Let us denote the semimajor axes at the boundaries by $a_{L}$ and $a_{R}$. Thus, the resonant width measured by the variation of semimajor axis can be expressed by $\Delta a = a_{R} - a_{L}$. About the resonant width shown in Fig. \ref{Fig5}, the left boundary corresponds to the locus of $(a_{L}, e_0)$ and the right boundary is for the locus of $(a_{R}, e_0)$ when the motion integral is varied in a given interval. Note that this type of representation method about resonant width is adopted by \citet{malhotra2020divergence}.

From Fig. \ref{Fig5}, we can observe that (a) with the eccentricity approaching zero, the characteristic curves of the pericentric and apocentric families of libration centers diverge away from the nominal resonance location (one is extending to the left side and the other one to the right side), (b) in the pericentric branch, the resonant width in terms of $\Delta a$ is a monotonically increasing function of the eccentricity, and (c) in the apocentric branch, the resonant width in terms of $\Delta a$ is first an increasing function and then a decreasing function of the eccentricity.

Regarding these first-order inner resonances (including the 2:1, 3:2 and 4:3 resonances), \citet{malhotra2020divergence} have numerically explored the resonant width by analyzing the Poincar\'e sections (non-perturbative analysis). Through comparing the analytical results given in the current work with the numerical results in \citet{malhotra2020divergence}, we can see that our analytical resonant widths shown in Fig. \ref{Fig5} are in perfect agreement with the numerical widths provided by \citet{malhotra2020divergence}, as expected. This means that our multi-harmonic Hamiltonian model (the second resonant model) is valid in predicting the location of libration center and the associated resonant width for those first-order mean motion resonances.

\subsection{Applications to outer resonances}
\label{Sect5-3}

In this section, we apply our multi-harmonic Hamiltonian model to the first-order outer resonances. In practical simulations, the outer 2:3 and 3:4 resonances are taken into consideration.

\begin{figure*}
\centering
\includegraphics[width=0.42\textwidth]{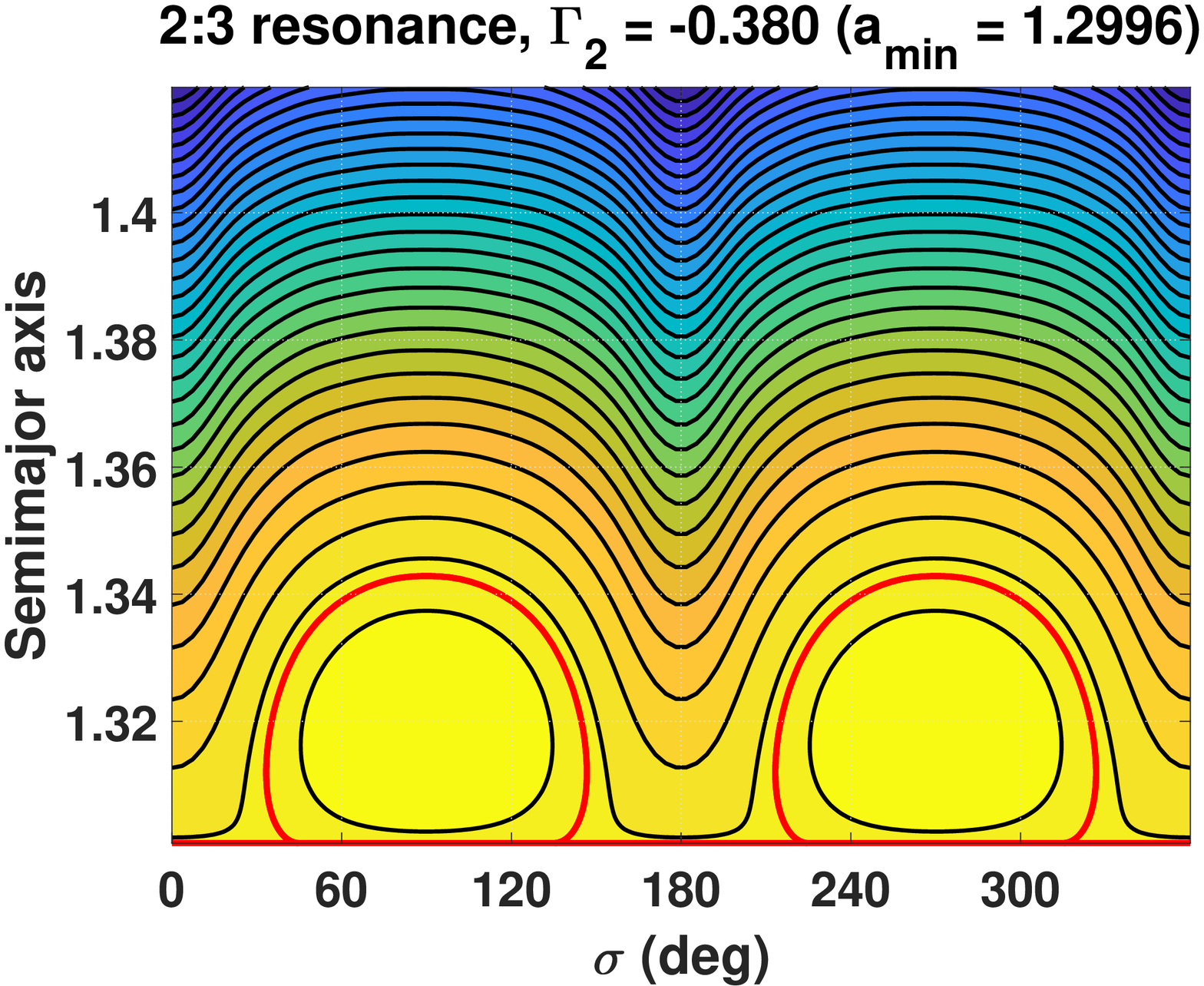}
\includegraphics[width=0.40\textwidth]{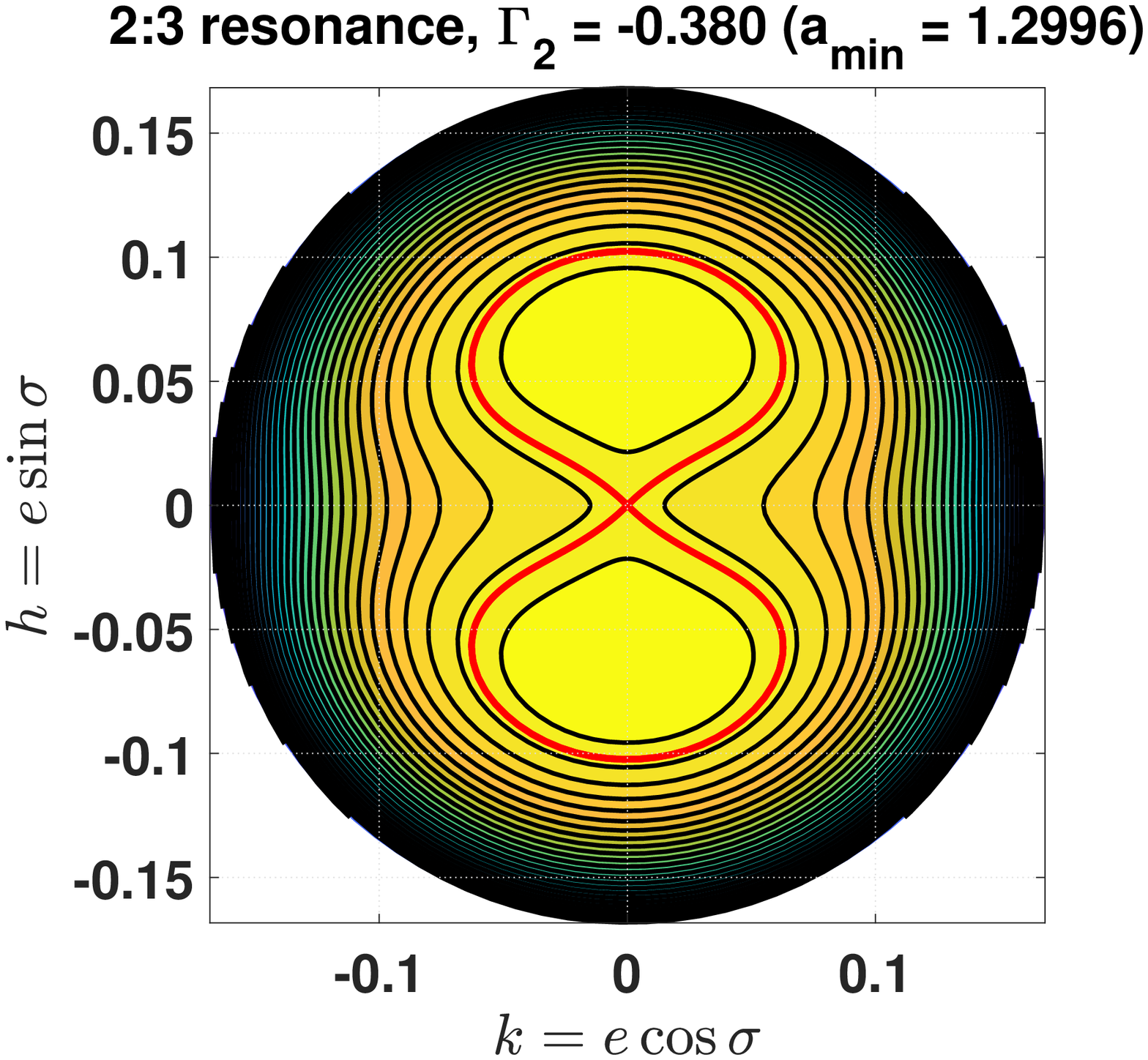}\\
\includegraphics[width=0.42\textwidth]{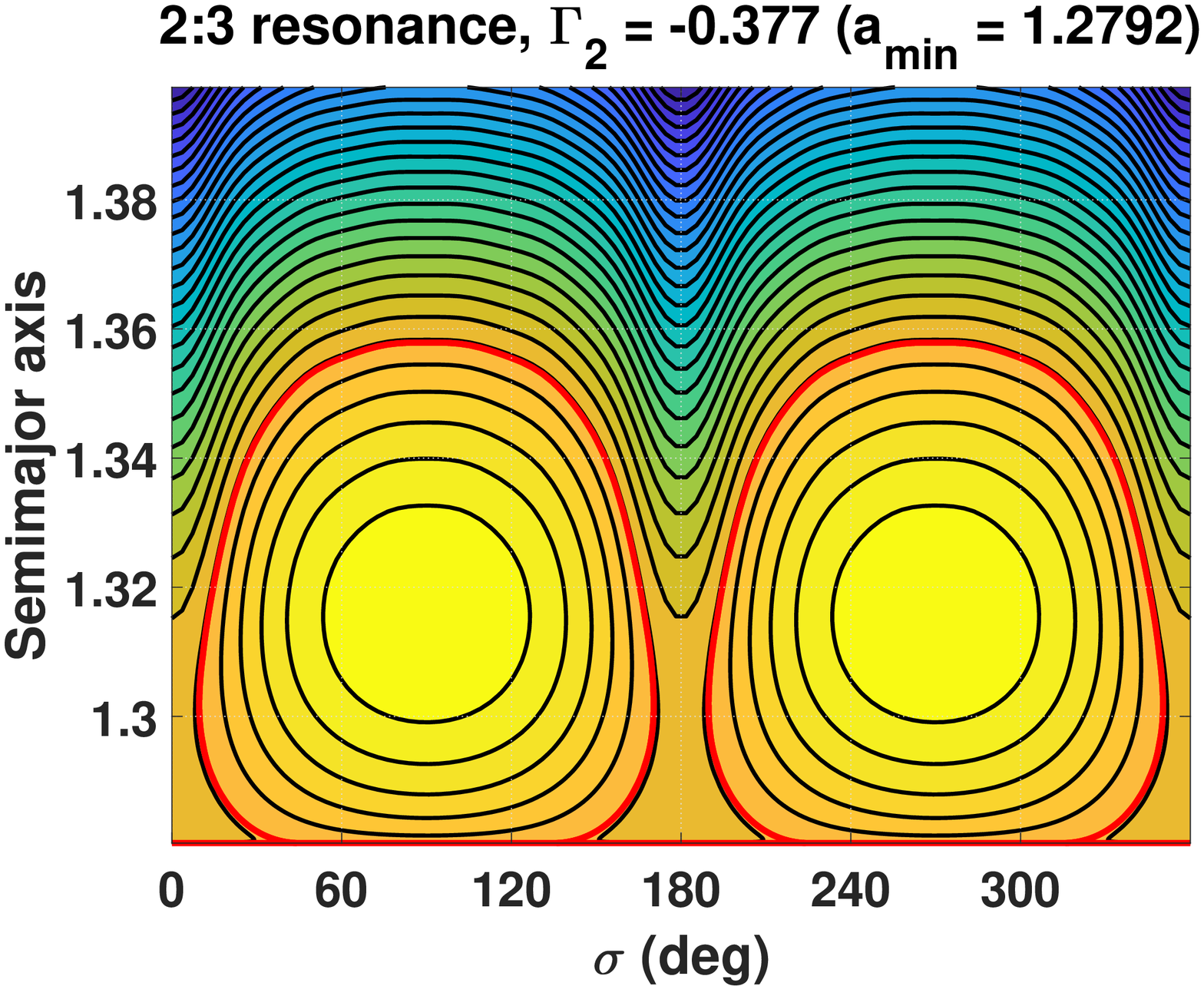}
\includegraphics[width=0.40\textwidth]{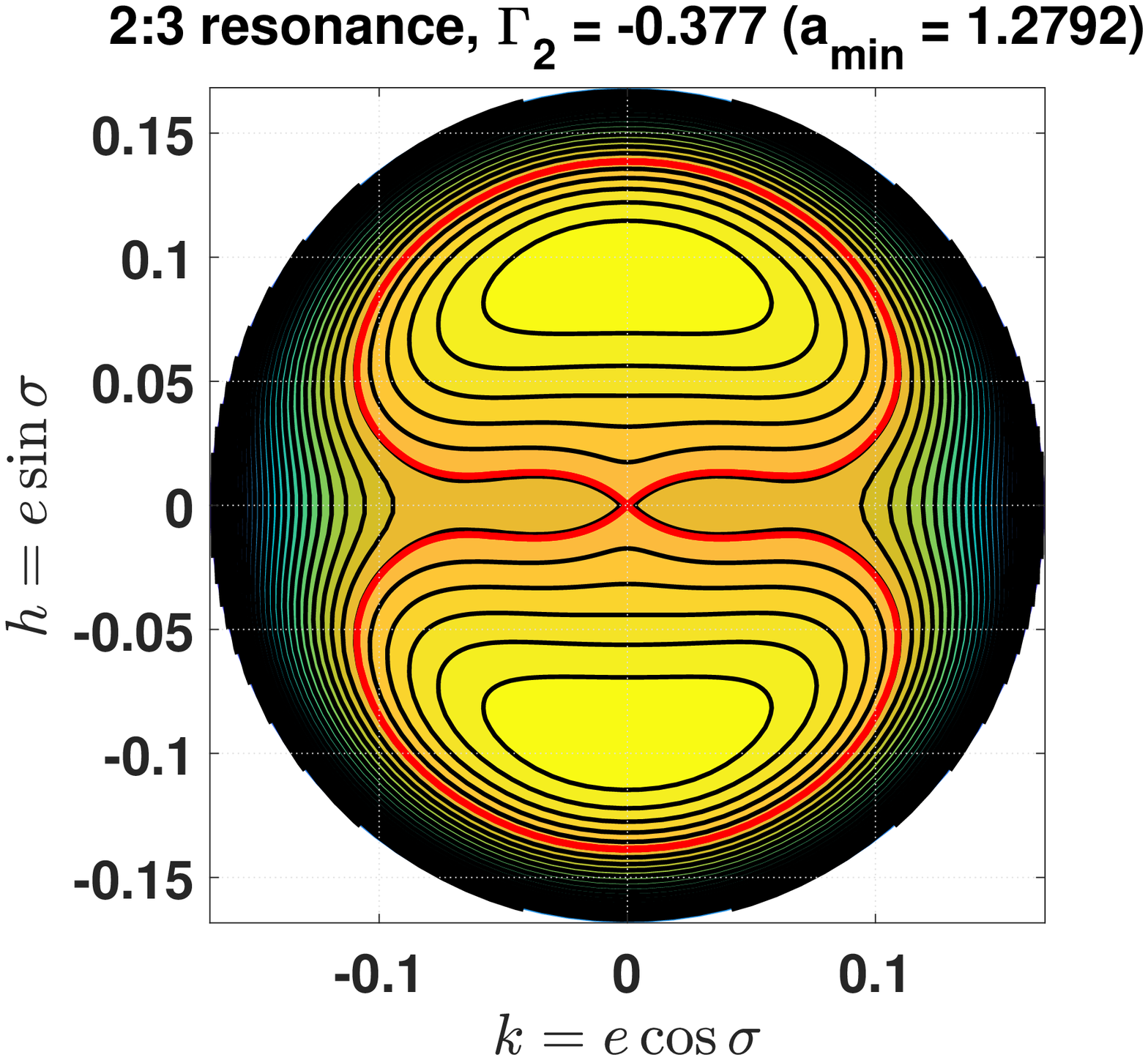}\\
\includegraphics[width=0.42\textwidth]{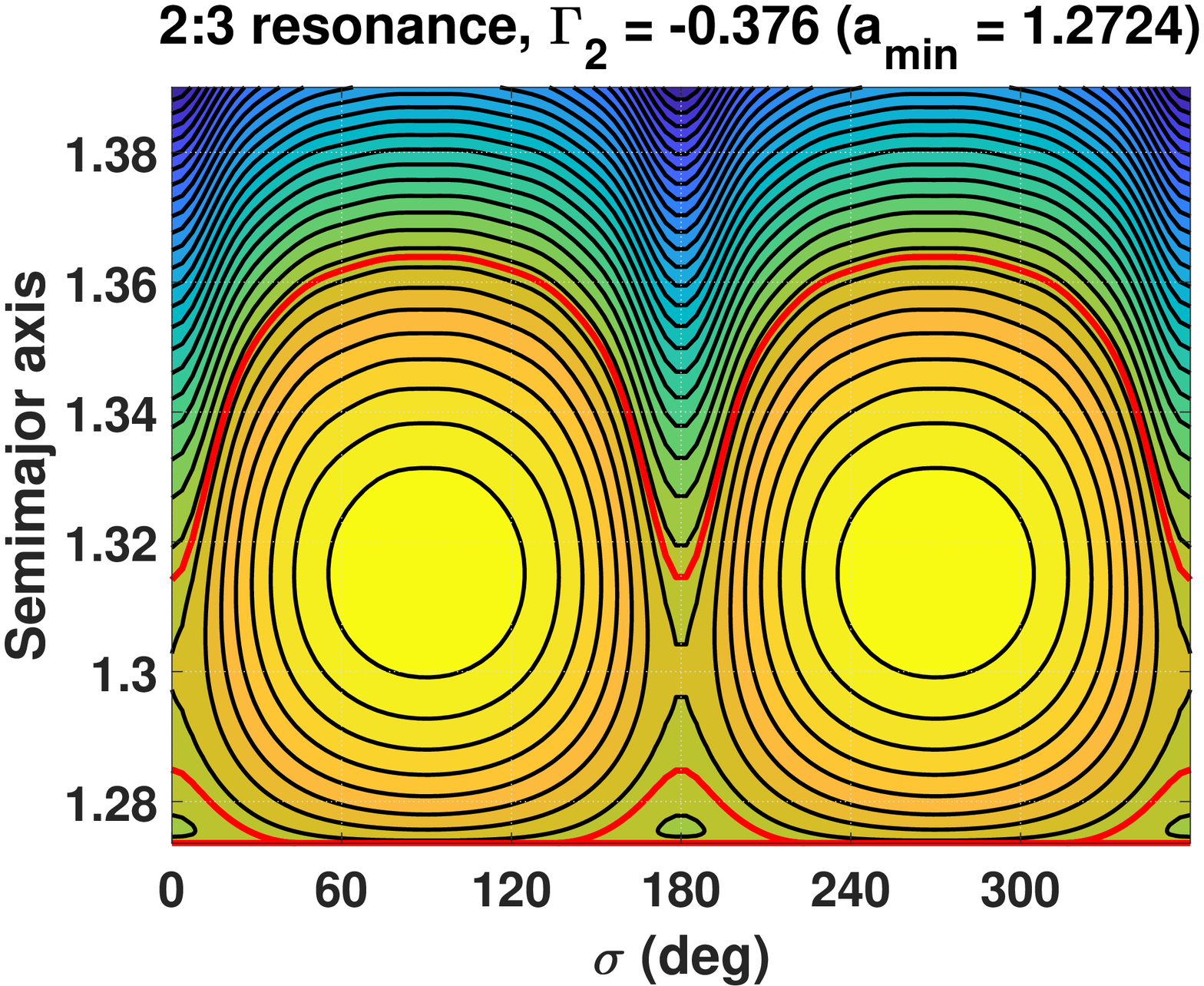}
\includegraphics[width=0.40\textwidth]{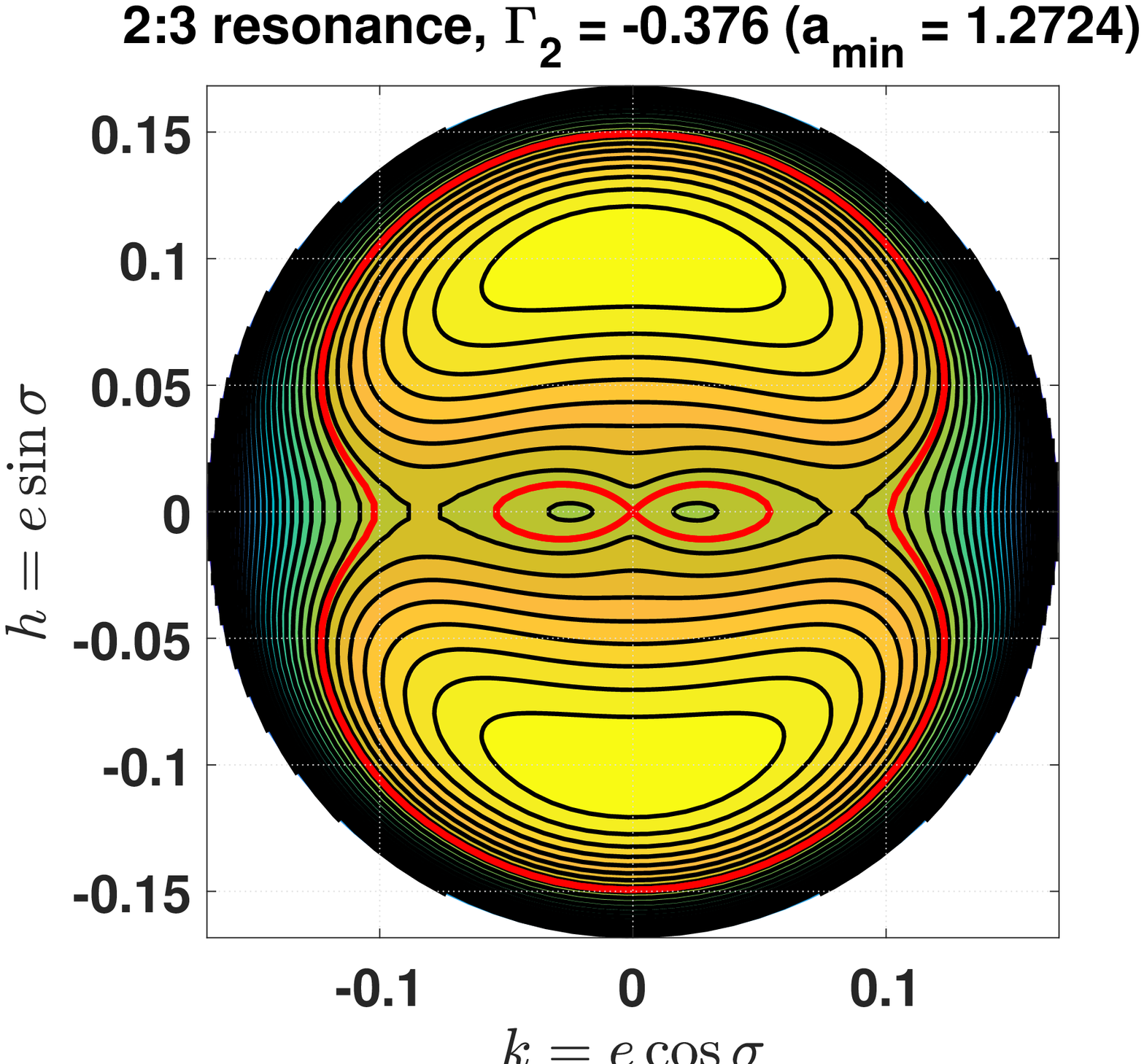}
\caption{Level curves of the resonant Hamiltonian associated with the outer 2:3 resonance specified by three values of the motion integral $\Gamma_2$ (i.e., three values of $a_{\min}$). The left panels display the phase structures shown in the $(\sigma, a)$ plane and the right panels in the $(k,h)=(e \cos{\sigma}, e \sin{\sigma})$ plane. For the outer 2:3 resonance, the critical motion integral is $\Gamma_2 = N_c = -0.377$ (corresponding to the value used in the middle panels). Normalized units are used for the semimajor axis.}
\label{Fig6}
\end{figure*}

Figure \ref{Fig6} reports the phase portraits of the 2:3 resonance for three values of the motion integral $\Gamma_2$. The panels in the left column are for the phase-space structures shown in the $(\sigma, a)$ plane, and the panels in the right column are for the phase-space structures shown in the $(e \cos{\sigma}, e \sin{\sigma})$ plane. For the 2:3 resonance, the critical motion integral is $N_c = -0.377$, and the associated phase portraits are shown in the middle row of Fig. \ref{Fig6}. In particular, when the motion integral satisfies $\Gamma_2 \le N_c$ (see the panels in the first two rows of Fig. \ref{Fig6}), there is one pair of stable equilibria located at $\sigma = \pm \pi/2$ (corresponding to $\varphi = \pi$ which belongs to the apocentric branch) and, when the motion integral is greater than $N_c$ (see the bottom-row panels), an additional pair of stable equilibria located at $\sigma = 0, \pi$ (corresponding to $\varphi = 0$ which belongs to the pericentric branch) appears. It is not difficult to conclude that the apocentric branch exists in the entire range of $\Gamma_2$, while the pericentric branch exists under the condition of $\Gamma_2 > N_c$.

From Fig. \ref{Fig6}, we can observe that (a) in the case of $\Gamma_2 \le N_c$, the apocentric libration zones centered at $\sigma = \pm \pi/2$ are bounded by the separatrices stemming from the zero-eccentricity saddle points, (b) in the case of $\Gamma_2 > N_c$, there are two branches of libration centers, where the apocentric libration zones centered at $\sigma = \pm \pi/2$ are bounded by the separatrices stemming from the saddle points with non-zero eccentricity located at $\sigma = 0$ or $\pi$ and the pericentric libration zones centered at $\sigma = 0, \pi$ are bounded by the separatrices stemming from the zero-eccentricity saddle point.

Regarding the outer 3:4 resonance, the phase-space structures are presented in the $(e \cos{\sigma}, e \sin{\sigma})$ plane for three values of motion integral $\Gamma_2$, as shown in Appendix \ref{A_2} (see Fig. \ref{FigA4} for more details).

\begin{figure*}
\centering
\includegraphics[width=0.45\textwidth]{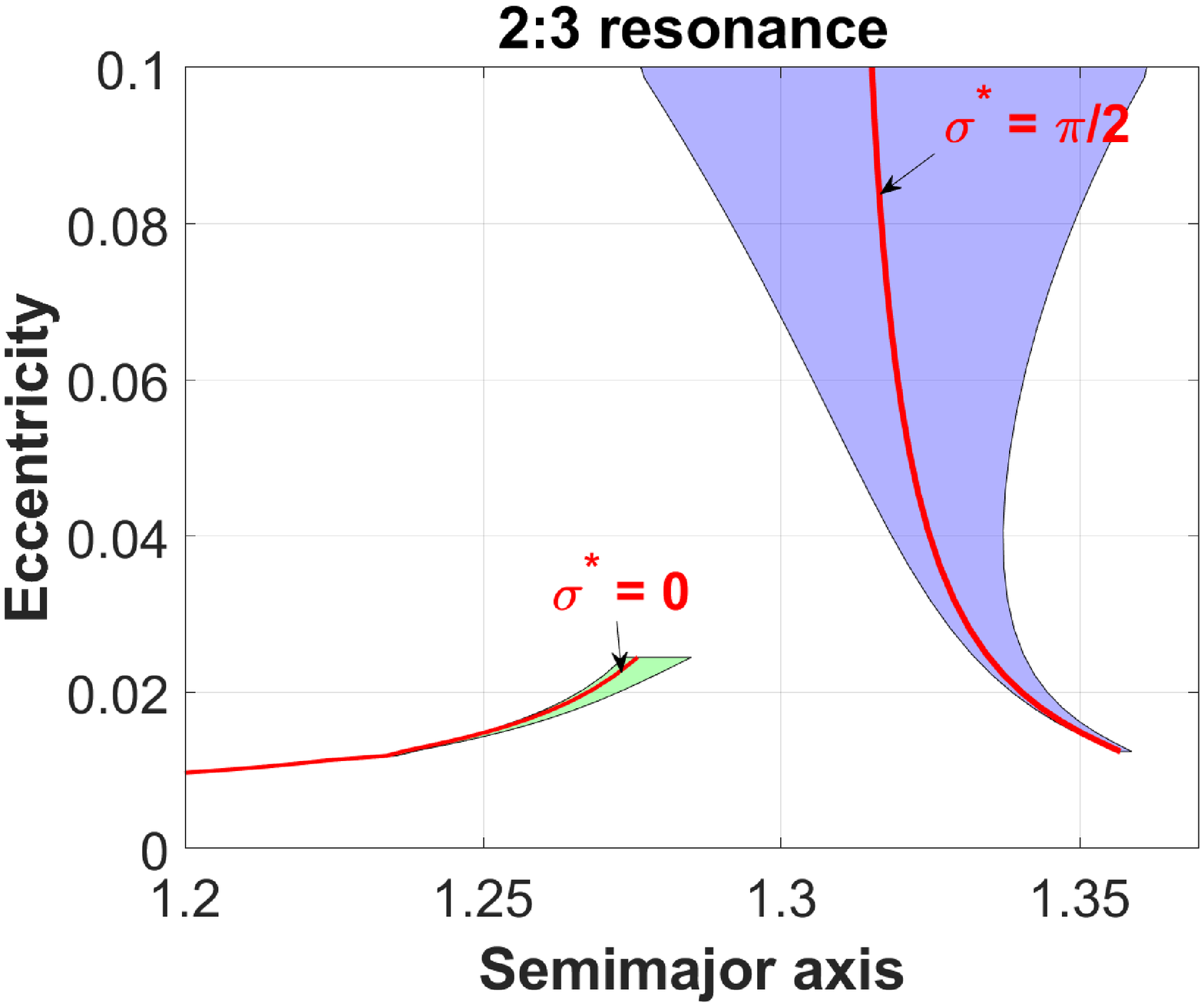}
\includegraphics[width=0.45\textwidth]{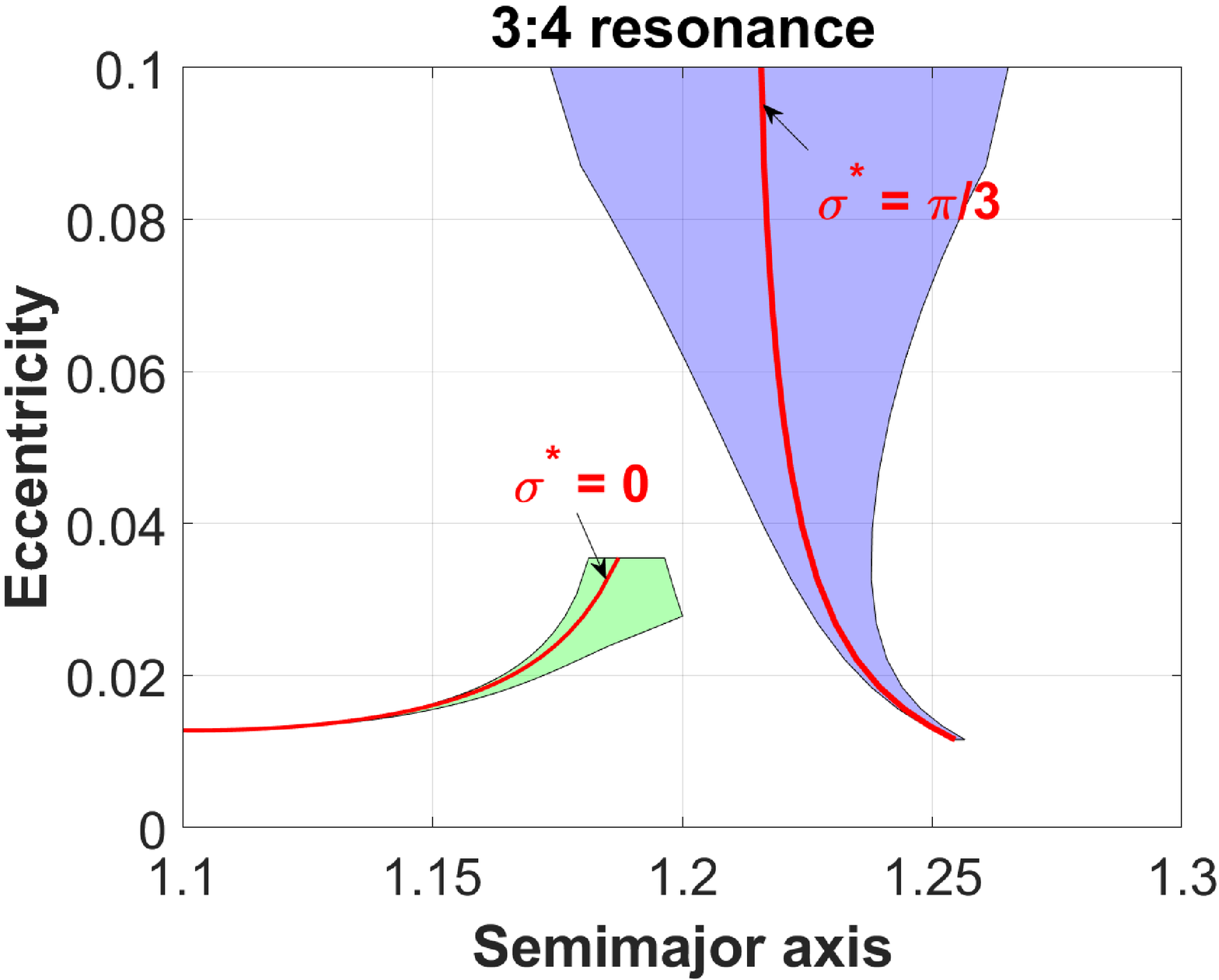}
\caption{The libration center and resonant width in terms of the variation of semimajor axis ($\Delta a$) of the outer 2:3 and 3:4 resonances for eccentricities covering from zero to 0.1. For each resonance, there are two branches of libration centres, in the pericentric branch the libration centres have $\sigma^* = 0$ corresponding to the usual critical argument at $\varphi^* = 0$ and, in the apocentric branch, the libration centres have $\sigma^* = \pi/{k_p}$ corresponding to the usual critical argument at $\varphi^* = \pi$. Normalized units are used for the semimajor axis.}
\label{Fig7}
\end{figure*}

Next, let us analyze the resonant widths associated with outer resonances. Similar to the case of inner resonances discussed in the previous subsection, the location of libration center and the associated resonant width in terms of variation of semimajor axis can be determined by means of the method presented in Section \ref{Sect5-1}. There are two branches of libration centers, corresponding to the pericentric and apocentric resonance zones. In the pericentric branch, the libration center at $\sigma^* = 0$ is considered and, in the apocentric branch, the one at $\sigma^* = \pi/k_p$ is taken into consideration. Figure \ref{Fig7} presents the characteristic curves of the libration centers and the boundaries of the pericentric and apocentric libration zones (the distance between the boundaries stands for the resonant width in terms of variation of semimajor axis). The left panel of Fig. \ref{Fig7} is for the 2:3 resonance, and the right panel is for the 3:4 resonance. It is observed that, as the eccentricity approaches zero, the centers in the pericentric and apocentric branches diverge away from the nominal resonance location and the resonant width decreases to zero.

It is to be noted that, in Figs. \ref{Fig5} and \ref{Fig7}, only the information of resonant width in terms of the variation of semimajor axis is provided, while the information of resonant width in terms of the variation of eccentricity is absent. This is a feature of this type of representation method about resonant width. In the coming subsection, an alternative representation will be introduced to contain the information of both $\Delta a$ and $\Delta e$.

\subsection{An alternative representation of resonant width}
\label{Sect5-4}

In this section, we will adopt an alternative representation for showing resonant width, which contains the variations of both semimajor axis and eccentricity. This type of representation for resonant width has been used in \citet{winter1997resonanceI, morbidelli2002modern} and \citet{ramos2015resonance}.

\begin{figure*}
\centering
\includegraphics[width=0.33\textwidth]{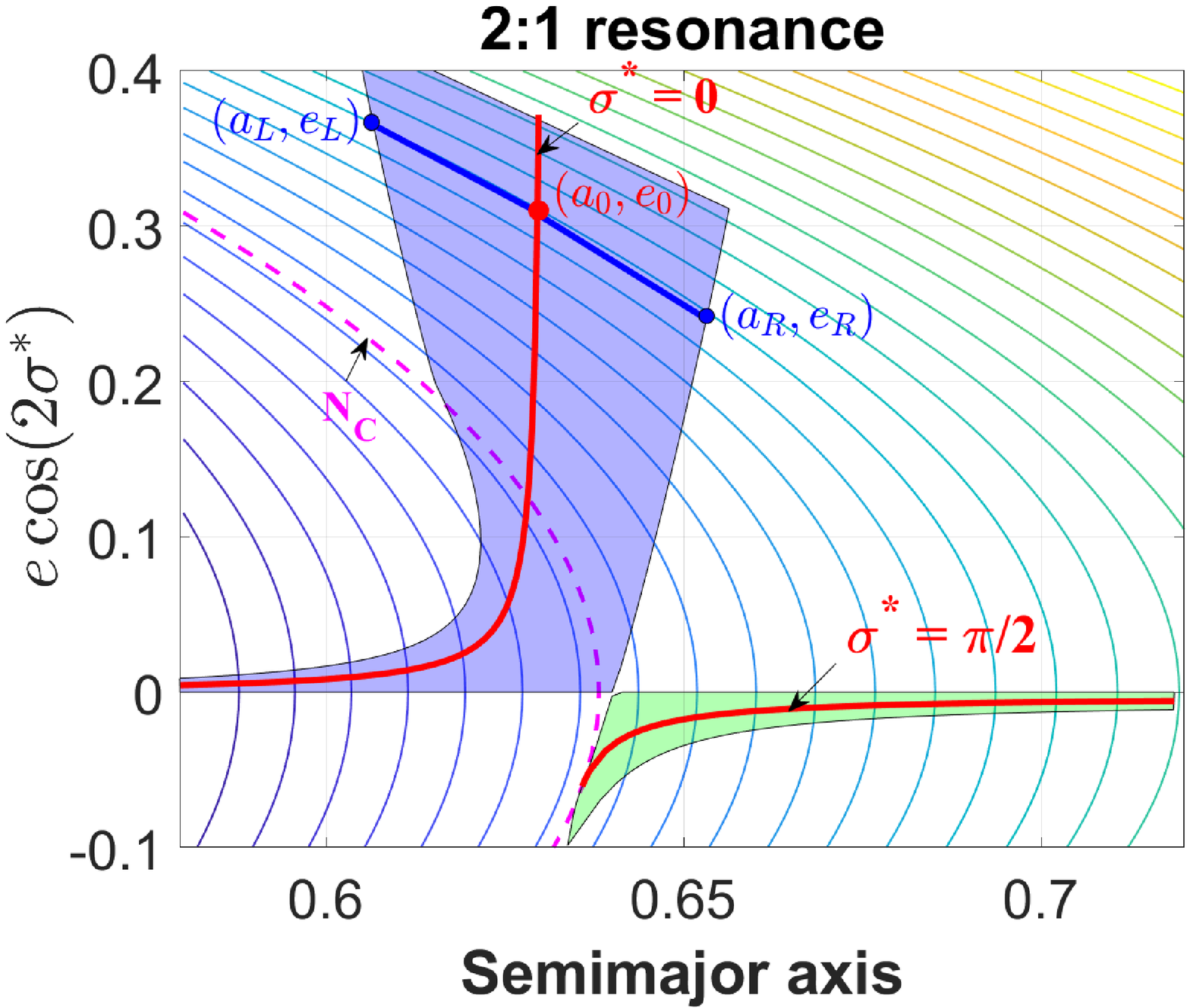}
\includegraphics[width=0.33\textwidth]{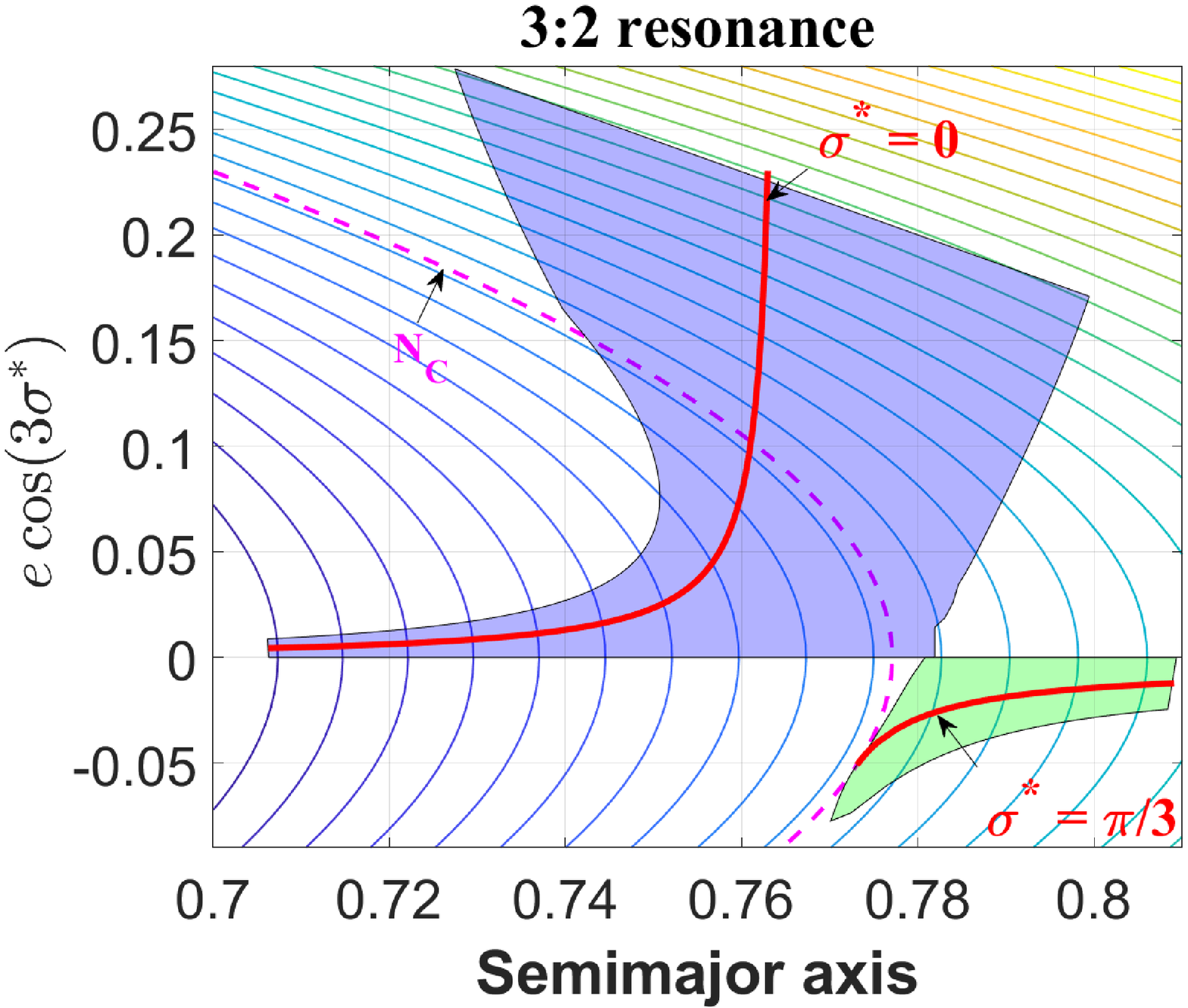}
\includegraphics[width=0.33\textwidth]{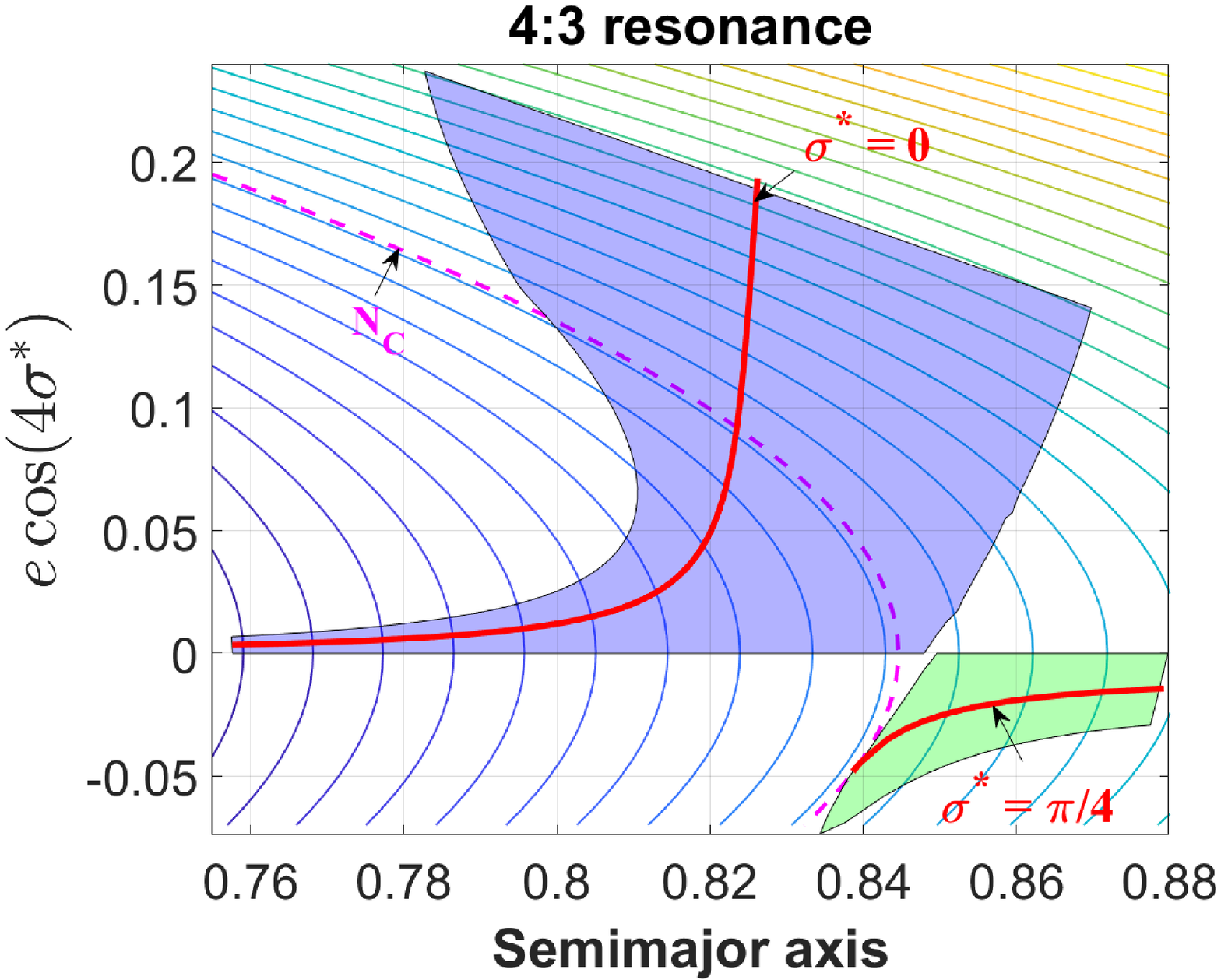}
\caption{Characteristic curves of libration centers in the pericentric and apocentric branches (red lines) and the associated libration zones bounded by the left and right boundaries (shaded areas) for the inner 2:1 (\emph{left panel}), 3:2 (\emph{middle panel}) and 4:3 (\emph{right panel}) resonances. The libration centers located at $\sigma^* = 0$ belong to the pericentric branch, and the ones located at $\sigma^* = \pi/k_p$ belong to the apocentric branch. In the left panel, the points of $(a_L,e_L)$, $(a_0,e_0)$ and $(a_R,e_R)$ for a certain value of $\Gamma_2$ are shown. Some level curves of the motion integral $\Gamma_2$ are also shown, and the dashed lines represent the critical level curves with $\Gamma_2 = N_c$ (it is noted that the test particles can only move along their respective level curves of $\Gamma_2$ specified by the initial conditions, so that the resonant width should be measured along the level curve of $\Gamma_2$, as shown by the blue line in the left panel). Normalized units are used for the semimajor axis shown in the $x$-axis.}
\label{Fig8}
\end{figure*}

It is known that an island of resonance centered at $(a_0, e_0)$ is bounded by the dynamical separatrix stemming from the nearby saddle point, which provides the left and right boundaries for the associated libration zone. For convenience, let us denote the left boundary point by $(a_{L}, e_{L})$ and the right boundary point by $(a_{R}, e_{R})$. It is noted that all these points including the boundary points $(a_{L}, e_{L})$ and $(a_{R}, e_{R})$ and the libration center $(a_0, e_0)$ share the same motion integral $\Gamma_2$. In other words, the following equality satisfies,
\begin{equation*}
\begin{aligned}
{\Gamma _2} &= \sqrt {\mu {a_0}} \left( {\frac{{{k_p}}}{k} - \sqrt {1 - e_0^2} } \right),\\
& = \sqrt {\mu {a_{L}}} \left( {\frac{{{k_p}}}{k} - \sqrt {1 - e_{L}^2} } \right),\\
& = \sqrt {\mu {a_{R}}} \left( {\frac{{{k_p}}}{k} - \sqrt {1 - e_{R}^2} } \right).
\end{aligned}
\end{equation*}
As the motion integral $\Gamma_2$ changes, the locus of $(a_0, e_0)$ corresponds to the characteristic curves of libration centers, the locus of $(a_{L},e_{L})$ provides the left boundary and the locus of $(a_{R},e_{R})$ stands for the right boundary. Please see the left panel of Fig. \ref{Fig8} for the detailed definition.

For the inner resonances including the 2:1, 3:2 and 4:3 resonances, Fig. \ref{Fig8} reports the location of resonant centers shown in red lines and the associated libration zones represented by shaded regions (including the pericentric libration zones centered at $\sigma^* = 0$ and the apocentric libration zones centered at $\sigma^* = \pi/k_p$). For clarity, some level curves of the motion integral $\Gamma_2$ are also shown and, in particular, the dashed line in each plot represents the curve with the critical motion integral $\Gamma_2 = N_c$. In particular, the critical motion integral is $N_c \approx 0.7984555$ for the 2:1 resonance, $N_c \approx 0.4405524$ for the 3:2 resonance and $N_c \approx 0.3061776$ for the 4:3 resonance. For all these inner resonances, it is observed that the pericentric branch of libration centers exists in the entire range of $\Gamma_2$, while the apocentric branch of libration centers (with $\sigma^* = \pi/k_p$) appears under the condition of $\Gamma_2 > N_c$. In particular, when $\Gamma_2 < N_c$ (in the left side of the critical curve with $\Gamma_2 = N_c$ in Fig. \ref{Fig8}), there is only the pericentric branch (the apocentric branch vanishes), in which the island of resonance is bounded by the dynamical separatrix stemming from the zero-eccentricity saddle point.

It is noted that the pericentric libration zones for the inner 2:1, 3:2 and 4:3 resonances shown in Fig. \ref{Fig8} are in agreement with the results given by \citet{winter1997resonanceI} (see Fig. 6 in their work).

\begin{figure*}
\centering
\includegraphics[width=0.45\textwidth]{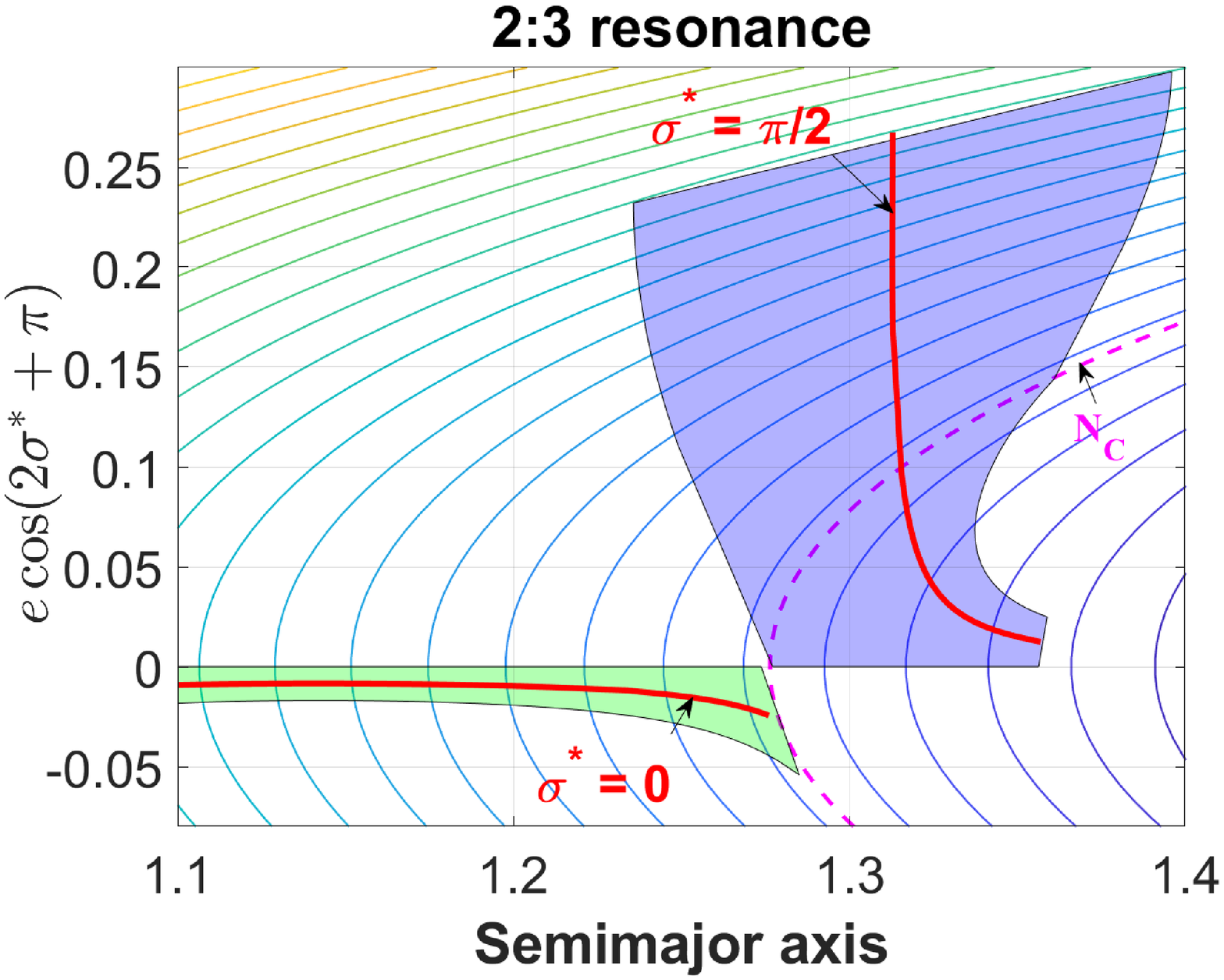}
\includegraphics[width=0.45\textwidth]{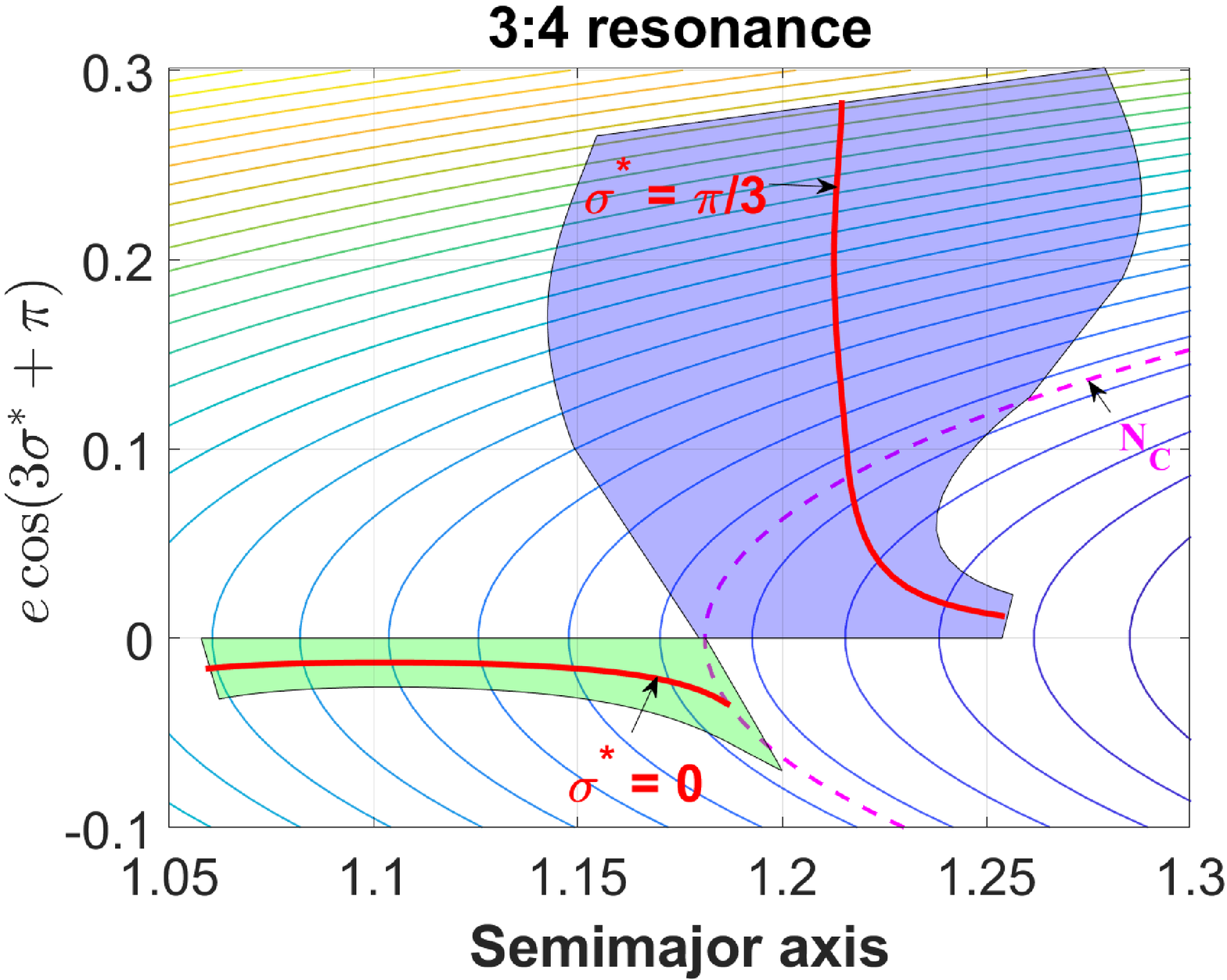}
\caption{Characteristic curves of libration centers in the pericentric and apocentric branches (red lines) and the associated libration zones bounded by the left and right boundaries (shaded areas) for the outer 2:3 (\emph{left panel}) and 3:4 (\emph{right panel}) resonances. The libration centers with $\sigma^* = \pi/k_p$ belong to the apocentric branch, and the ones with $\sigma^* = 0$ belong to the pericentric branch. Some level curves of the motion integral $\Gamma_2$ are shown and, in particular, the dashed lines represent the critical level curves with $\Gamma_2 = N_c$. Normalized units are used for the semimajor axis shown in the $x$-axis.}
\label{Fig9}
\end{figure*}

Regarding the outer 2:3 and 3:4 resonances, Fig. \ref{Fig9} shows the associated libration centers and resonant widths in the $(a,e)$ plane. Also, the dashed lines in Fig. \ref{Fig9} stand for the curve with the critical motion integral (i.e. $\Gamma_2 = N_c$). In particular, the critical motion integral is $N_c \approx -0.377$ for the 2:3 resonance and $N_c \approx -0.2715583$ for the 3:4 resonance. For the outer resonances, the apocentric branch of libration centers (with $\sigma^* = \pi/k_p$) exists in the entire range of $\Gamma_2$, while the pericentric branch (with $\sigma^* = 0$) appears under the condition of $\Gamma_2 > N_c$. In particular, when $\Gamma_2 < N_c$ (in the right side of the critical curve with $\Gamma_2 = N_c$ in Fig. \ref{Fig9}), there is only the apocentric branch (the pericentric branch vanishes), in which the island of resonance is bounded by the dynamical separatrix stemming from the zero-eccentricity saddle point.

\section{Summary and discussion}
\label{Sect6}

Based on the Laplacian expansions of planetary disturbing function, we have formulated two multi-harmonic Hamiltonian models for mean motion resonances and then applied them to the first-order inner and outer resonances with a Jupiter-mass planet. In the first Hamiltonian model, we adopt the usual critical argument, denoted by $\varphi = k\lambda - k_p \lambda_p + (k_p - k)\varpi$, as the resonant angle, while, in the second Hamiltonian model, the angle given by $\sigma = \varphi / k_p$ is taken as the new resonant angle. Based on canonical transformations, the resonant Hamiltonian associated with these two resonant models have been formulated and, in particular, both of them are totally integral (both models are of one degree of freedom).

It is known that, during every libration period of $\sigma$, there is only one point appearing in the Poincar\'e section. This feature allows us to make a direct correspondence between the phase-space structures in the resonant model with $\sigma$ as the resonant angle and the Poincar\'e sections numerically produced in the full model. By plotting the level curves of the resonant Hamiltonian, it is possible for us to produce the associated phase-space structures, from which the global dynamical behaviors of mean motion resonances can be identified.

Some important conclusions of the present work are summarized below.
\begin{itemize}
  \item For the $k_p$:$k$ resonance, it is found that one libration center (or one saddle point) arising in the first resonant model with $\varphi$ as the resonant angle are split into $k_p$ libration centers (or $k_p$ saddle points) in the second resonant model (with $\sigma$ as the resonant angle).
  \item In the phase portraits of the first resonant model, the zero-eccentricity point is not a visible saddle point, but, in the second resonant model, the zero-eccentricity point is a saddle point.
  \item A perfect consistency is found between the phase portraits in the second resonant model and the Poincar\'e sections numerically produced by \citet{malhotra2020divergence} (please compared Fig. \ref{Fig4} in the current work with Fig. 2 in the work of \citet{malhotra2020divergence}). This allows us to identify the location of libration centers and determine the associated resonant widths in an analytical manner.
  \item In the analytical models with $N=2$, the asymmetric libration centers are found in the phase portraits of the first-order inner and outer resonances (these analytical structures with asymmetric libration centers are incorrect because, in the numerical model, there are no asymmetric libration centers). This feature has been found by \citet{beauge1994asymmetric} in the case of the outer 2:3 resonance. According to our simulations, the problem with incorrect topology is caused by the poor approximation of the disturbing function truncated at order $N=2$ in eccentricity (instead of the convergence problem of the Laplacian expansion of disturbing function). In other words, this problem can be avoided if we truncate the disturbing function at a higher order in eccentricity.
  \item The number of stationary points is determined by the motion integral $\Gamma_2$ (i.e., $a_{\max}$ for inner resonances and $a_{\min}$ for outer resonances). There is a critical value of the motion integral denoted by $N_c$ for a certain resonance, at which the stationary point bifurcates. In particular, the saddle points with nonzero eccentricity can be found under the condition of $\Gamma_2 > N_c$, and the zero-eccentricity saddle point exists in the entire range of $\Gamma_2$. The dynamical separatrices stemming from the zero- and/or nonzero-eccentricity saddle points could provide boundaries for libration zones. Thus, the dynamical separatrix stemming from the zero-eccentricity saddle point will never vanish for arbitrary motion integral.
  \item For a first-order resonance, there are two branches in the phase portraits, including the pericentric and apocentric libration zones. In particular, when $\Gamma_2 > N_c$, both the pericentric and apocentric branches of libration centers can be found in the phase portraits and, when $\Gamma_2 \leq N_c$, only the pericentric (or apocentric) branch of libration centers can be found for the inner (or outer) resonances.
  \item As the eccentricity is approaching zero, the centers of the pericentric and apocentric libration zones diverge away from the nominal resonance location for both the first-order inner and outer resonances.
  \item Based on our Hamiltonian model, the resonant widths are analytically determined. For a given motion integral $\Gamma_2$, the associated resonant width can be measured by the variation of $\Gamma_1$, namely $\Delta \Gamma_1$. Alternatively, the resonant width can be equivalently represented by the variation of the semimajor axis ($\Delta a$) and the variation of eccentricity ($\Delta e$). In the present work, two types of representation are adopted to show resonant widths. The first type of presentation used in \citet{malhotra2020divergence} contains only the information of $\Delta a$ (the information of $\Delta e$ is absent). The second type of representation used in \citet{morbidelli2002modern} measures the resonant width along the isoline of $\Gamma_2$, so that this representation contains the information of both $\Delta a$ and $\Delta e$.
  \item As the eccentricity approaches zero, the resonant width in terms of the variation of semimajor axis decreases to zero for both the inner and outer (first-order) resonances. This means that the resonant strength becomes very weak at low eccentricities and, thus, it becomes relatively difficult to capture test particles inside mean motion resonances at low-eccentricity regions.
  \item For the inner resonances including the 2:1, 3:2 and 4:3 resonances, it is interesting to observe that the resonant widths obtained from our analytical resonant model are in perfect agreement with the resonant widths numerically determined in \citet{malhotra2020divergence} by analyzing the Poincar\'e sections. Please compare Fig. \ref{Fig5} in the current work with Fig. 4 in the work of \citet{malhotra2020divergence}.
\end{itemize}

Regarding the first-order inner resonances, \citet{morbidelli2002modern} adopted the usual critical argument $\varphi$ as the resonant angle to formulate the resonant model (the same as the first resonant model discussed in the present study) and they described that, when the motion integral `$N$' (corresponding to $\Gamma_2$ in the present work) is smaller than a threshold value $N_c$, no unstable equilibria and no separatrices are visible. Then, \citet{morbidelli2002modern} stated that, in the case of the 2:1 resonance, one of the two separatrices vanishes under the condition of $\Gamma_2 < N_c$ (or, equivalently, $e \leq 0.2$), so that the resonant widths are not defined at low eccentricities. Please refer to Fig. 9.11 in the textbook of \citet{morbidelli2002modern} for more details.

On the same topic, \citet{ramos2015resonance} also described that, in the case of the 2:1 resonance, there is no outer branch of separatrix for the eccentricity $e$ smaller than 0.18 (see Fig. 2 in their study). Due to the absence of the outer separatrix, \citet{ramos2015resonance} concluded that the motion with semimajor axis $a<a_{\rm res}$ ($a_{\rm res}$ is the semimajor axis of libration center) and $e < 0.18$ will be of the type of circulation.

In addition, for the first-order inner resonances at low eccentricities, the absence of one branch of separatrix can be found in \citet{lemaitre1990origin} (see Fig. 18), in \citet{morbidelli1993secular} (see Figs. 2, 3, 9 and 10 in their work), in \citet{henrard1996note} (see Fig. 1 in his work), in \citet{nesvorny1997asteroidal} (see Figs. 1 and 7 in their work), in \citet{deck2013first} (see Figs. 2 and 3 in their work), in \citet{hadden2018criterion} (see Fig. 12 in their work) and in \citet{beauge2019shannon} (see the first panels of Figs. 10 and 11 in their work).

Concerning the absence of dynamical separatrix for first-order inner resonances at low eccentricities, we hold a different opinion from the previous conclusions. According to the phase portraits shown in Fig. \ref{Fig4} and the resonant widths shown in Figs. \ref{Fig5} and \ref{Fig8}, we can observe that, when $\Gamma_2 > N_c$, both the pericentric and apocentric libration zones exist and both the zero-eccentricity and nonzero-eccentricity saddle points exist, so that the separatrices stemming from these zero-eccentricity and nonzero-eccentricity saddle points provide the boundaries for pericentric and apocentric libration zones. On the other hand, when $\Gamma_2 \le N_c$ (in the low-eccentricity regions), only the pericentric branch of libration centers exists, and only the zero-eccentricity saddle point exists (the nonzero-eccentricity saddle points disappear from the phase portraits), so that the separatrices stemming from the zero-eccentricity saddle point will provide the boundaries for pericentric libration zones. Thus, no matter whether the motion integral is greater than $N_c$ or smaller than $N_c$, the zero-eccentricity point (the coordinate center in the phase portrait) is a visible saddle point of our resonant model. The same discussions can be applied to the first-order outer resonances. To conclude, for the first-order (inner and outer) resonances, the dynamical separatrices bounding the libration zones will never vanish in the total range of motion integral $\Gamma_2$ (or, equivalently, in the entire range of eccentricity). In other words, there are always dynamical separatrices stemming from nonzero- and/or zero-eccentricity saddle points, which provide inner and outer boundaries for libration zones at arbitrary eccentricities.

In addition, from the Poincar\'e sections shown by Fig. 2 of \citet{malhotra2020divergence}, it is observed that the zero-eccentricity point is an unstable fixed point (a saddle point of the dynamical model) regardless of the Jacobi constant (or the motion integral in the present work), so that the separatrix stemming from such a zero-eccentricity point does not vanish at low eccentricities, as pointed out by \citet{malhotra2020divergence}. Thus, in the case of inner resonances, our result obtained from analytical approach is in agreement with that obtained from numerical approach given in \citet{malhotra2020divergence}.

It should be noted that, in the analytical models discussed in this work, only one resonance is considered to dominate the long-term dynamics in a phase space considered (this is a common feature of analytical models). Thus, in those regions where two or more resonances have comparable influences, our analytical models may have some deviation to the exact model. This is why we cannot observe smooth ``bridges'' between adjacent first-order resonances (at low eccentricities, the apocentric libration zone smoothly extends towards the pericentric libration zone of the nearby first-order resonance), as observed in \citet{malhotra2020divergence} by analyzing the Poincar\'e sections. However, in the regime of ``bridge'', it is observed that the resonance strength is very weak, so that it is difficult to capture test particles inside the libration zone. We could understand the regime near ``bridge'' corresponds to an overlapping region between two neighboring first-order resonances, so that the nearby area of ``bridge'' is filled with chaotic motion, as shown by Fig. 5 in \citet{malhotra2020divergence}.

\section*{Acknowledgments}

Hanlun Lei acknowledges helpful discussions with C. Beaug{\'e} about asymmetric libration centers appearing in the resonant model `F2', and Jian Li wishes to thank Prof. Zhihong Jeff Xia for providing an understanding about the stationary points arising in the multi-harmonic Hamiltonian model as period-$k_p$ fixed points in Poincar\'e sections. We also thank the anonymous reviewer and editor for providing useful suggestions. This work is performed with the financial support of the National Natural Science Foundation of China (Nos. 12073011, 11973027, 11933001, 41774038, 11603011), the National Key R\&D Program of China (No. 2019YFA0706601).

\appendix

\section{The motion integral $\Gamma_2$ (or $\Phi_2$)}
\label{A_0}

In \citet{malhotra2020divergence}, the Poincar\'e surfaces of section are characterized by the Jacobi constant, given by (see Eq. 4 in their work)
\begin{equation}\label{A1}
C_J = \frac{\mu}{a} + 2 \sqrt{\mu a (1-e^2)} + O(\mu_p)
\end{equation}
where $\mu = {\cal G} m_0$ and $\mu_p = {\cal G} m_p$. It is noted that the elements shown in Eq. (\ref{A1}) are osculating elements. In our multi-harmonic Hamiltonian models, the disturbing function has been averaged by means of Eq. (\ref{Eq7}), so that the elements used in the resonant model are mean elements. The difference between the osculating and mean elements is on the order of $\mu_p$.

We ignore the difference in the process of discussing the relationship between the Jacobi constant used in \citet{malhotra2020divergence} and the motion integral adopted in this work.

According to Eq. (\ref{A1}), the angular momentum of the test particle can be approximated as
\begin{equation}\label{A2}
\sqrt{\mu a (1-e^2)} \approx \frac{1}{2} C_J - \frac{\mu}{2a}.
\end{equation}
According to the expression of the motion integral $\Gamma_2$ in Eq. (\ref{Eq19}), the angular momentum can be written as
\begin{equation}\label{A3}
\sqrt{\mu a (1-e^2)} = \frac{k_p}{k} \sqrt{\mu a} - \Gamma_2.
\end{equation}
Equating Eq. (\ref{A2}) and Eq. (\ref{A3}) leads to the relationship between $C_J$ and $\Gamma_2$ as follows:
\begin{equation}\label{A4}
C_J + 2\Gamma_2  \approx  2 \frac{k_p}{k} \sqrt{\mu a} + \frac{\mu}{a}.
\end{equation}
According to the resonant Hamiltonian represented by Eq. (\ref{Eq18}), we have
\begin{equation}\label{A5}
\frac{k_p}{k} \sqrt{\mu a} + \frac{\mu}{2a} = - {\cal H}^* - {\cal R}^*,
\end{equation}
where the resonant disturbing function ${\cal R}^*$ is given by Eq. (\ref{Eq8}) or (\ref{Eq15}). Substituting Eq. (\ref{A5}) into Eq. (\ref{A4}) yields
\begin{equation}\label{A6}
C_J + 2\Gamma_2 + 2 {\cal H}^* \approx  - 2 {\cal R}^*,
\end{equation}
It is known that the magnitude of ${\cal R}^*$ is on the order of $\mu_p$. Thus, we have the following relationship:
\begin{equation}\label{A7}
C_J = - 2\Gamma_2 - 2 {\cal H}^* + O(\mu_p),
\end{equation}
which tells us that the Jacobi constant adopted by \citet{malhotra2020divergence} is an approximated linear combination of $\Gamma_2$ and ${\cal H}^*$. In our multi-harmonic Hamiltonian model, both the motion integral $\Gamma_2$ and the resonant Hamiltonian ${\cal H}^*$ are conserved quantities.

\section{Phase portraits in different resonant models}
\label{A_1}

\begin{figure*}
\centering
\includegraphics[width=0.33\textwidth]{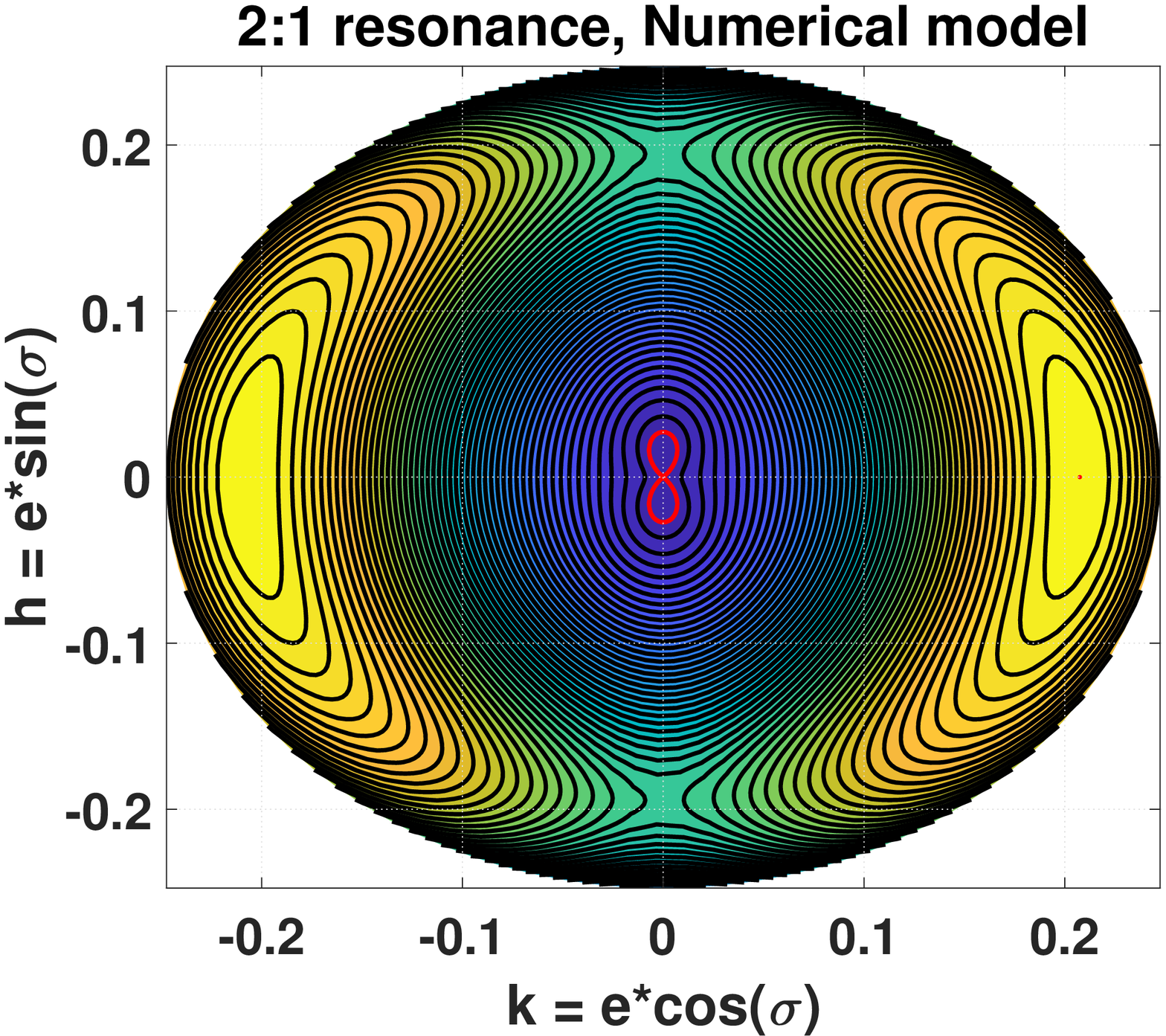}
\includegraphics[width=0.33\textwidth]{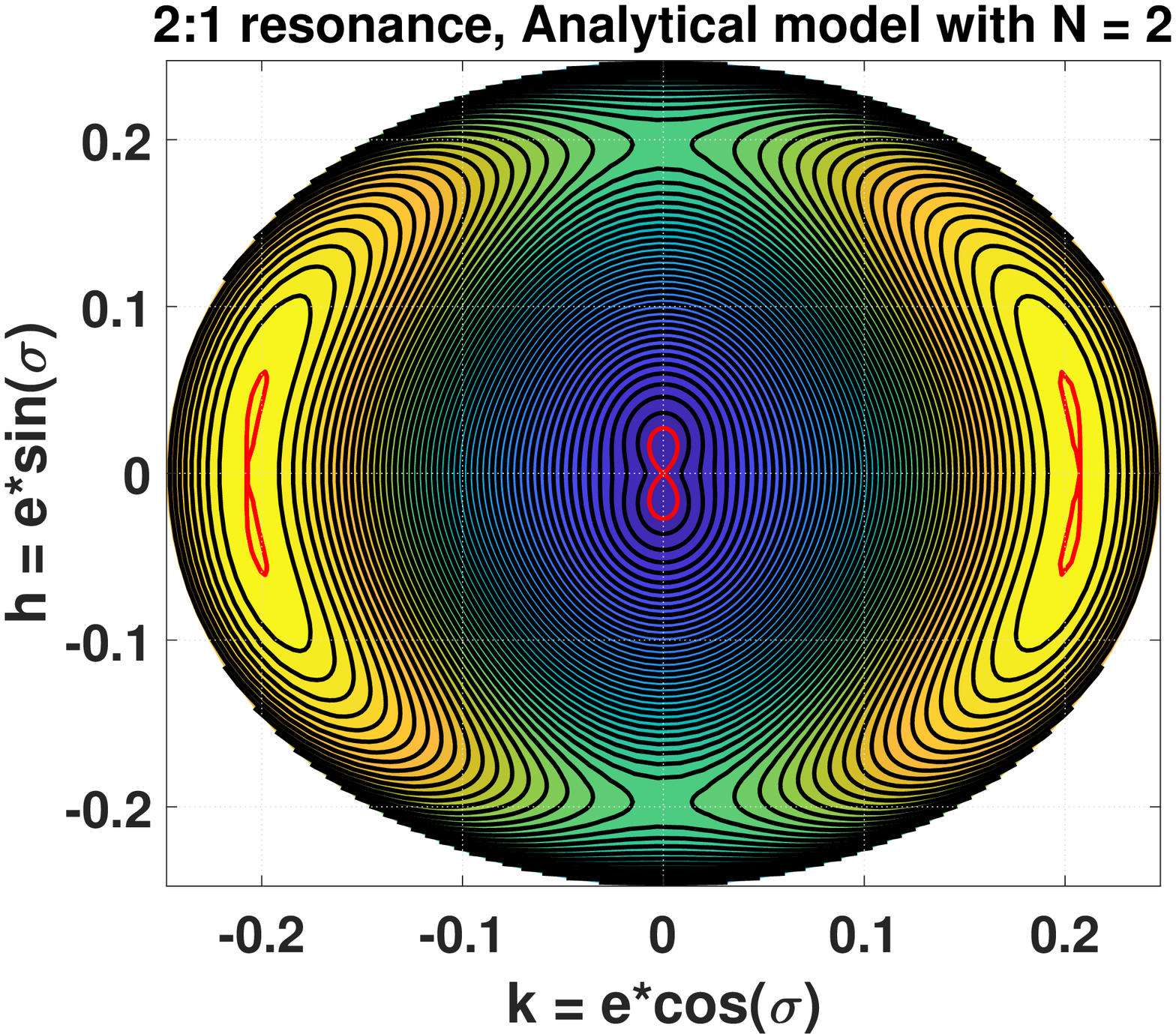}
\includegraphics[width=0.33\textwidth]{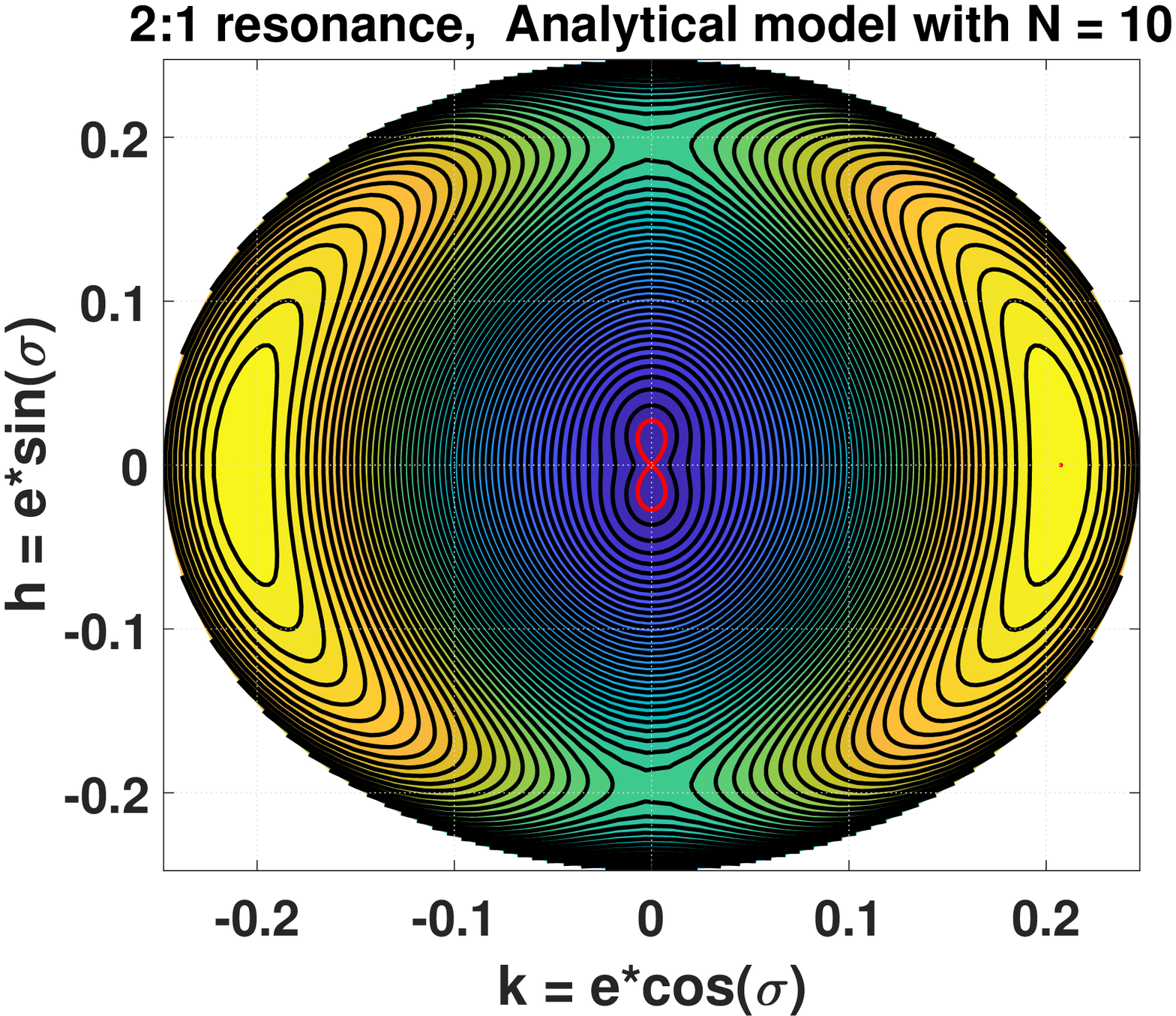}\\
\includegraphics[width=0.33\textwidth]{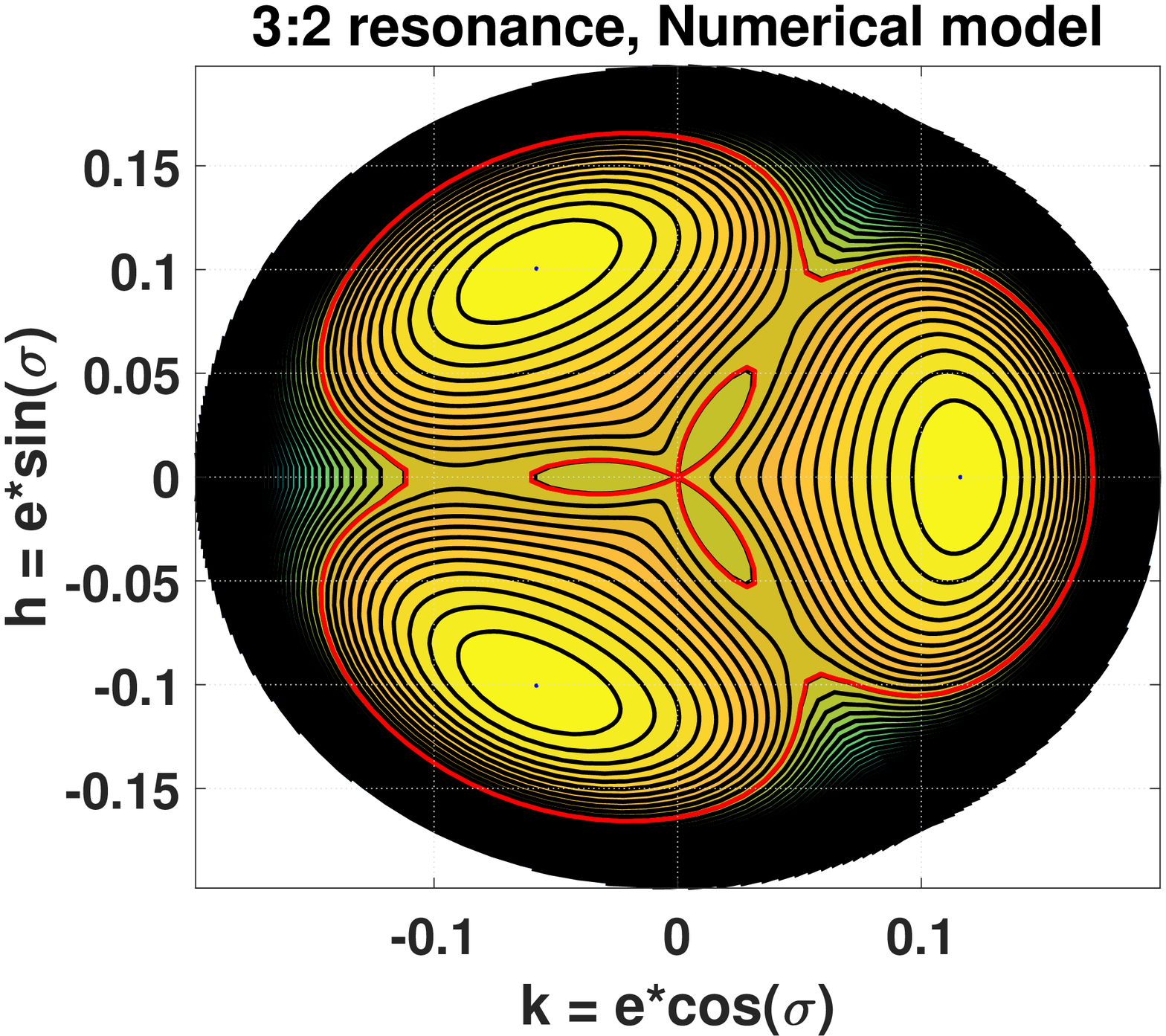}
\includegraphics[width=0.33\textwidth]{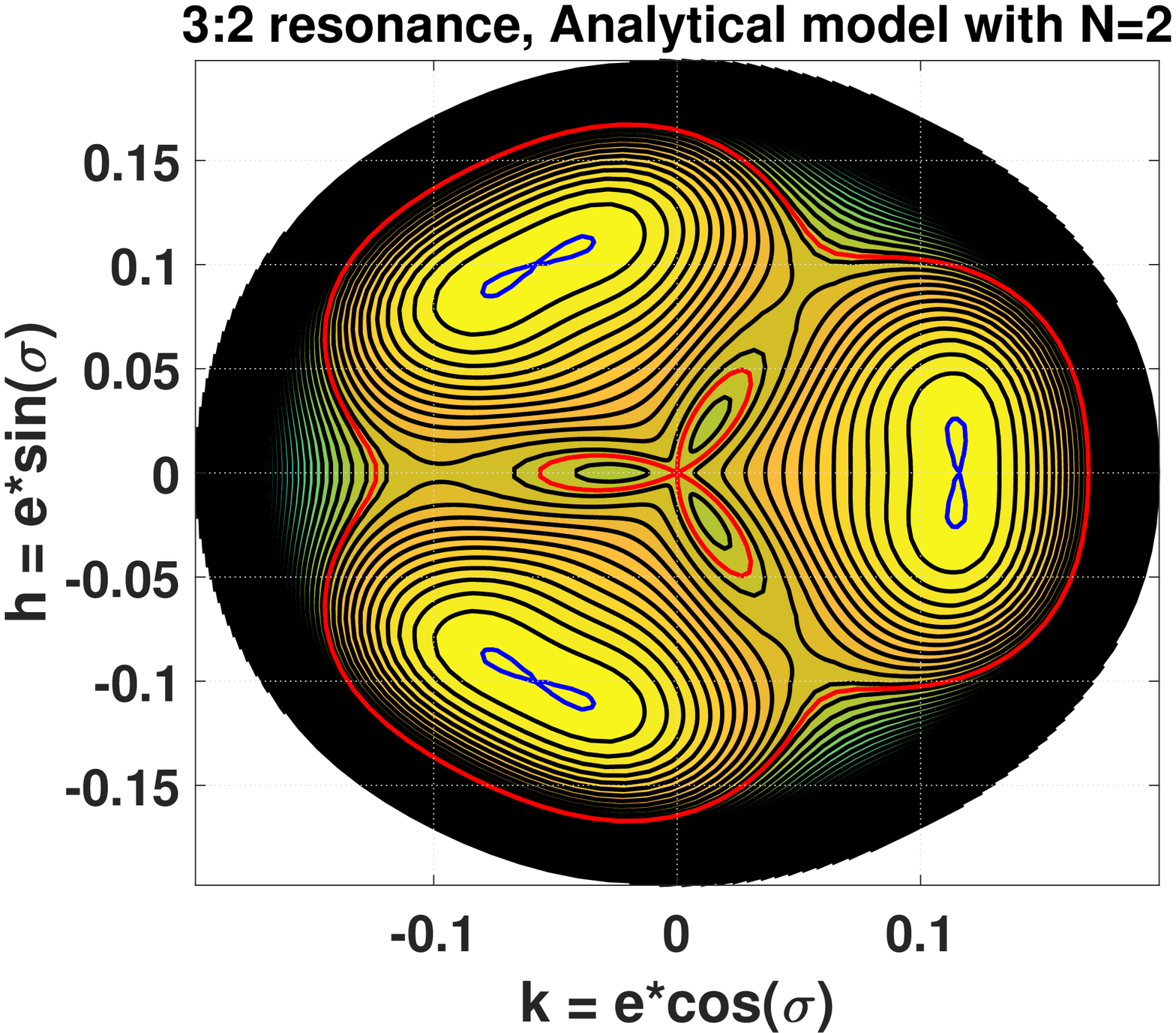}
\includegraphics[width=0.33\textwidth]{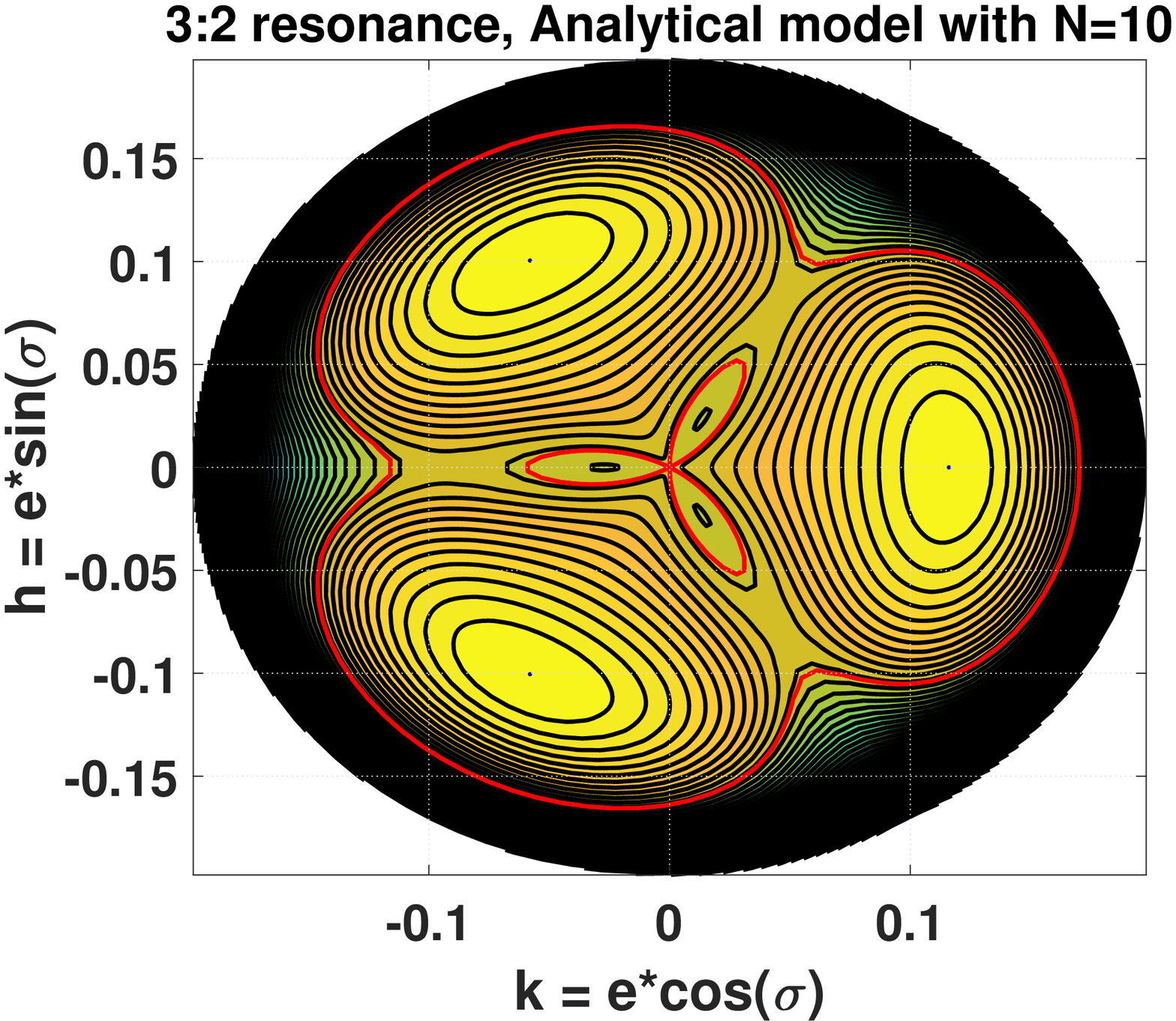}\\
\includegraphics[width=0.33\textwidth]{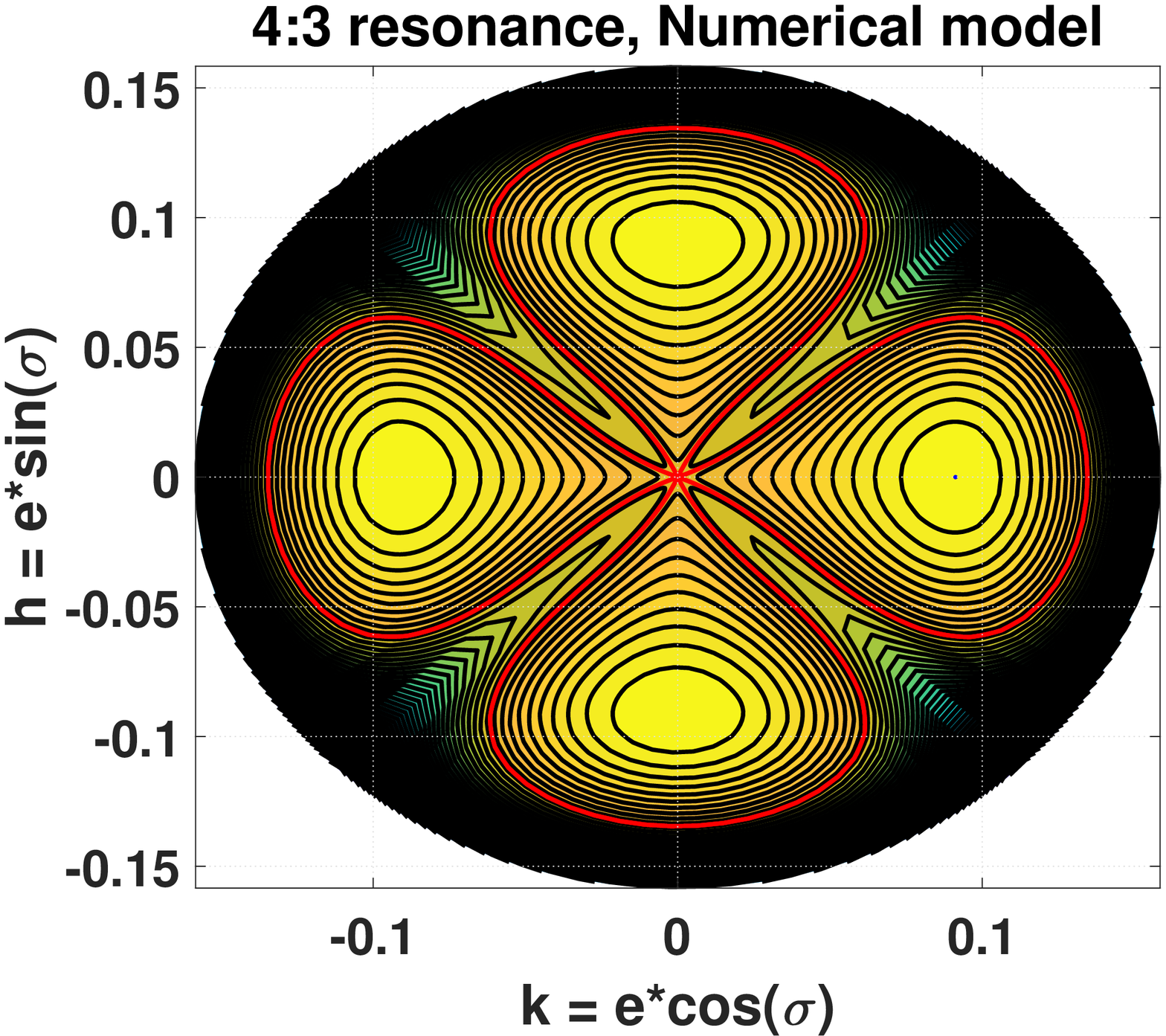}
\includegraphics[width=0.33\textwidth]{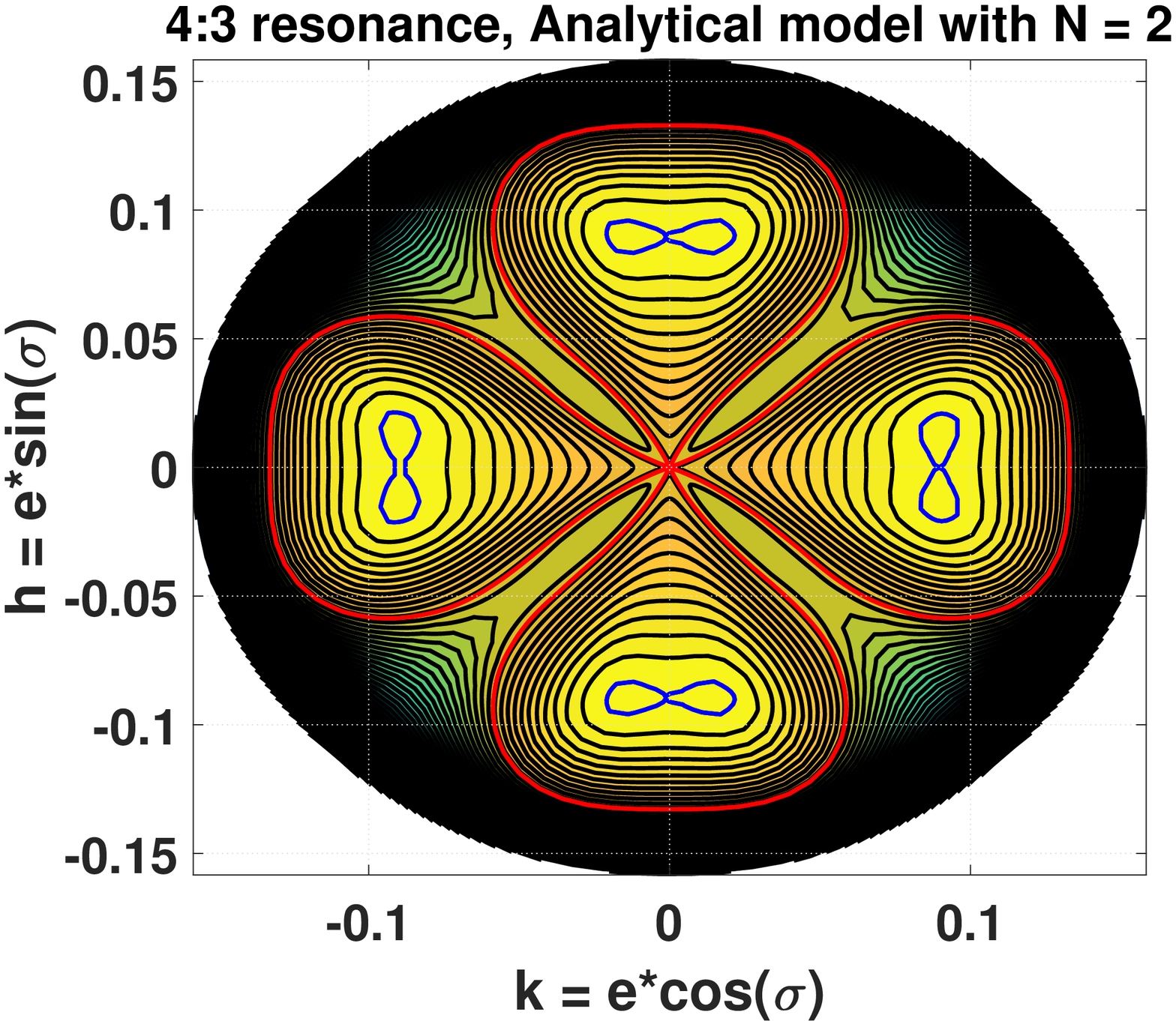}
\includegraphics[width=0.33\textwidth]{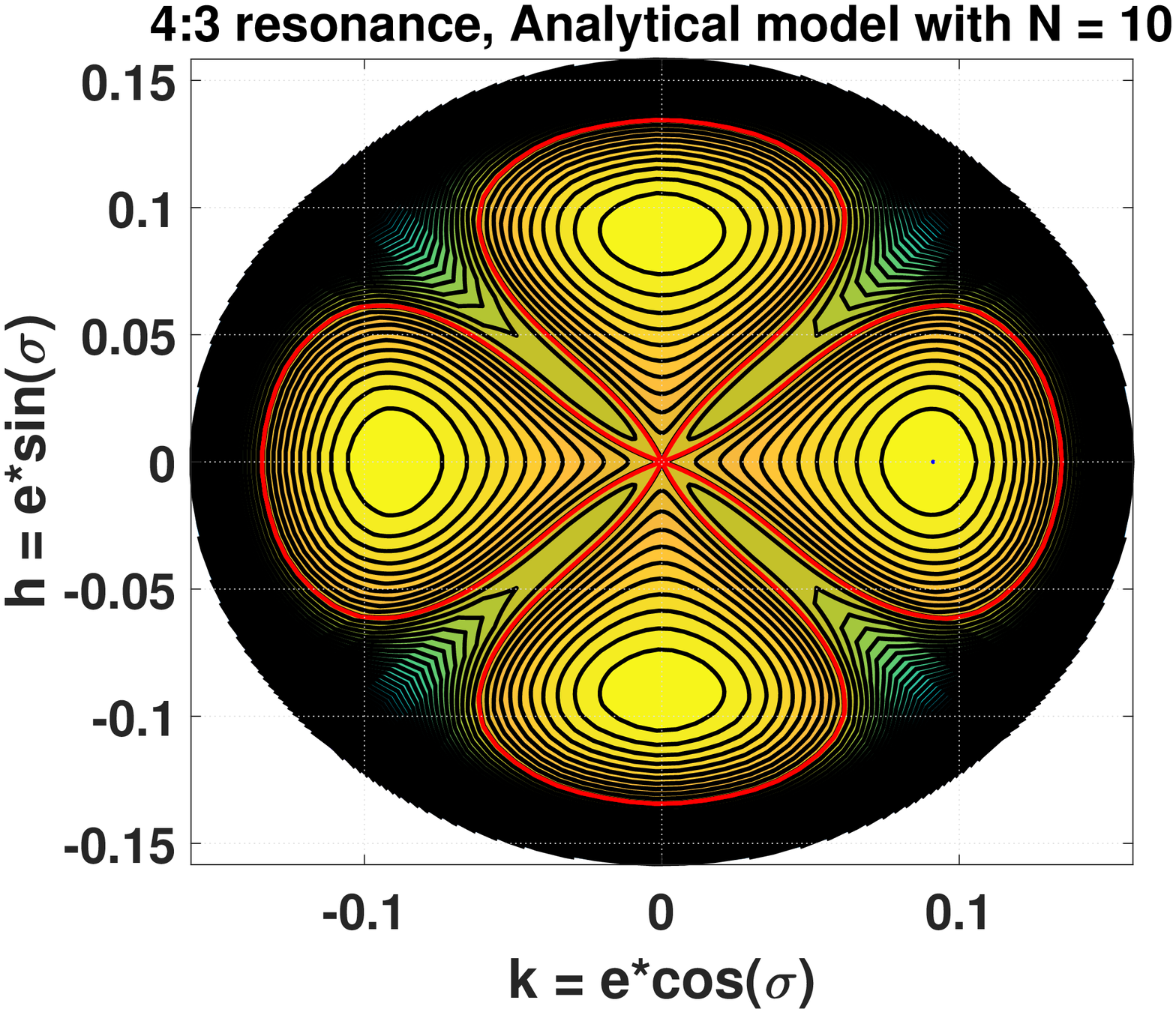}
\caption{Level curves of the resonant Hamiltonian associated with the inner 2:1 resonance specified by $\Gamma_2 = 0.81$ (the panels in the top row), the 3:2 resonance specified by $\Gamma_2 = 0.4419873$ (the panels in the middle row) and the 4:3 resonance specified by $\Gamma_2 = 0.306$ (the panels in the bottom row). The panels in the left column correspond to the numerical model, the panels in the middle column are for the analytical models with $N=2$, and the panels in the right column are for the analytical models with $N=10$. In the analytical models with $N=2$, asymmetric libration centres are observed. Evidently, the structures in the analytical model with $N=10$ are identical to the ones in the numerical model.}
\label{FigA1}
\end{figure*}

The phase-space structures in the numerical model and in the analytical models with $N=2$ and $N=10$ are presented in Fig. \ref{FigA1} for the 2:1, 3:2 and 4:3 resonances. The phase portraits associated with the 2:1 resonance is specified by $\Gamma_2 = 0.81$, the ones associated with the 3:2 resonance is specified by $\Gamma_2 = 0.4419873$ and the ones corresponding to the 4:3 resonance is specified by $\Gamma_2 = 0.306$. It is observed from Fig. \ref{FigA1} that, for all the inner resonances considered, asymmetric libration centers appear in the phase portraits produced from the analytical model with $N=2$. However, the asymmetric libration centers disappear from the phase portraits in both the numerical model and the analytical model with $N=10$.

\section{Phase portraits of mean motion resonances}
\label{A_2}

\begin{figure*}
\centering
\includegraphics[width=0.33\textwidth]{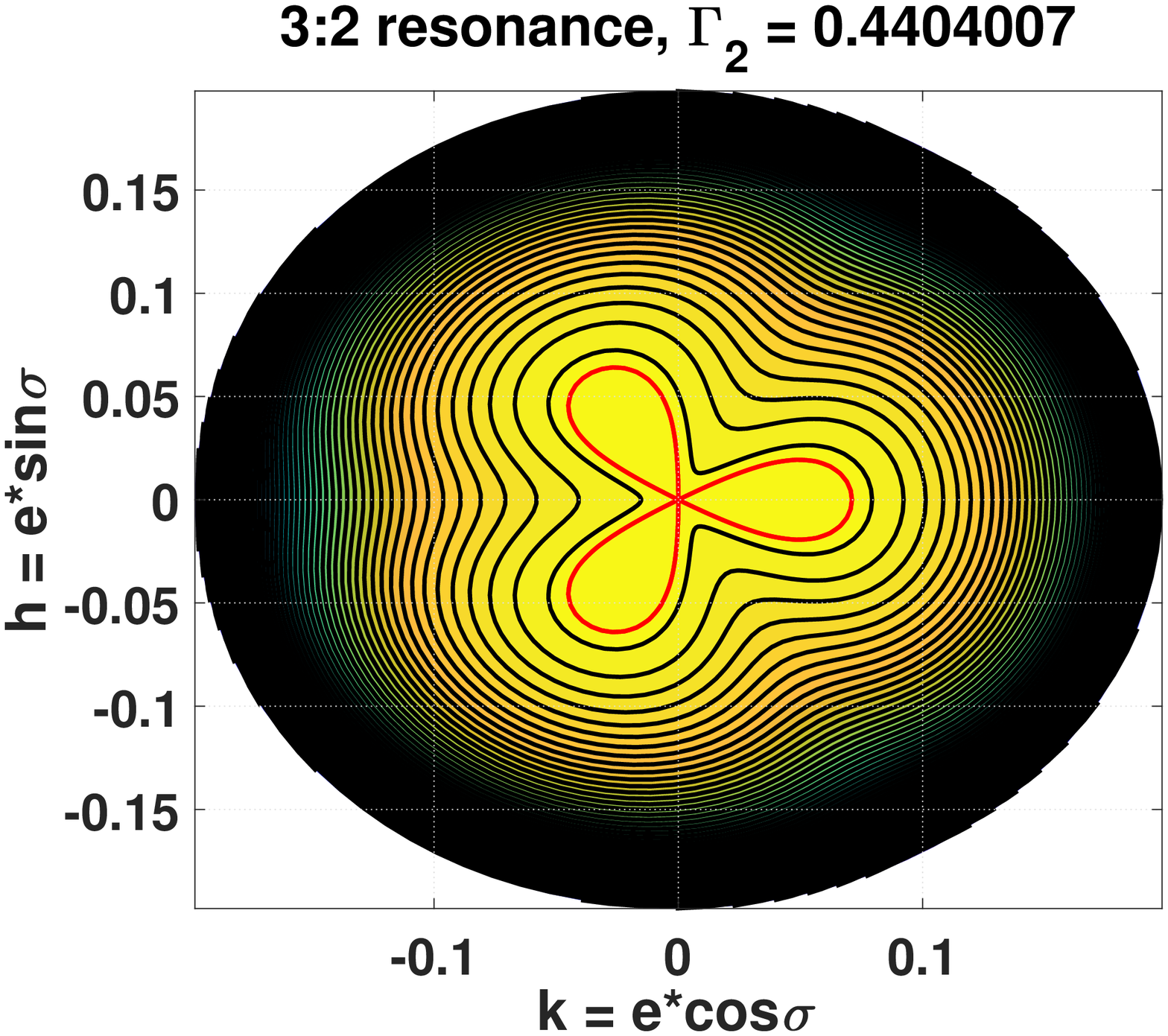}
\includegraphics[width=0.33\textwidth]{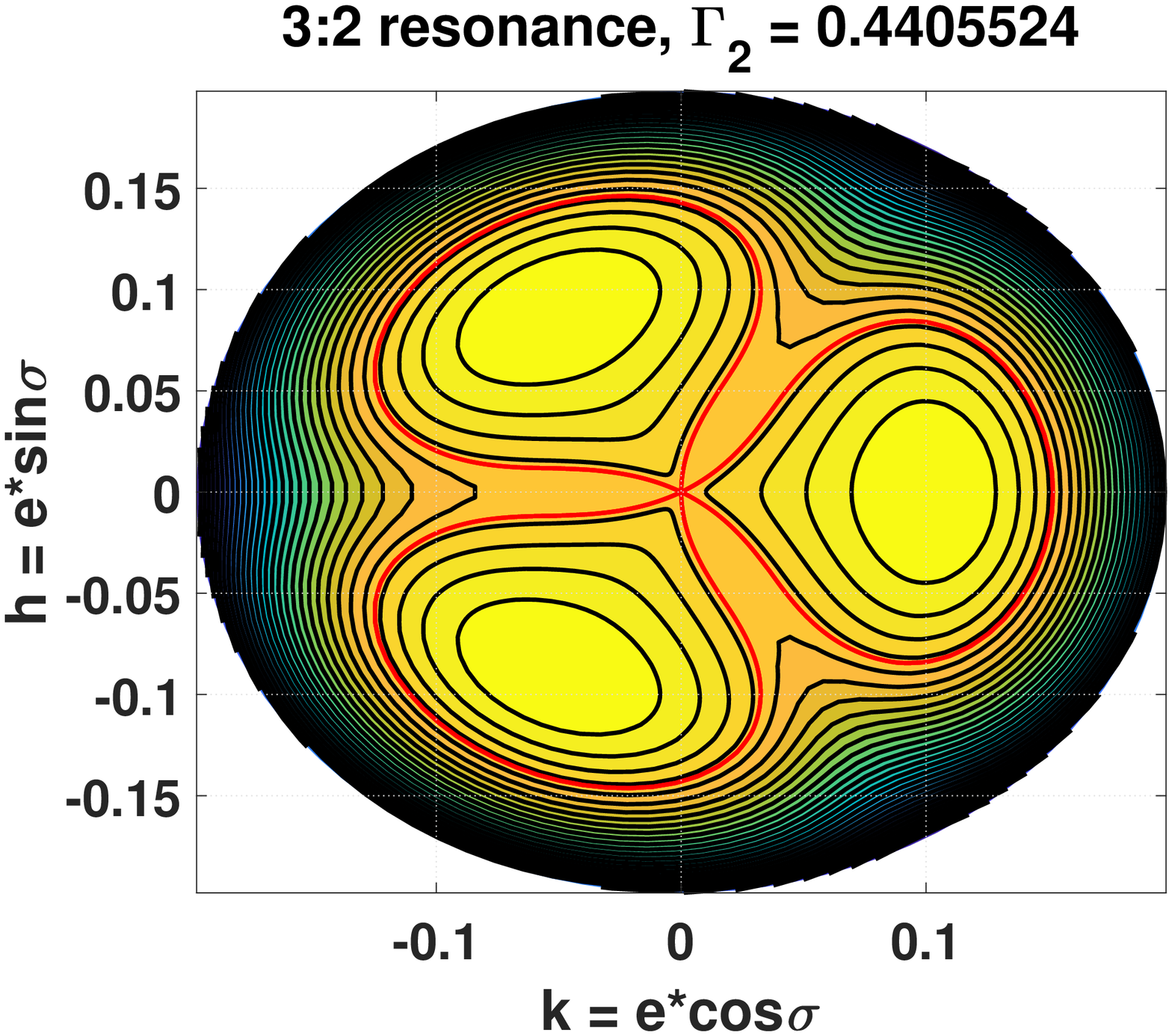}
\includegraphics[width=0.33\textwidth]{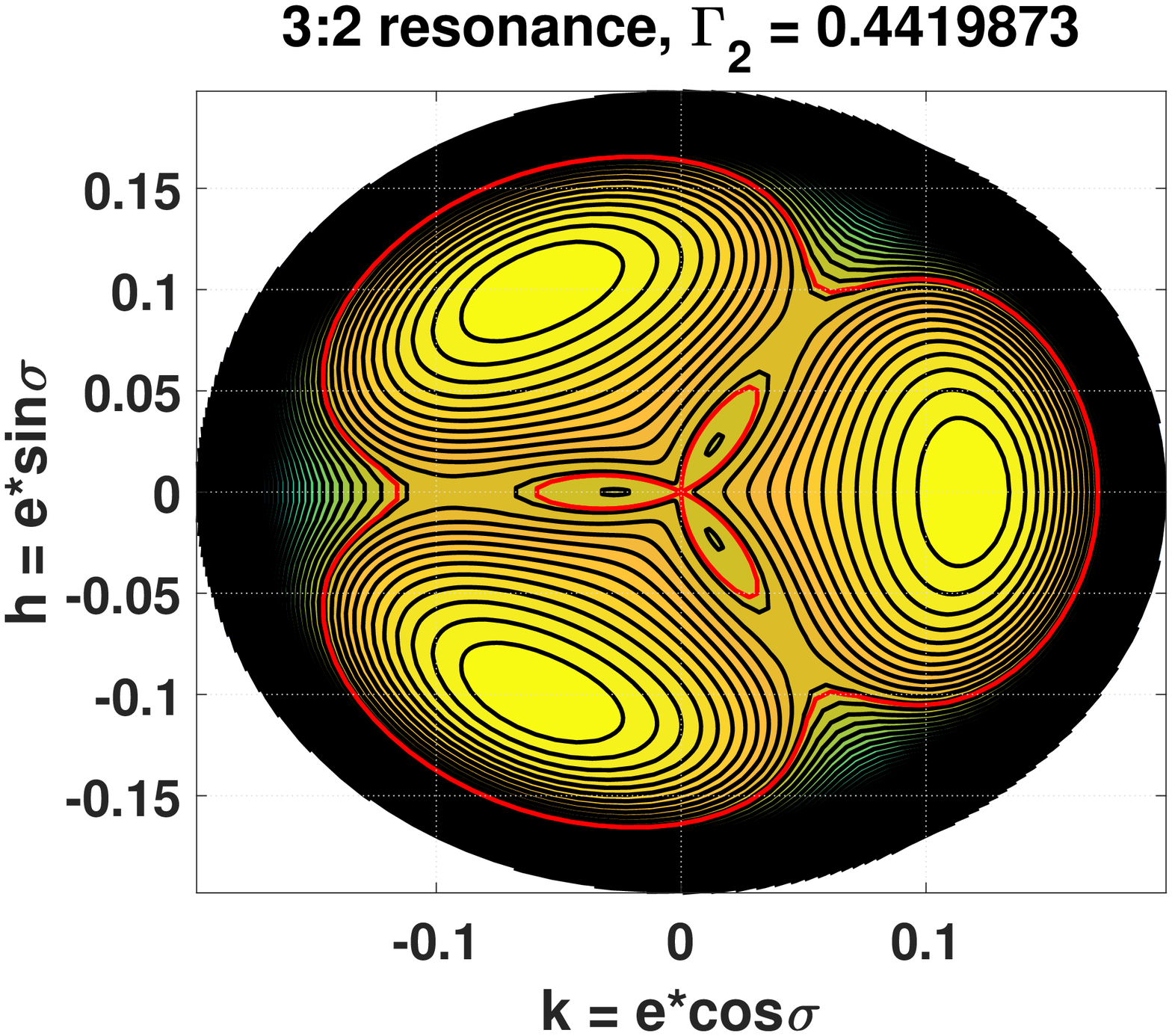}
\caption{Phase-space structures for the inner 3:2 resonances with three different values of the motion integral.}
\label{FigA2}
\end{figure*}

\begin{figure*}
\centering
\includegraphics[width=0.33\textwidth]{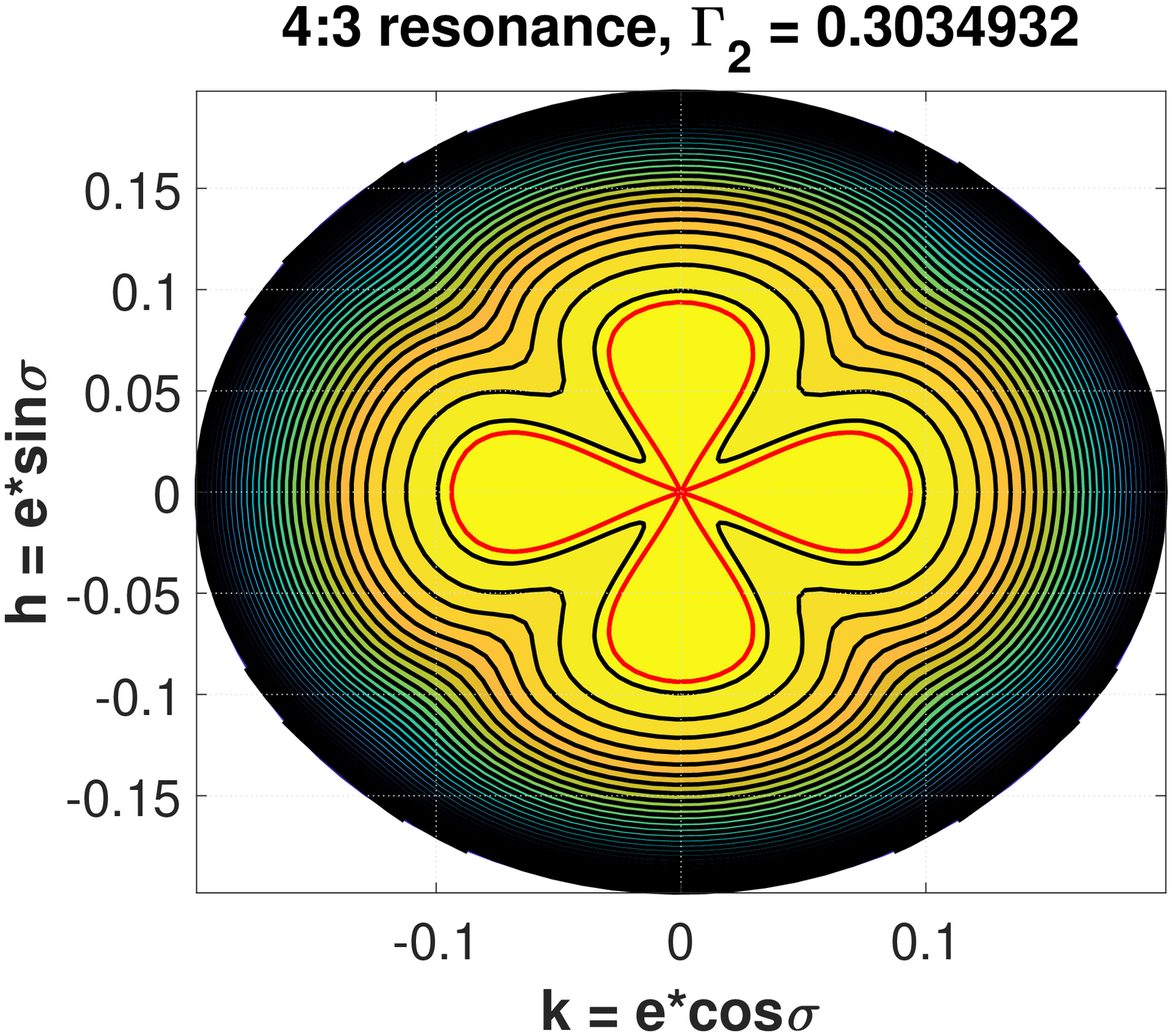}
\includegraphics[width=0.33\textwidth]{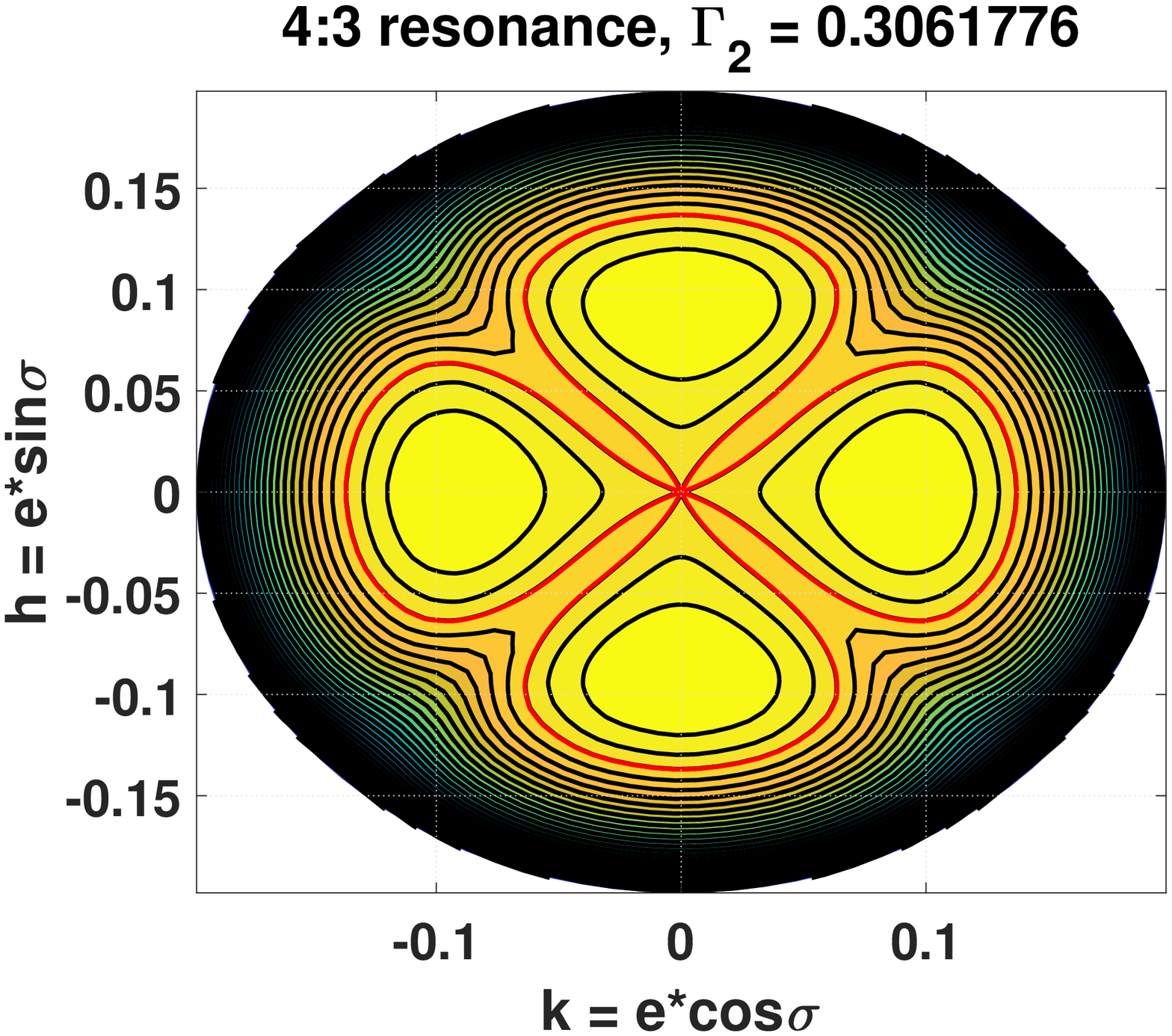}
\includegraphics[width=0.33\textwidth]{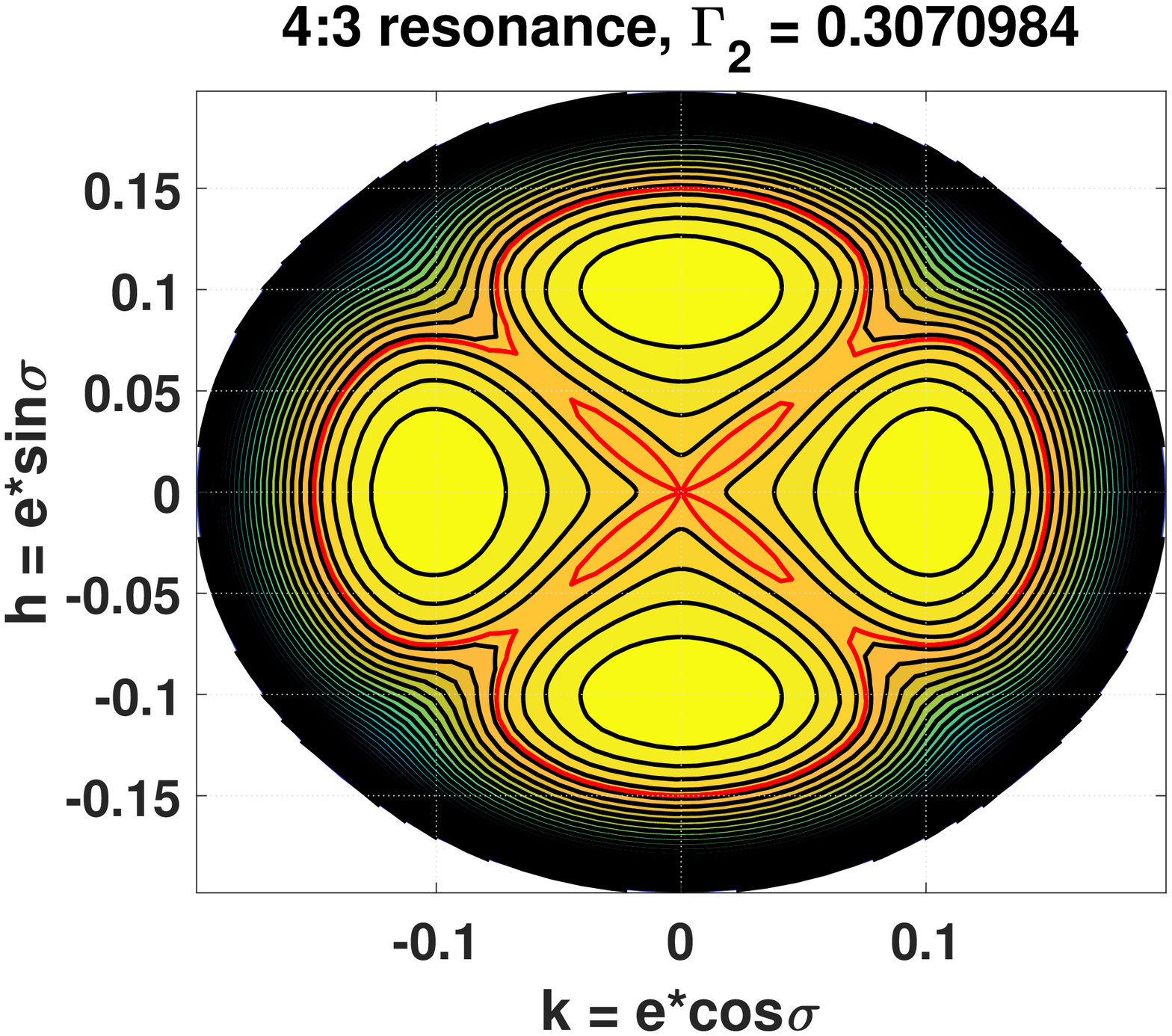}
\caption{Phase-space structures for the inner 4:3 resonances with three different values of the motion integral.}
\label{FigA3}
\end{figure*}

\begin{figure*}
\centering
\includegraphics[width=0.33\textwidth]{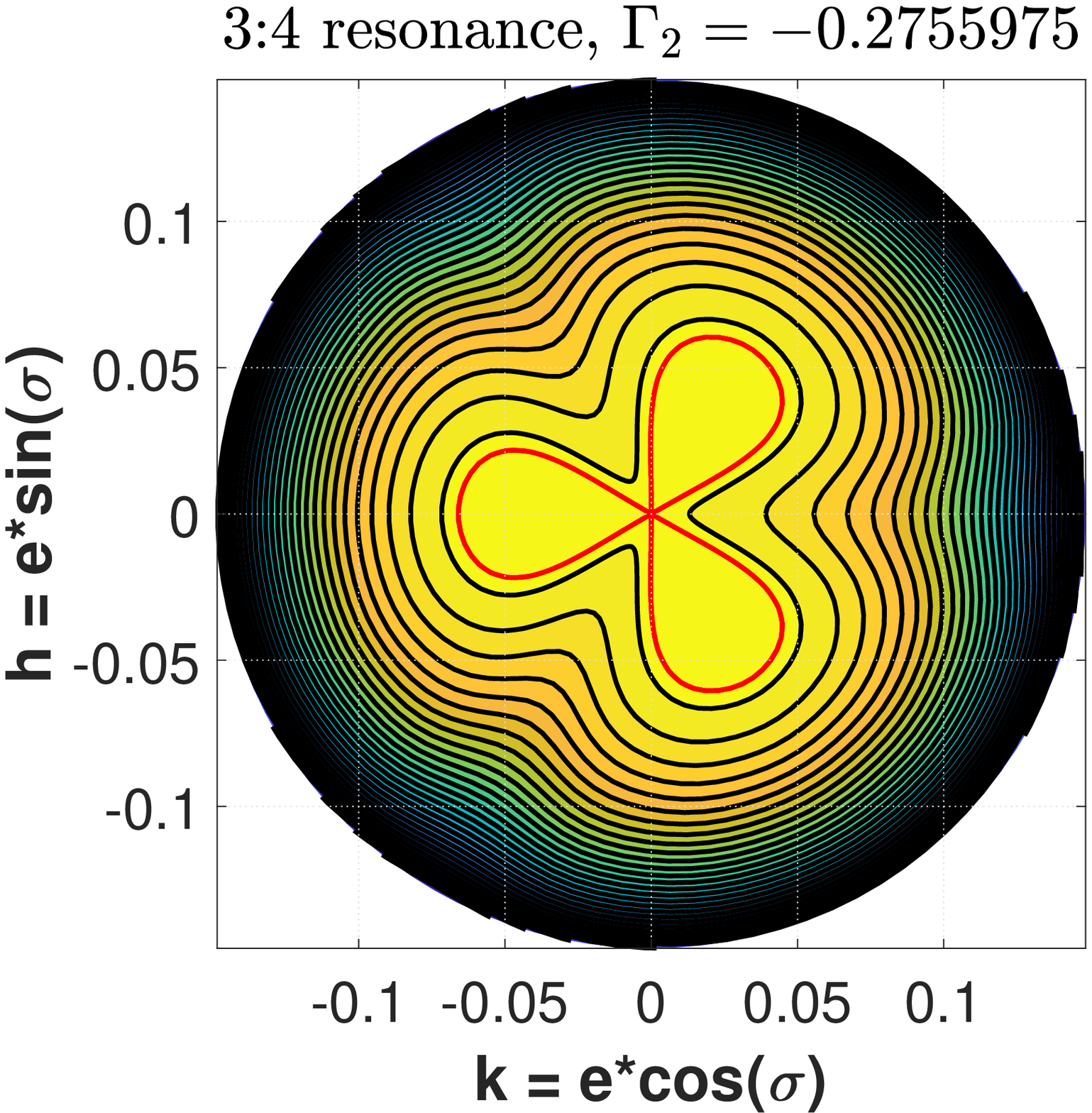}
\includegraphics[width=0.33\textwidth]{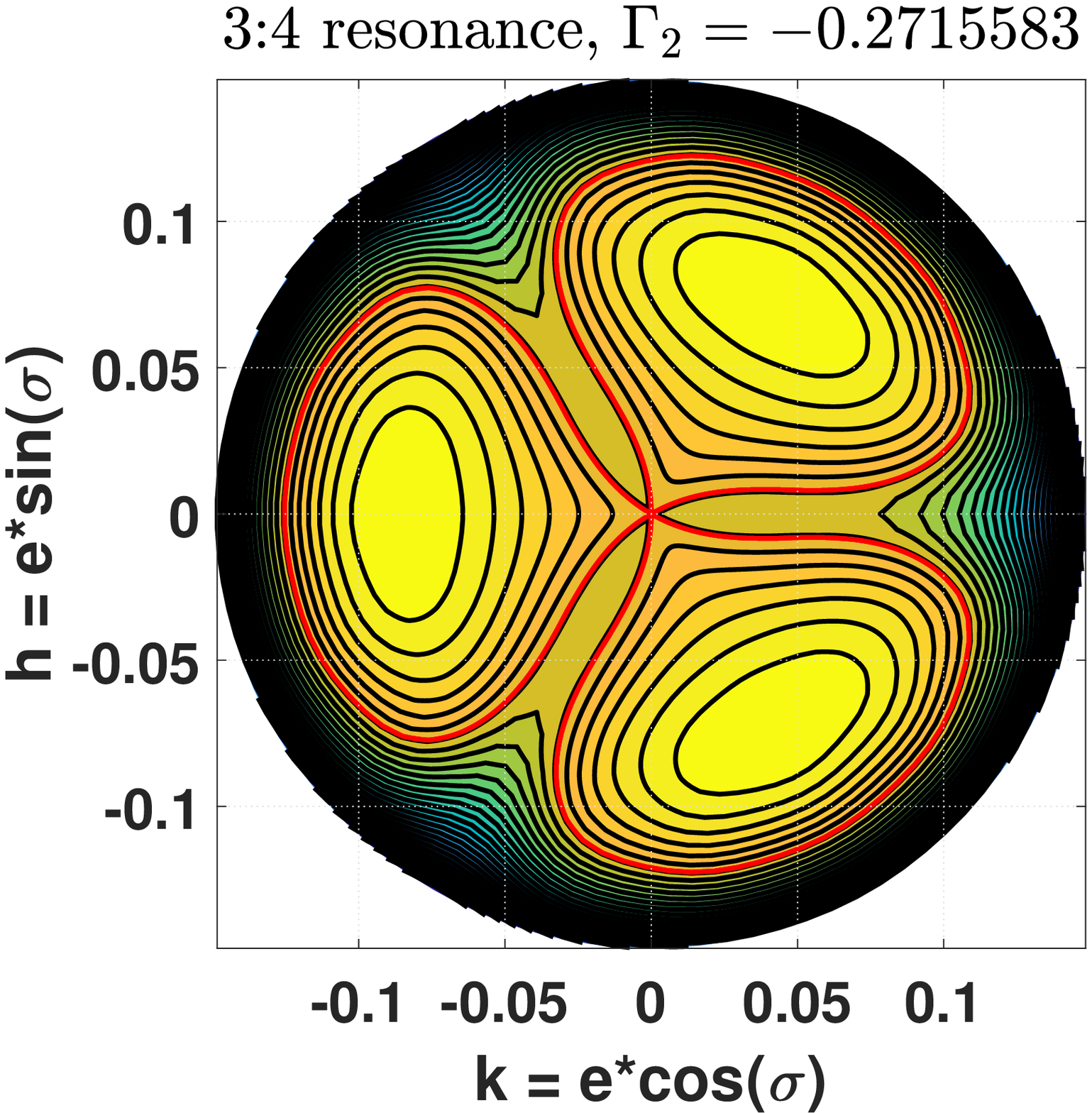}
\includegraphics[width=0.33\textwidth]{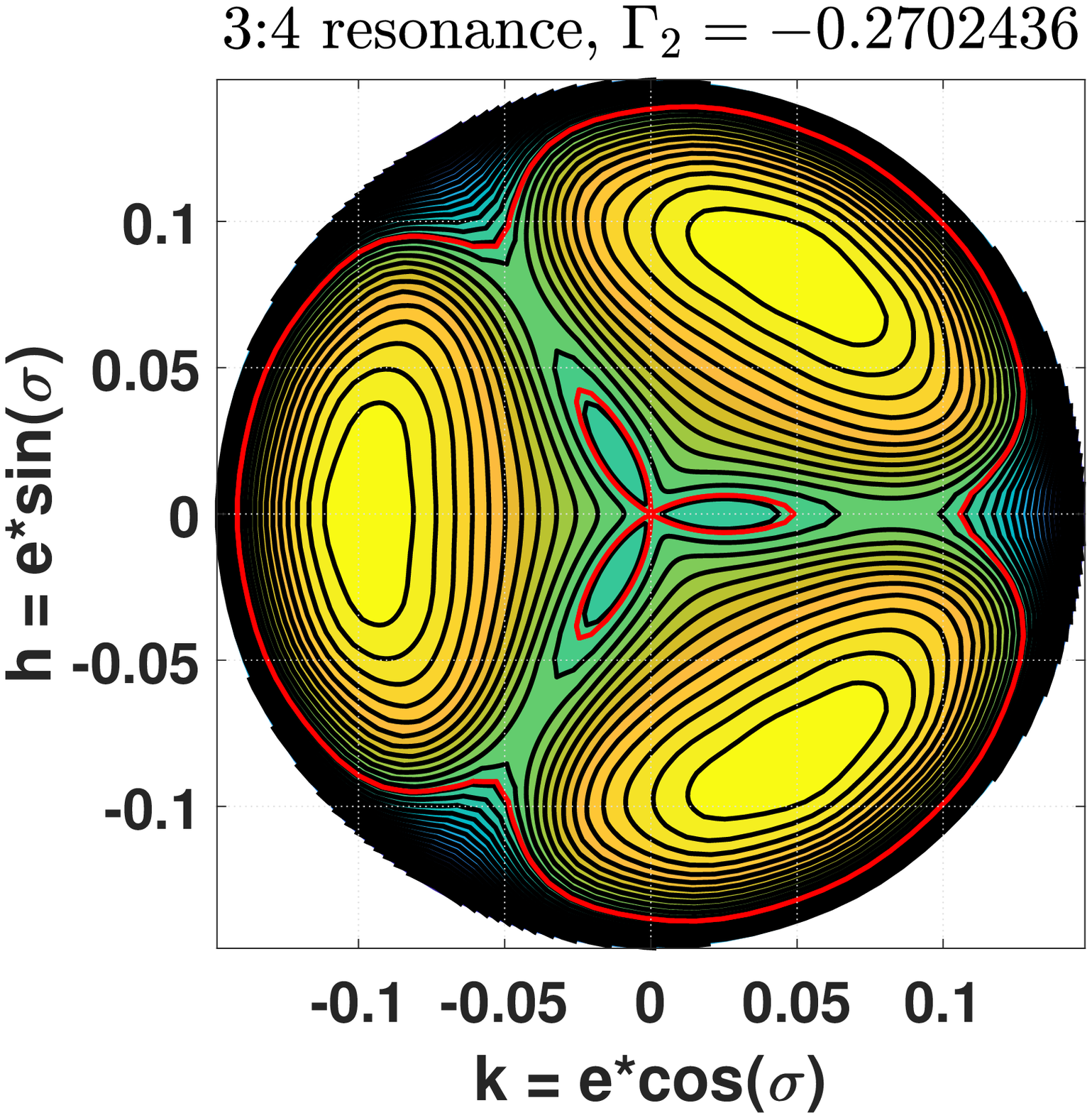}
\caption{Phase-space structures for the outer 3:4 resonances with three different values of the motion integral.}
\label{FigA4}
\end{figure*}

Based on the second Hamiltonian model with $\sigma$ as the resonant angle, phase-space structures with three values of motion integral are reported in Fig. \ref{FigA2} for the inner 3:2 resonance, in FIg. \ref{FigA3} for the inner 4:3 resonance and in Fig. \ref{FigA4} for the outer 3:4 resonance.

\section*{Data availability}
The data underlying this article are available in the article and in its online supplementary material.

\bibliographystyle{mn2e}
\bibliography{mybib}

\bsp
\label{lastpage}
\end{document}